\RequirePackage{silence}
\documentclass[11pt]{article}
\usepackage[letterpaper, portrait, margin=1in]{geometry}
\usepackage{amsmath}
\usepackage{amssymb}
\usepackage{amsthm}
\usepackage{mathtools}
\usepackage{mathrsfs}
\usepackage{thmtools}
\usepackage[noend]{algpseudocode}

\usepackage[dvipsnames]{xcolor}
\usepackage{graphicx}
\usepackage{pgf}
\usepackage{subcaption}
\usepackage{microtype} % Improves general justification and spacing

\usepackage{algorithm}
\usepackage[indLines=true]{algpseudocodex}

\usepackage{lmodern}
\usepackage{anyfontsize}
\usepackage{enumitem}
\usepackage{capt-of}
    
\usepackage{natbib}

\newcommand{\red}[1]{\textcolor{red}{#1}}

\usepackage{xspace}

\usepackage{minitoc,todonotes,tcolorbox}
\usepackage{soul}
\graphicspath{{figures/}}
\usepackage{nicefrac}
\usepackage{lineno}
\newcommand{\pr}{\mathbb{P}}
\newcommand{\vol}{\operatorname{Vol}}
\newcommand{\E}{\mathbb{E}}
\newcommand{\R}{\mathbb{R}}

\newcommand{\pois}{\mathtt{Pois}}

\newcommand{\bigO}{\mathcal{O}}
\newcommand{\define}{:=}

\newcommand{\err}{\varepsilon}
\newcommand{\conf}{\delta}
\newcommand{\length}{M}
\newcommand{\body}{K}
\newcommand{\stream}{\mathcal{K}}
\newcommand{\inrad}{r}
\newcommand{\circrad}{R}
\newcommand{\empinrad}{d_0}
\newcommand{\dm}{n}

\newcommand{\unionvol}{F_0}
\newcommand{\goodunionvol}{F_0'}

\newcommand{\core}{\body_\cap}
\newcommand{\corevol}{V_\cap}

\newcommand{\bucket}{\mathcal{X}}
\newcommand{\bucketsize}{|\mathcal{X}|}
\newcommand{\thresh}{\texttt{threshold}}
\newcommand{\probthresh}{\mathcal{L}}

\newcommand{\sampgood}{\eta}
\newcommand{\sampbad}{\beta}

\newcommand{\volerr}{\kappa}
\newcommand{\volconf}{\delta_v}
\newcommand{\prob}{p_\mathrm{final}}
\newcommand{\empvol}{\widetilde{V}}

\newcommand{\ttc}{\texttt{ttc}\xspace}
\newcommand{\algo}{\hyperref[algo:alpona]{\texttt{Alpona}\xspace}}
\newcommand{\gensamp}{\textsc{GenerateSamples}\xspace}
\newcommand{\compvol}{\textsc{ComputeVolume}\xspace}

\newcommand{\algomod}{{\texttt{Alpona-modified}}}

\newcommand{\threshval}{\left\lceil\frac{576}{\err^2}\log\frac{16\length}{\conf}\right\rceil}

\newcommand{\volerrval}{\frac{\err}{32}}
\newcommand{\volconfval}{\frac{\conf}{4\length}}
\newcommand{\sampgoodval}{\frac{\err}{32}}
\newcommand{\sampbadval}{\frac{\err}{32\length}}

\newcommand{\timevolcv}{\frac{\max\left\{\dm^3, \dm^2 (\circrad / \inrad)^2\right\}}{\err^2} \log^2\dm \log^2 \frac{32}{\err} \log^2\frac{32\dm}{\err} \log\frac{4\length}{\conf}}

\newcommand{\timesamplv}{\dm^3 (\circrad / \inrad)^2 \left[ \log\frac{2\circrad}{\empinrad} + \log\frac{32}{\err} + \log\length \right]}
\newcommand{\timesampkv}{\dm^3 (\circrad / \inrad)^2 \left[ \log\frac{32\circrad}{\empinrad} + \log\frac{32}{\err} + \log\length\right]}
\newcommand{\timesampus}{\dm^2 (\circrad / \inrad)^2 \left[ \dm\log\frac{32\circrad}{\empinrad} + \log\frac{32}{\err} + \log\length \right]}

\newcommand{\thr}{\texttt{threshold}}
\newcommand{\gooderr}{\frac{15\err}{16}}

\newcommand{\TV}{\operatorname{TV}}
\newcommand{\dist}{\operatorname{dist}}

\newcommand{\last}{\hat{K}}
\newcommand{\bad}{\text{Bad}}
\newcommand{\bd}[1]{\last_{#1}\cap \bad_{#1}}
\newcommand{\good}{\text{Good}}
\newcommand{\gd}[1]{\last_{#1}\backslash \bad_{#1}}
\newcommand{\distr}{\mathcal{D}}
\newcommand{\density}{\rho}
\newcommand{\unif}{\pi}

\newcommand{\mean}{\Lambda}
\newcommand{\meangood}{\Lambda^Y}
\newcommand{\meanbad}{\Lambda^Z}

\newcommand{\out}{\text{Out}}

\newcommand{\estgood}{\mathrm{Est}_{\mathrm{good}}}
\newcommand{\estbad}{\mathrm{Est}_{\mathrm{bad}}}

\newcommand{\failvol}{\textbf{Fail}_\text{vol}}
\newcommand{\failprob}{\textbf{Fail}_{\text{prob}}}
\newcommand{\failgood}{\textbf{Fail}_{\text{good}}}
\newcommand{\failbad}{\textbf{Fail}_{\text{bad}}}

\newcommand{\tempmean}[1]{\mean_{(#1)}}
\newcommand{\tempbucket}[1]{\bucket_{(#1)}}
\newcommand{\tempprobthresh}[1]{\probthresh_{(#1)}}
\newcommand{\tempfailprob}[1]{\failprob^{(#1)}}

\newcommand{\freqbound}{\tau}

\theoremstyle{plain}
\newtheorem{theorem}{Theorem}

\newtheorem{lemma}[theorem]{Lemma}
\newtheorem{corollary}[theorem]{Corollary}
\newtheorem{problem}{Problem}
\newtheorem*{conjecture}{Conjecture}

\theoremstyle{definition}
\newtheorem{definition}{Definition}

\newtheorem{question}{Question}
\newtheorem{cl}[theorem]{Claim}

\theoremstyle{remark}
\newtheorem{remark}{Remark}
\newtheorem{properties}{Properties}

\newcommand{\Var}{\operatorname{Var}}

\newcommand{\diam}{\operatorname{diam}}

\newcommand{\one}{\mathbf{1}}

\newcommand{\norm}[1]{\lVert #1\rVert}

\newcommand{\dd}{\,\mathrm{d}}
\newcommand{\dimension}{n}
\newcommand{\nbodies}{M}
\newcommand{\uve}{Union Volume Estimation}

\newcommand{\har}{Hit-and-Run}

\usepackage[
    colorlinks=true,
    linkcolor=RoyalBlue,
    citecolor=ForestGreen,
    urlcolor=MidnightBlue,
    filecolor=Mulberry,
    pdfpagelabels, unicode=true, psdextra
]{hyperref}

\usepackage[capitalize, noabbrev, nameinlink]{cleveref}

\crefname{thm}{Theorem}{Theorems}
\Crefname{thm}{Theorem}{Theorems}

\crefname{lem}{Lemma}{Lemmas}
\Crefname{lem}{Lemma}{Lemmas}

\crefname{proposition}{Proposition}{Propositions}
\Crefname{proposition}{Proposition}{Propositions}

\crefname{corollary}{Corollary}{Corollaries}
\Crefname{corollary}{Corollary}{Corollaries}

\crefname{cl}{Claim}{Claims}
\Crefname{cl}{Claim}{Claims}

\crefname{p}{Problem}{Problems}
\Crefname{p}{Problem}{Problems}

\crefname{defn}{Definition}{Definitions}
\Crefname{defn}{Definition}{Definitions}

\crefname{algorithm}{Algorithm}{Algorithms}
\Crefname{algorithm}{Algorithm}{Algorithms}
\title{Estimating the Volume of a Union of Convex Bodies}
\author{Eeshan Banerjee\thanks{Indian Statistical Institute, Kolkata, India. Email: eeshanbanerjee02@gmail.com}    \and
  Aranya Kumar Bal\thanks{Indian Statistical Institute, Kolkata, India. Email: hayatea90@gmail.com} 
  \and Debabrota Basu\thanks{Univ. Lille, Inria, CNRS, Centrale Lille, UMR 9189 – CRIStAL, France. Email: debabrota.basu@inria.fr}
    \and
  Sourav Chakraborty\thanks{Indian Statistical Institute, Kolkata, India. Email: chakraborty.sourav@gmail.com} 
  \and 
  Debarshi Chanda\thanks{Indian Statistical Institute, Kolkata, India. Email: debarshi.chanda.1997@gmail.com}
  \and
  Uddalok Sarkar\thanks{Indian Statistical Institute, Kolkata, India. Email: uddaloksarkar@gmail.com}
}
\date{}

\newcommand{\deb}[1]{\todo[inline,color=white]{\textcolor{teal}{\textbf{DB: #1}}}}

\newcommand{\highlight}[1]{{\color{red} #1}}

\begin{document}

\maketitle

\begin{abstract}
    The \textit{Union Volume Estimation problem}, i.e. yielding an $(\varepsilon, \delta)$-approximation of the volume of the union of convex bodies, is a classical problem in computational geometry and formal methods. 
    In this paper, we first detect where and why the efficient union volume estimators with approximate sampling and PAC volume estimation oracles fail.
    Specifically, we show that having only TV-distance guarantees on the approximate sampler does not suffice, and demonstrate this bottleneck by constructing the sunflower families of convex bodies.
    In this context, we propose the concept of \textit{tail-controlled samplers} that can yield with high-probability almost-$\ell_{\infty}$ closeness. 
    
    Secondly, we leverage this concept to propose a generic algorithmic framework \texttt{Alpona} that properly combines a PAC volume computation and approximate point sampling oracles with tail-controls to yield an $(\varepsilon, \delta)$-estimate of the union volume for any collection of convex bodies.  
    We show that \texttt{Alpona} needs only linear (in number of convex bodies) oracle calls to be $(\varepsilon, \delta)$-correct. 

    Finally, we provide a tail-controlled analysis of the Hit-and-Run sampler~\citep{lovasz2006fast}. We further provide a novel geometric analysis to improve the required walk length of Hit-and-Run.
    We further instantiate \texttt{Alpona}'s correctness for the sparse frequency streams of convex bodies: a special case of the union volume estimation problem where no convex body intersects more than $\tau$ many other bodies.
\end{abstract}
\thispagestyle{empty}
% \deb{our abstract is still ttc centric. let's discuss before editing!}
\doparttoc
\tableofcontents
\thispagestyle{empty} 
\clearpage

\pagenumbering{arabic}
\section{Introduction}
\textit{Volume computation} is a classical problem at the intersection of computational geometry, randomized
algorithms, and approximate counting~\citep{dyer1988complexity,khachiyan1989,lovasz1990mixing,cousins-vempala18}. The challenge is to yield an estimate $\hat{V}$ of the volume $\vol(K)$ of a convex body $K \subset \mathbb{R}^\dimension$, when we have only sampling access to it. 
Exact volume computation is \#P-hard even for a single polytope~\citep{dyer1988complexity,khachiyan1989}.
Thus, we retreat to randomized algorithms, and aim to yield an estimate $\widehat{V}$ that is an $(\err, \conf)$-approximation of $\vol(K)$, i.e.,
\begin{align*}
    \Pr\left[(1-\err)\vol(K) \leq \widehat V
       \leq (1+\err)\vol(K)\right] \ge 1- \conf\,,
\end{align*}
for some $\err >0$ and $\conf \in (0,1]$.
In this direction, a celebrated line of work beginning with~\citet*{dyer1991random} established polynomial-time randomized approximation algorithms for convex bodies. 
A unifying theme across this extensive line of work is the central role of random walks and Markov Chain Monte Carlo (MCMC) methods. 
Specifically, rapidly mixing random walks have emerged as the predominant algorithmic paradigm for both volume approximation and approximately uniform sampling from high-dimensional convex bodies.
These developments have led to powerful techniques based on geometric random walks, approximate sampling, and rapidly mixing Markov chains.
Most notably, the recent work of \citet{cousins-vempala18} solves $(\err, \conf)$-approximate volume computation of $(\inrad, \circrad)$-fat convex body\footnote{An $(\inrad, \circrad)$-fat convex body contains an sphere of radius at most $\inrad > 0$ and is circumscribed by a sphere of radius at least $\circrad \ge \inrad > 0$.} in $\bigO(\nicefrac{n^3\circrad^2}{\inrad^2})$ by expanding upon the idea of generating sequences of interpolating convex bodies via resorting to Gaussian distributions: 
a technique referred to as Gaussian cooling.
\vspace{.5em}

In this paper, we take a step further, and study how to \textit{efficiently estimate the volume of a union of convex bodies} with provable guarantees. 
\begin{problem}[$(\err, \conf)$-\textbf{Union Volume Estimation}]\label{pr:union-volume-estimation}
Given $M$ convex bodies $K_1,\ldots,K_{\nbodies}\subset \mathbb{R}^{\dimension}$ yield an estimate of the volume of their union
\[
    V \define \vol\left(\bigcup_{i=1}^{\nbodies}K_i\right)
\]
within a multiplicative factor of $1\pm\err$,
with probability at least $1-\conf$.
\end{problem}

Estimating the volume of a union of convex bodies presents additional difficulties. While each individual body is convex, their union need not be, and substantial overlaps can make the sum of the individual volumes a poor estimate of the volume of their union. 

For discrete convex bodies, the \uve{} problem manifests to the enumeration of union size. 
Randomized approximation algorithms for union size, beginning with the classical work of~\citet*{karp-luby-1983}, provide a natural framework for addressing this problem. 
Subsequent works~\citep{karp-luby-madras89,meel2017hashing,soos-sarkar-24,bringmann2024approximating} have extended this framework to realistic implementations with approximate size enumeration and sampling oracles.
Thus, our central challenge is not proving only the existence of an efficient approximation algorithm, 
rather \textit{understanding which oracle guarantees suffice to provably solve the \uve{} problem, and how these guarantees can be realized efficiently for general convex bodies}.

\paragraph{Motivations.} Beyond its theoretical significance, the \uve{} problem naturally arises in automated reasoning and probabilistic verification~\citep{meel-chakraborty-vinodchandran22}. 
If the convex bodies are axis-parallel hypercubes, the \uve{} reduces to the Klee's measure problem, which was first proposed by~\citet{klee1977can} in one dimension.
Its multidimensional version is specifically useful for DNF (Disjunctive Normal Forms) counting and has been addressed by~\citet{meel-vinodchandran-chakraborty21}, when one has access to `exact' sampling and size enumeration oracles for each hypercube (known as a Delphic set). But the algorithms used in practice are deployed with approximate samplers~\citep{soos-sarkar-24}.

In a similar spirit, the satisfying assignments of a formula in Satisfiability Modulo Theories over Linear Real Arithmetic (SMT-LRA) can be represented as a union of convex polytopes. Estimating the volume of this union provides quantitative information about the space of satisfying assignments, with applications to reasoning about continuous systems.
The recent work by~\citet{shaw-sarkar-meel25} introduced an algorithm for estimating the volume of regions represented by SMT-LRA formulas.
The algorithm builds on randomized union-size estimation and demonstrates impressive empirical performance, significantly improving upon earlier approaches on the benchmarks considered.
Their implementation uses, in the background, the volume and sampling oracles with sound theoretical guarantees assumed in their analysis. However none of these guarantees are retained in practice.
In practice, the volume estimators (such as, discretize-and-count, Gaussian cooling~\citep{cousins-vempala18}) and geometric random walks (in particular, Hit-and-Run~\citep{lovasz2006fast})- run only up to a fixed walk length, since the walk lengths required by the theoretical guarantees are prohibitively long. 
Although these are natural and computationally effective choices, they do not satisfy the assumptions used in the analysis. 
This exposes a theory-to-practice gap in the \uve{} problem, which is not unique to the work of \citep{shaw-sarkar-meel25} but rather exhibited by any \uve{} algorithm using approximate samplers. 

\textit{We aim to understand and algorithmically close this gap for general convex bodies that include all the aforementioned problems as special cases.} To this end, we ask the following three questions.
\begin{tcolorbox}
\begin{enumerate}[leftmargin=*]
    \item Does an \uve{} algorithm with an approximate sampler and a PAC volume computation oracle fail to be $(\err, \conf)$-approximate for certain sets of convex bodies?
    \item Can we design a generic and efficient algorithmic framework that can incorporate the errors of the volume computation and approximate sampling oracles to \textit{provably} yield an $(\err, \conf)$-union volume estimate?
    \item Can this algorithmic framework be computationally efficient, i.e. what are the time and sample complexities of this algorithmic framework when instantiated with well-known and computationally-efficient approximate volume computation algorithms, like Gaussian cooling, and approximate samplers, like Hit-and-Run?
\end{enumerate}
\end{tcolorbox}

\subsection{Overview of the Contributions}
By addressing the aforementioned questions, our work establishes a generic framework and the necessary properties of sampling and volume computation oracles to ensure the reliability of \uve{} algorithms when the idealized volume and uniform sampling oracles are replaced by computationally efficient approximations. %, and characterizes the necessary properties for efficiently and provably solving \uve{}. 
The contributions are fivefold.

\paragraph{(C1) Characterizing the limitations due to efficient but approximate samplers}\,

We investigate the efficiency-correctness dilemma encountered by the \uve{} algorithms. 
First, we show that the approximate sampling guarantees for each convex body in terms of TV-distance does not suffice to control the total error accumulated over the union of convex bodies (Theorem~\ref{thm:worst-case-benchmarks}). 
TV-distance based guarantees control the overall discrepancy between two
distributions, but do not ensure that their probability densities are multiplicatively close at every point.
In \uve, the locations of sampled points matter crucially: samples lying in overlapping regions between multiple bodies are treated differently from those lying in regions belonging to only one body. 
Consequently, distributional errors may interact with the geometry of the input in ways that are not captured by the TV distance. 
Motivated by this geometric insight, we construct the sunflower family (Lemma~\ref{lem:convex-sunflower}), where every convex body is of same size with a common overlapping region and individual petals, and show that any \uve{} algorithm with TV-distance close sampler would fail to be correct in this sunflower family.
%\deb{should we say something more about \ttc{} here or just ignore? not essential.}
% Then, we verify another efficiency claim of \ttc{} that fixed number of walk steps of $\Theta(\dimension)$ \textit{should} suffice to obtain an $(\err, \conf)$-\uve{}. \dc{Does \ttc{} actually `claim' this?}
% Along with the sunflower family, we numerically test accuracy of \ttc{} for collections of thin boxes, where the ratio between the circumscribed and inscribed radii, say $R$ and $r$, affects the mixing behavior of \har.
% As explained before, \ttc{} with fixed length walks fail for the sunflower family as $M$ increases. On the other hand, \ttc{} fails for boxes as $R/r$ increases, i.e., as each box become thinner and sharper. 
% These results demonstrate that \ttc's estimation error can be substantial when the walk is not run sufficiently long depending on the geometry and number of the convex bodies. 
% Thus, it is imperative to account for the approximation quality of sampling along with the geometry to ensure correctness of an \uve{} algorithm while retaining efficiency.

% For efficiency in practice, \ttc{} uses \har{} with fixed number of walk steps as the approximate sampling oracle.
% The hope is that since \har{} yields approximate samples that are $\theta$-close in TV-distance to the uniform distribution over a convex body in TV-distance, \ttc{} will retain correctness with appropriately modified approximation errors. We show that this is not true due to two reasons.
% \dc{Do we add a point of implications being more generic than just \ttc{}?}

\paragraph{(C2) A Tail-sensitive Distance for Approximate Samplers}\,

% \Eeshan{Since we are being technical here, should the definition be explicitly mentioned here as well as the tail-control section?}\deb{better to refer than repeat}

Instead of TV-distance close approximate samplers, if we focus on only using $\ell_{\infty}$-close approximate samplers, we can design correct \uve{} algorithm but that becomes too restrictive as most of the efficient samplers are \textit{not} provably $\ell_{\infty}$-close (e.g. \har, Ball walk~\citep{kannan1997random}, Billiard walk~\citep{polyak2014billiard}).
We propose a new $(\eta,\beta)$-closeness measure to measure divergence between two distributions.
\begin{definition}[$(\eta,\beta)$-closeness]
\label{def:almost-linf-informal}
    Given a probability space $(S,\unif)$, and a distribution $\distr$ which is absolutely continuous with respect to the base measure $\unif$, let $h \define \frac{\dd\distr}{\dd\unif}$ denote its relative density. We call $\distr$ to be $(\sampgood,\sampbad)$-close to $\unif$, if we have 
    \begin{align}
        \unif(\bad_\sampgood)\le \sampbad \text{ and }\distr(\bad_\sampgood) \le 2\sampbad, \text{ where }
        \bad_\sampgood := \{x\in S: |h(x) - 1|>\sampgood \}\,,\nonumber
    \end{align}
    for some $(\eta,\beta)\in[0,1]^2$.
\end{definition}
$(0,0)$-closeness implies the distribution is exactly uniform, except possibly on a measure zero subset. The $(\eta, \beta)$-tailed samplers are the approximate samplers that generate samples exhibiting empirical distributions differing from uniform distribution by $\beta$, i.e. the support mass with $\ell_{\infty}$ distance more than $\eta$.
Informally, the $(\eta, \beta)$-tailed sampler produces samples whose density is multiplicatively close to the uniform density on a majority fraction of the set. Outside of this region (hereby referred as the \textit{exceptional} region), we have no information as to how the density behaves.

We prove that \har, which is known to be an efficient approximate sampler, yields $(\eta, \beta)$-close samples to uniform for any $(\inrad, \circrad)$-fat convex body in polynomial number of steps.
\begin{lemma}[$(\eta,\beta)$-closeness of \har{} Samples, Informal version of Theorem~\ref{thm:improved-tail-hit-and-run-informal}]
    For any $1>\beta,\eta>0,$ if \har{} is run on a $(\inrad, \circrad)$-fat convex body $\body\subset\R^\dm$ starting from any boundary point, we have $\unif_\body(\bad_{t,\sampgood})\le \sampbad$ if $$t= \omega\left(\mathrm{poly}\left(\dm, \frac{\circrad}{\inrad}, \log\frac{1}{\sampbad}, \log\frac{1}{\sampgood}\right)\right)\,.$$ 
    Here, $\unif_K$ denotes the uniform distribution over $K$. Moreover, $\distr_t(\bad_{t,\sampgood})\le 2\sampbad$, i.e., a point sampled from $\distr_t$ has density $(1\pm \eta)$ close to that of $\unif$ with probability at least $1-2\sampbad$.
\end{lemma}
% Next, we construct an algorithmic framework, \algo, that shows having $(\sampgood,\sampbad)$-tailed samplers are sufficient to estimate the volume of the union of $M$ general convex bodies (Problem~\ref{pr:union-volume-estimation}).
%We also envision that such a probabilistic closeness measure will be, in general, useful to analyse and understand practical efficiency of randomized algorithms that deploy approximate but efficient samplers underneath.

\paragraph{(C3) Designing a correct and efficient framework for \uve{}}\,

Now, we propose a provably correct algorithmic framework, \algo , for \uve{} that access two oracles: (i) a PAC~(Probably Approximately Correct) volume estimator for each convex body, and (ii) an $(\eta, \beta)$-tailed sampler.  Under appropriate choices of the oracle parameters, $\algo$ returns a multiplicative approximation to the union volume with high probability. 

\begin{theorem}[Informal version of Theorem~\ref{thm:main-theorem}]
Let $K_1,\ldots,K_M\subset \mathbb{R}^{\dimension}$ be convex bodies satisfying the geometric assumptions required by the PAC volume estimation and approximate sampling oracles.
For every $\varepsilon,\delta\in(0,1)$, there is
a randomized algorithm $\algo$ that, using PAC
volume estimators and $(\eta, \beta)$-tailed
samplers, outputs an estimate $\widehat V$ satisfying
\[
    \Pr\left[
       (1-\varepsilon)V
       \leq \widehat V
       \leq (1+\varepsilon)V
    \right]\geq 1-\delta\,,
\]
where $V\define \vol\left(\bigcup_{i=1}^{M}K_i\right)$. 
\algo{} requires $\bigO(M)$ calls to the PAC volume estimation oracle and $\widetilde{\bigO}(M/\varepsilon^2)$ calls to $(\eta, \beta)$-tailed samplers.
\end{theorem}

% \dc{Keep this theorem in terms of oracle costs? The oracle realisation issue is in the next contribution.}
% \highlight{The algorithm builds on representative-sampling
% techniques for union-size estimation, but uses a
% direct Poisson-based formulation for continuous
% geometric domains. Each body contributes a Poisson
% number of representative points proportional to its
% estimated volume. Overlapping points are accounted
% for by assigning them to the last body containing
% them in a fixed ordering. An adaptive thinning
% procedure controls the number of retained samples,
% without requiring a discretization of the geometric
% space.}\dc{Let's elaborate a bit more?}

The algorithm is a adaptive sampling based algorithm that aims to retain a Poisson number of representative points from each body proportional to its estimated volume.   
The main technical challenge is to analyse the interaction between three sources of uncertainty:
errors in volume estimation, non-uniformity in the sampling distributions, and adaptive changes in the sampling intensity.
Our analysis separates the regions where the sampling density is approximately uniform from the exceptional regions where this guarantee may fail. 
The former contribute a controlled approximation to the union volume, while the contribution from the latter is bounded using their small measure and sampling probability.
Together with concentration arguments for the sampling procedure, this establishes the correctness guarantee.

\paragraph{(C4) Connecting geometric sampling guarantees to implementable algorithms}\,

Finally, we propose a practically efficient instantiation of \algo{} with \har{} as $(\eta, \beta)$-tailed approximate sampler and Gaussian cooling as the PAC volume estimator.
We show that space and time complexity of  this algorithm are 
\[\widetilde{\bigO}\Bigl(\frac{\dm}{\err^2}\log\frac{\length}{\conf}\Bigr)\,, \text{ and } \widetilde{\bigO}\left(\length \frac{\max\left\{\dm^3, \dm^2 (\circrad / \inrad)^2\right\}}{\err^2}\right).\]

The resulting complexity analysis separates the number of oracle calls made by the main algorithm from the cost of implementing each oracle. 
This distinction is important: the number of bodies affects both the number of operations and the accuracy parameters required by the subroutines, while the geometry of each body influences the complexity of the individual approximate volume computation and sampling queries.
Thus, the analysis provides a quantitative connection between the desired approximation guarantee, the number of convex bodies, their geometric parameters, and the computational resources required by the implementation.

In this process, we derive in Theorem~\ref{thm:exceptional-application} an improved walk length that \har{} needs to take to generate an $(\eta, \beta)$-close sample is $\widetilde{\bigO}\!\left(\dm^3\frac{\circrad^2}{\inrad^2}\right)$ from $\widetilde{\bigO}\!\left(\dm^4\frac{\circrad^2}{\inrad^2}\right)$~\citep{lovasz2006fast}.
The proof passes through two geometric claims: %\Eeshan{shouldn't we mention what $\gamma_{\mathrm{HR}}$ is, if we are mentioning it here?. agreed}
\begin{enumerate}
    \item For any $0<\sampgood,\sampbad<1$, \har{} taking at least
    \[1+\dm/2+\bigO\left(\frac{\dm\log(32\circrad/d)+\log(1/(\sampgood^2\sampbad))}{\gamma_{\mathrm{HR}}}\right)\]
    suffices to guarantee that $\unif(\bad_{t,\sampgood})\le\sampbad$ and $\distr_t(\bad_{t,\sampgood})\le2\sampbad$, where $\gamma_{\mathrm{HR}}$ is the spectral gap for \har.
    \item  We show contraction of \har's sampling operator $P$ for functions $f$ with mean $0$:
    \[\norm{P^sf}_{L^2(\unif)}\le e^{-\gamma_{\mathrm{HR}}s}\|f\|_2\,,\]
    for $s > 1+ \frac{n}{2}$.
\end{enumerate}

% Finally, we complement the abstract algorithmic framework
% by investigating how its required sampling
% guarantees can be obtained using geometric random
% walks.
% In particular, we study almost-$\ell_\infty$
% closeness as a suitable guarantee for the
% Hit-and-Run sampling procedure. By relating
% the desired density approximation to the mixing
% behavior of the random walk, we obtain explicit
% conditions under which the sampling subroutine
% can be used in $\algo$.

% The resulting complexity analysis separates the
% number of oracle calls made by the main algorithm
% from the cost of implementing each oracle. This
% distinction is important: the number of bodies
% affects both the number of operations and the
% accuracy parameters required by the subroutines,
% while the geometry of each body influences the
% complexity of approximate volume computation
% and sampling.
% Thus, the analysis provides a quantitative
% connection between the desired approximation
% guarantee, the number of convex bodies, their
% geometric parameters, and the computational
% resources required by the implementation.

% \dc{This reads essentially like an extension of C3. Rewrite in terms of Hit-and-Run proofs and move the oracle abstraction point to C3?}

\paragraph{(C5) Bounded Frequency Streams}\,

Till now, we have worked with a pre-fixed number of convex bodies $M$. But we can adapt \algo{} further to streams of convex bodies with bounded frequency. 
Here, a bounded frequency stream of convex bodies means that each point of the union of these bodies has a bounded frequency $1 \le \tau \le M$.
In this case, we show (Theorem~\ref{thm:bounded-frequency}) that time and space complexities of \algo{} will be dependent on the frequency $\tau$ instead of $M$.
% \deb{Eeshan: add an organisation paragraph here.}

\paragraph{Organisation.} Rest of the paper is organised as follows: in \cref{sec:motivation}, we demonstrate and highlight the weakness in sampling from distributions with only TV-control. In \cref{sec:tail-controlled}, we propose the notion of tail-controlled distributions and discuss its properties. The paper's algorithmic contribution is described in \cref{sec:union-algorithm}, which also includes a short discussion on the related works. Correctness of the algorithm and it's complexity requirements are discussed in \cref{sec:correctness} and in \cref{sec:union-complexity} respectively. We cover the special case of sparse streams in \cref{sec:bounded-frequency}. We finally conclude in \cref{sec:conclusion}.

\section{From Worst-Case Sampling Guarantees to Practical Performance}\label{sec:motivation}

Estimating the volume of a union of convex bodies from approximate samples presents a subtle difficulty: a small sampling error for each body need not translate into a small error for the union. We first demonstrate this limitation through a sunflower construction. 
We then examine its relevance to the practical algorithm of \cite{shaw-sarkar-meel25}. Although adversarial sampling errors can cause substantial failures, the practical performance of their algorithm is often considerably better than this pessimistic scenario suggests. Understanding this discrepancy motivates our study of more informative sampling guarantees.

\subsection{A Fundamental Limitation of Approximate Sampling}

Suppose an estimator has sampling-oracle access to convex bodies $\body_1,\ldots,\body_\length$. Denote by $\unif_{i}$ the uniform distribution on $\body_i$. For each $i$, the oracle returns independent samples from a distribution $\distr_i$ satisfying $\|\distr_i-\unif_{i}\|_1\leq\theta$. We ask whether this assumption alone suffices for reliable estimation of the union volume. The following theorem gives a negative answer in the sampling-only oracle model. 
Here, $\|\cdot\|_1$ denotes the $\ell_1$ distance between probability distributions, which is twice their TV distance.

\begin{theorem}[Limitations of approximate sampling]
\label{thm:worst-case-benchmarks}

Fix $\err,\theta\in(0,1)$, and let
$\mathcal A$ be any randomized algorithm whose
access to the input convex bodies is restricted
to the following oracles:
\begin{itemize}
    \item A sampling oracle for each body $\body_i$
    that returns independent samples from a
    distribution $\distr_i$ satisfying
    \[
        \|\distr_i-\unif_{i}\|_1
        \leq \theta.
    \]
    \item A volume oracle for each body $\body_i$
    that returns a $(1\pm\theta)$-approximation
    to $\operatorname{Vol}(\body_i)$.
\end{itemize}
Then there exists an instance consisting of
$\body=\bigO_\err(1/\theta)$ convex bodies,
together with oracles satisfying the above
guarantees, on which $\mathcal A$ fails to
return a $(1\pm\err)$-approximation
to the volume of their union with probability
at least $1/2$.
\end{theorem}

\begin{proof}
Let $a=\theta/(2-\theta)\in(0,1)$.
By Lemma~\ref{lem:convex-sunflower} (proved later),
there exist $\length$ convex bodies
$\body_1,\ldots,\body_\length\subseteq\mathbb R^{\length-1}$
forming a sunflower family with common
core $\core$ and pairwise disjoint petals
$C_1,\ldots,C_\length$, such that
\[
    \body_i\define \core \sqcup C_i,
    \qquad
    \operatorname{Vol}(C_i)
    \define a\operatorname{Vol}(\core).
\]
Write $\corevol=\operatorname{Vol}(\core)>0$.

For each $i$, define the sampling
distribution $\distr_i=\unif_{\core}$.
Thus, the sampler returns uniformly random points from the common core $\core$ and never
samples from the petals. Then:
\[
\begin{aligned}
\|\distr_i-\pi_i\|_1
&=
\int_{\core}
\left|
\frac{1}{\corevol}-\frac{1}{(1+a)\corevol}
\right|dx
+
\int_{C_i}\frac{dx}{(1+a)\corevol}\\
&=\frac{2a}{1+a}
=\theta.
\end{aligned}
\]
Hence, every sampling oracle satisfies
the required $\ell_1$-accuracy guarantee,
despite completely ignoring its petal.

Next, let the volume oracle for every
body return the value $\corevol$. This is a
valid $(1\pm\theta)$-factor approximation of the corresponding volumes since
\[
\begin{aligned}
(1-\theta)\operatorname{Vol}(\body_i)
&=(1-\theta)(1+a)\corevol\\
&=\frac{2(1-\theta)}{2-\theta}\corevol
\leq \corevol
\leq (1+\theta)\operatorname{Vol}(\body_i).
\end{aligned}
\]

Now consider a second instance consisting
of $\length$ identical copies of the core $\core$,
each equipped with an exact uniform
sampling oracle and an exact volume
oracle returning $\corevol$.

The two instances are indistinguishable
to $\mathcal A$: in both cases, every
sampling oracle returns independent
uniform samples from $\core$, and every
volume oracle returns $V_{\cap}$.
Consequently, the complete oracle
transcripts, and hence, the output
distributions of $\mathcal A$ are
identical on the two instances,
regardless of the number of adaptive
oracle queries.

However, the union volumes of the
two instances differ. For the second
instance, the union volume is
$V_0=\corevol$,
whereas for the sunflower instance,
\[
    V_1
    =\corevol+\sum_{i=1}^{\length}\vol(C_i)
    =\corevol(1+\length a).
\]

Choose an integer $\length$ satisfying
\[
    \length>
    \frac{2\err}{(1-\err)a}
    =
    \frac{2\err(2-\theta)}
         {(1-\err)\theta}.
\]
Then,
\[
    (1-\err)V_1
    >(1+\err)V_0,
\]
so the intervals of valid
$(1\pm\err)$-approximations
to $V_0$ and $V_1$ are disjoint.

Since $\mathcal A$ has identical output
distributions on the two instances,
the sum of its success probabilities
on these instances is at most one.
Therefore, $\mathcal A$ must fail
with probability at least $1/2$
on at least one of them.

Finally, for every fixed $\err$,
the above choice of $\length$ satisfies
$\length=\bigO_\err(1/\theta)$,
completing the proof.
\end{proof}

Now, we construct a specific family of convex bodies, namely sunflower families, where the union volume estimator $\mathcal{A}$ fails to yield an $(\err, \conf)$-estimate.

\begin{lemma}[Existence of convex sunflower families]
\label{lem:convex-sunflower}
For every integer $\length\geq 2$ and every $a\in(0,1)$,
there exist $\length$ full-dimensional convex polytopes
$\body_1,\ldots,\body_\length\subseteq\mathbb R^{\length-1}$
and a convex body $\body$ of positive volume such that,
writing $C_i \define \body_i\setminus \body$, we have
\[
    \body_i=\body\sqcup C_i,\qquad
    \operatorname{Vol}(C_i)
       =a\operatorname{Vol}(\body)
       \quad\text{for every }i\in[\length],
\]
and the petals $C_1,\ldots,C_\length$ are pairwise disjoint.
\end{lemma}

\begin{proof}
Let $n\define \length-1$, and let $\body\subseteq\mathbb R^n$
be a regular $n$-dimensional simplex with
vertices $v_1,\ldots,v_\length$.
For each $i\in[\length]$, let $F_i$ be the facet
opposite to $v_i$, and let $z_i$ denote
the centroid of $F_i$.

By symmetry, the segment joining $v_i$ to
$z_i$ is perpendicular to $F_i$, and its
length $h>0$ is independent of $i$.
Let us define
\[
    w_i\define z_i+a(z_i-v_i)
\]
and
\[
    \body_i\define \operatorname{conv}(\body\cup\{w_i\}).
\]

We claim that $\body_i$ is obtained by attaching
the pyramid
\[
    P_i\define \operatorname{conv}(F_i\cup\{w_i\})
\]
to $\body$ along the facet $F_i$.

To see this, consider the barycentric
coordinates of $w_i$ with respect to
$v_1,\ldots,v_\length$. Since
\[
    z_i=\frac1{\length-1}\sum_{j\neq i}v_j,
\]
we have
\[
    w_i
    =-av_i+\frac{1+a}{\length-1}
      \sum_{j\neq i}v_j.
\]
Thus, the barycentric coordinate
corresponding to $v_i$ is negative,
whereas all the remaining coordinates
are positive. Consequently, $w_i$
lies strictly beyond $F_i$ but satisfies
all the other facet-defining inequalities
of $\body$.

It follows that
\[
    \body_i=\body\cup P_i,
    \qquad
    \body\cap P_i=F_i.
\]
Moreover, every point of $P_i\setminus \body$
violates the facet inequality corresponding
to $F_i$, while satisfying all the other
facet inequalities. Hence
\[
    (P_i\setminus \body)\cap(P_j\setminus \body)
    =\varnothing
    \qquad\text{for }i\neq j.
\]

It remains to verify the volumes.
Let $A$ denote the common
$(n-1)$-dimensional volume of the
facets of $\body$. Since $\body$ has altitude
$h$ over each facet,
\[
    \operatorname{Vol}(\body)=\frac{Ah}{n}.
\]
The pyramid $P_i$ has base $F_i$
and altitude $ah$, giving
\[
    \operatorname{Vol}(P_i)
    =\frac{A(ah)}{n}
    =a\operatorname{Vol}(\body).
\]

Taking $C_i=P_i\setminus \body$,
we obtain
\[
    \operatorname{Vol}(C_i)
    =a\operatorname{Vol}(\body)
\]
for every $i\in[\length]$.
Since the petals are pairwise disjoint,
the desired sunflower family follows.
\end{proof}

% \noindent\emph{Oracle-model qualification.} The indistinguishability conclusion applies when the bodies are accessible only through the stated sampling oracles. If individual-volume, membership, or explicit-description oracles are also supplied, the two instances above may be distinguishable. The geometric sunflower family used in the proof should be instantiated by an explicit convex construction in the final version.

\subsection{Beyond Worst-Case Sampling Guarantees}
\label{subsec:beyond-worst-case}

While the preceding theorem establishes that any algorithm
$\mathcal{A}$ can fail on a worst-case instance involving
only $\bigO(\nicefrac{1}{\theta})$ convex bodies, practical
algorithms often perform considerably well.
This discrepancy raises a natural question:
is this limitation inherent to approximate sampling,
or does it arise from the overly pessimistic assumptions
underlying the worst-case analysis?
To understand this discrepancy, we take a closer look
at the behavior of practical sampling procedures.

The lower bound via sunflower construction relies on highly adversarial samplers
that sample uniformly from a common core while
completely ignoring the remaining regions of the bodies.
However, practical geometric samplers, such as those
based on Hit-and-Run random walks, need not exhibit
such extreme behavior. Instead, we expect their
sampling distributions to be much closer to uniform
over most of the body, with significant deviations
potentially concentrated in regions that are harder
to explore. In particular, rather than assuming only
an $\ell_1$-distance guarantee, it may be more
appropriate to seek a multiplicative $\ell_\infty$
guarantee over a large portion of the body.

%\Eeshan{Must we pitch $(\sampgood, \sampbad)$-sampling here? Or can we save it up for the next section?}

More precisely, for a convex body $K_i$ with uniform
density $\pi_i$, we seek a measurable good region
$G_i\subseteq K_i$ satisfying
\[
  \operatorname{Vol}(K_i\setminus G_i)
  \leq \beta_i\operatorname{Vol}(K_i),
\]
such that the sampling distribution $D_i$ satisfies
\[
  (1-\eta_i)\pi_i(x)
  \leq D_i(x)
  \leq (1+\eta_i)\pi_i(x)
  \qquad\text{for almost every }x\in G_i.
\]
Here, $\eta_i$ measures the multiplicative distortion
of the sampling density over the good region, while
$\beta_i$ measures the relative volume of the
exceptional region. This provides a more refined
description of sampling quality than a global
$\ell_1$-distance bound.

A second important consideration is the geometry of
the underlying convex bodies. The accuracy of a
practical sampler is influenced by how easily it can
explore different regions of the body, which in turn
depends on its geometric conditioning. In particular,
the ratio $R_i/r_i$, where $R_i$ and $r_i$ are the
radii of suitable enclosing and enclosed balls,
can affect the convergence of random-walk-based
samplers. Poorly conditioned bodies, with large
$R_i/r_i$, may contain long or thin regions that
are more difficult to explore accurately.

These observations suggest two complementary ways
to understand and potentially overcome the pessimism
of the worst-case bound. One possibility is to require
more accurate samplers as the number of bodies grows.
Indeed, the sunflower construction suggests that
an $\ell_1$-accuracy of order $\bigO(\varepsilon/m)$
is a natural conservative target for controlling
the accumulation of sampling errors. However,
achieving such accuracy may significantly increase
the computational cost.

An alternative is to exploit the more favorable
behavior of practical samplers by incorporating
multiplicative approximation guarantees over
large good regions, together with geometric
information about the bodies. Such guarantees
may provide a more realistic characterization
of sampling quality without requiring uniformly
stringent $\ell_1$-accuracy.

% These considerations identify two natural parameters
% for experimental investigation: the number $m$
% of bodies, which determines the potential
% accumulation of sampling errors, and the geometric
% conditioning ratio $R/r$, which can influence
% the quality of the individual samplers.

% We next investigate the effect of these two
% parameters on the performance of \ttc, aiming
% to understand when the theoretical limitations
% manifest themselves and why the algorithm often
% performs substantially better in practice.

\subsection{The Gap between Worst-case Guarantees and \texorpdfstring{$\ttc$}{} in Practice}

The algorithm of \citet{shaw-sarkar-meel25}, namely \ttc{}, provides a useful case study. It estimates volumes of regions represented by SMT-LRA formulas, building on the Delphic-set framework for union-size estimation \citep{meel-vinodchandran-chakraborty21}. At a high level, $\ttc$ combines approximate individual-volume estimates with samples from the constituent convex bodies. Its adaptive, Poisson-based representative-sampling procedure maintains a bounded bucket of sampled points and adjusts the sampling rate as the bodies are processed. The theoretical analysis requires sufficiently accurate sampling oracles; its practical implementation uses geometric random walks, notably Hit-and-Run, with much shorter walks than conservative worst-case mixing analyses may prescribe.

\begin{figure}[t!]
    \centering
    \begin{subfigure}[b]{0.25\textwidth}
        \centering
        \resizebox{\textwidth}{!}{\input{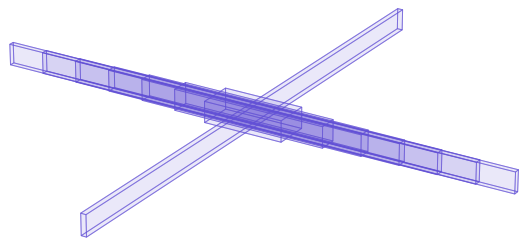}}
        \caption{Boxes, $\max_i \tfrac{R_i}{r_i}=1000$}
        \label{fig:rdhr_box_10}
    \end{subfigure}
    \hfill
    \begin{subfigure}[b]{0.3\textwidth}
        \centering
        \resizebox{\textwidth}{!}{\input{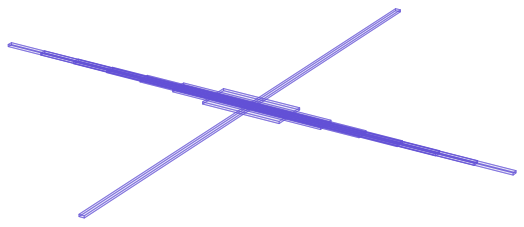}}
        \caption{Boxes, $\max_i \tfrac{R_i}{r_i}=10,000$}
        \label{fig:rdhr_box_100}
    \end{subfigure}
    \hfill
    \begin{subfigure}[b]{0.3\textwidth}
        \centering
        \resizebox{\textwidth}{!}{\input{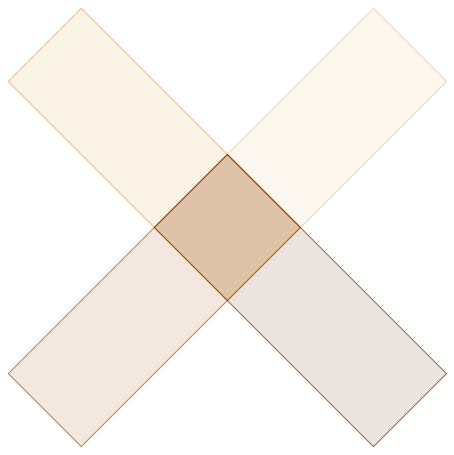}}
        \caption{Boxes, $\max_i \tfrac{R_i}{r_i}=100,000$}
        \label{fig:rdhr_box_1000}
    \end{subfigure}

    \medskip
    \begin{subfigure}[b]{0.32\textwidth}
        \centering
        \resizebox{0.6\textwidth}{!}{\begingroup%
\makeatletter%
\begin{pgfpicture}%
\pgfpathrectangle{\pgfpointorigin}{\pgfqpoint{6.800000in}{6.800000in}}%
\pgfusepath{use as bounding box, clip}%
\begin{pgfscope}%
\pgfsetbuttcap%
\pgfsetmiterjoin%
\definecolor{currentfill}{rgb}{1.000000,1.000000,1.000000}%
\pgfsetfillcolor{currentfill}%
\pgfsetlinewidth{0.000000pt}%
\definecolor{currentstroke}{rgb}{1.000000,1.000000,1.000000}%
\pgfsetstrokecolor{currentstroke}%
\pgfsetdash{}{0pt}%
\pgfpathmoveto{\pgfqpoint{0.000000in}{0.000000in}}%
\pgfpathlineto{\pgfqpoint{6.800000in}{0.000000in}}%
\pgfpathlineto{\pgfqpoint{6.800000in}{6.800000in}}%
\pgfpathlineto{\pgfqpoint{0.000000in}{6.800000in}}%
\pgfpathlineto{\pgfqpoint{0.000000in}{0.000000in}}%
\pgfpathclose%
\pgfusepath{fill}%
\end{pgfscope}%
\begin{pgfscope}%
\pgfpathrectangle{\pgfqpoint{0.166667in}{0.166667in}}{\pgfqpoint{6.466667in}{6.466667in}}%
\pgfusepath{clip}%
\pgfsetbuttcap%
\pgfsetmiterjoin%
\definecolor{currentfill}{rgb}{0.964706,0.788235,0.545098}%
\pgfsetfillcolor{currentfill}%
\pgfsetfillopacity{0.130000}%
\pgfsetlinewidth{0.803000pt}%
\definecolor{currentstroke}{rgb}{0.964706,0.788235,0.545098}%
\pgfsetstrokecolor{currentstroke}%
\pgfsetstrokeopacity{0.750000}%
\pgfsetdash{}{0pt}%
\pgfpathmoveto{\pgfqpoint{3.400000in}{2.363675in}}%
\pgfpathlineto{\pgfqpoint{5.990812in}{4.954487in}}%
\pgfpathlineto{\pgfqpoint{6.508974in}{5.472650in}}%
\pgfpathlineto{\pgfqpoint{5.990812in}{5.990812in}}%
\pgfpathlineto{\pgfqpoint{5.472650in}{6.508974in}}%
\pgfpathlineto{\pgfqpoint{2.881838in}{3.918162in}}%
\pgfpathlineto{\pgfqpoint{2.363675in}{3.400000in}}%
\pgfpathlineto{\pgfqpoint{2.881838in}{2.881838in}}%
\pgfpathlineto{\pgfqpoint{3.400000in}{2.363675in}}%
\pgfpathclose%
\pgfusepath{stroke,fill}%
\end{pgfscope}%
\begin{pgfscope}%
\pgfpathrectangle{\pgfqpoint{0.166667in}{0.166667in}}{\pgfqpoint{6.466667in}{6.466667in}}%
\pgfusepath{clip}%
\pgfsetbuttcap%
\pgfsetmiterjoin%
\definecolor{currentfill}{rgb}{0.929412,0.639216,0.301961}%
\pgfsetfillcolor{currentfill}%
\pgfsetfillopacity{0.130000}%
\pgfsetlinewidth{0.803000pt}%
\definecolor{currentstroke}{rgb}{0.929412,0.639216,0.301961}%
\pgfsetstrokecolor{currentstroke}%
\pgfsetstrokeopacity{0.750000}%
\pgfsetdash{}{0pt}%
\pgfpathmoveto{\pgfqpoint{3.400000in}{2.363675in}}%
\pgfpathlineto{\pgfqpoint{3.918162in}{2.881838in}}%
\pgfpathlineto{\pgfqpoint{4.436325in}{3.400000in}}%
\pgfpathlineto{\pgfqpoint{1.845513in}{5.990812in}}%
\pgfpathlineto{\pgfqpoint{1.327350in}{6.508974in}}%
\pgfpathlineto{\pgfqpoint{0.809188in}{5.990812in}}%
\pgfpathlineto{\pgfqpoint{0.291026in}{5.472650in}}%
\pgfpathlineto{\pgfqpoint{2.881838in}{2.881838in}}%
\pgfpathlineto{\pgfqpoint{3.400000in}{2.363675in}}%
\pgfpathclose%
\pgfusepath{stroke,fill}%
\end{pgfscope}%
\begin{pgfscope}%
\pgfpathrectangle{\pgfqpoint{0.166667in}{0.166667in}}{\pgfqpoint{6.466667in}{6.466667in}}%
\pgfusepath{clip}%
\pgfsetbuttcap%
\pgfsetmiterjoin%
\definecolor{currentfill}{rgb}{0.662745,0.325490,0.070588}%
\pgfsetfillcolor{currentfill}%
\pgfsetfillopacity{0.130000}%
\pgfsetlinewidth{0.803000pt}%
\definecolor{currentstroke}{rgb}{0.662745,0.325490,0.070588}%
\pgfsetstrokecolor{currentstroke}%
\pgfsetstrokeopacity{0.750000}%
\pgfsetdash{}{0pt}%
\pgfpathmoveto{\pgfqpoint{1.327350in}{0.291026in}}%
\pgfpathlineto{\pgfqpoint{1.845513in}{0.809188in}}%
\pgfpathlineto{\pgfqpoint{4.436325in}{3.400000in}}%
\pgfpathlineto{\pgfqpoint{3.918162in}{3.918162in}}%
\pgfpathlineto{\pgfqpoint{3.400000in}{4.436325in}}%
\pgfpathlineto{\pgfqpoint{2.881838in}{3.918162in}}%
\pgfpathlineto{\pgfqpoint{0.291026in}{1.327350in}}%
\pgfpathlineto{\pgfqpoint{0.809188in}{0.809188in}}%
\pgfpathlineto{\pgfqpoint{1.327350in}{0.291026in}}%
\pgfpathclose%
\pgfusepath{stroke,fill}%
\end{pgfscope}%
\begin{pgfscope}%
\pgfpathrectangle{\pgfqpoint{0.166667in}{0.166667in}}{\pgfqpoint{6.466667in}{6.466667in}}%
\pgfusepath{clip}%
\pgfsetbuttcap%
\pgfsetmiterjoin%
\definecolor{currentfill}{rgb}{0.443137,0.196078,0.035294}%
\pgfsetfillcolor{currentfill}%
\pgfsetfillopacity{0.130000}%
\pgfsetlinewidth{0.803000pt}%
\definecolor{currentstroke}{rgb}{0.443137,0.196078,0.035294}%
\pgfsetstrokecolor{currentstroke}%
\pgfsetstrokeopacity{0.750000}%
\pgfsetdash{}{0pt}%
\pgfpathmoveto{\pgfqpoint{5.472650in}{0.291026in}}%
\pgfpathlineto{\pgfqpoint{5.990812in}{0.809188in}}%
\pgfpathlineto{\pgfqpoint{6.508974in}{1.327350in}}%
\pgfpathlineto{\pgfqpoint{5.990812in}{1.845513in}}%
\pgfpathlineto{\pgfqpoint{3.400000in}{4.436325in}}%
\pgfpathlineto{\pgfqpoint{2.881838in}{3.918162in}}%
\pgfpathlineto{\pgfqpoint{2.363675in}{3.400000in}}%
\pgfpathlineto{\pgfqpoint{2.881838in}{2.881838in}}%
\pgfpathlineto{\pgfqpoint{5.472650in}{0.291026in}}%
\pgfpathclose%
\pgfusepath{stroke,fill}%
\end{pgfscope}%
\end{pgfpicture}%
\makeatother%
\endgroup%
}
        \caption{Sunflower, $M=4$}
        \label{fig:rdhr_urchin_4}
    \end{subfigure}
    \hfill
    \begin{subfigure}[b]{0.32\textwidth}
        \centering
        \resizebox{0.6\textwidth}{!}{\input{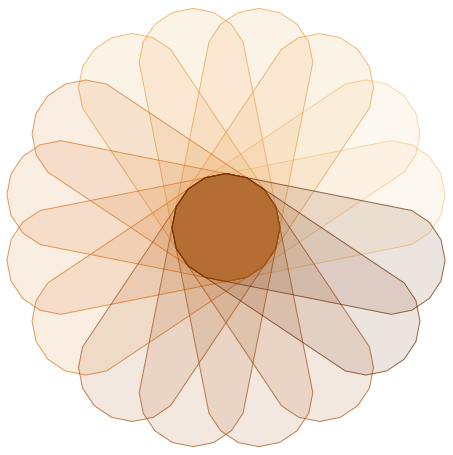}}
        \caption{Sunflower, $M=16$}
        \label{fig:rdhr_urchin_16}
    \end{subfigure}
    \hfill
    \begin{subfigure}[b]{0.32\textwidth}
        \centering
        \resizebox{0.6\textwidth}{!}{\input{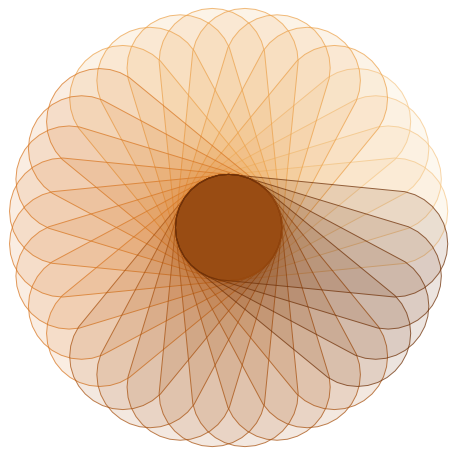}}
        \caption{Sunflower, $M=32$}
        \label{fig:rdhr_urchin_32}
    \end{subfigure}
    \caption{Illustrations of Boxes and Sunflower Family Benchmarks.}
    \label{fig:rdhr}
\end{figure}

Despite this compromise, $\ttc$ exhibits strong empirical performance on previously considered benchmarks. This raises the following genuine question: what are the properties of a benchmark that influence the corresponding empirical performances?
% The sunflower construction therefore raises the following genuine question: what all geometric properties of a benchmark 
% Can errors caused by short random walks actually be observed in the final union-volume estimate? And, if so, why does $\ttc$ often perform much better  on standard benchmark than the adversarial sampling constructions? \Eeshan{Can we rephrase the last question slightly?}
% %\subsection{Experimental evidence: vulnerabilities without worst-case behavior}
We investigate this question by proposing two families of geometric benchmarks, designed to stress the samplers in the backend, but in different ways:
\begin{enumerate}
\item \emph{Boxes benchmark.} Each instance is a union of axis-aligned boxes. By varying their widths, we control the geometric aspect ratio $\max_i R_i/r_i$, where $R_i$ and $r_i$ denote circumscribed and inscribed radii. Thin bodies can make accurate sampling more challenging.
\item \emph{Sunflower benchmark.} Each instance consists of $M$ convex bodies arranged around a common core. Increasing $M$ creates more distinct petal regions, making it possible for sampling biases to accumulate across bodies.
\end{enumerate}
We constructed 26 boxes instances and 20 sunflower instances. \Cref{fig:rdhr} illustrates representative examples.

\begin{figure}[t!]
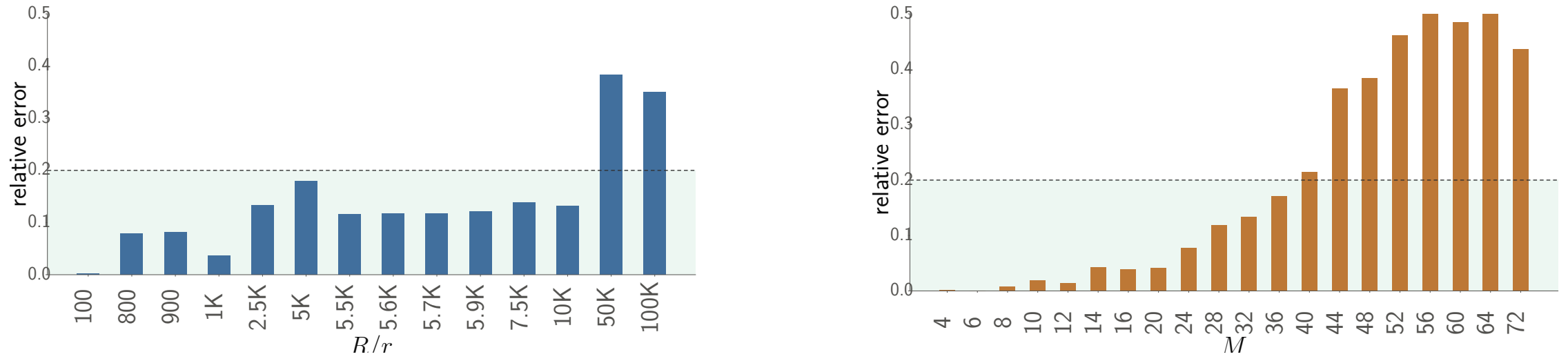

    \centering
    \begin{subfigure}[b]{0.4\textwidth}
        \centering
        \resizebox{\textwidth}{!}{\input{figures/boxes_fixed_walk_1000n_relative_error_2625803.pgf}}
        \caption{Boxes benchmarks}
        \label{fig:boxes_walk}
    \end{subfigure}
    \hspace{40pt}
    \begin{subfigure}[b]{0.4\textwidth}
        \centering
        \resizebox{\textwidth}{!}{\input{figures/radial_fixed_walk_1000n_relative_error_2625803.pgf}}
        \caption{Sunflower benchmarks}
        \label{fig:radial_walk}
    \end{subfigure}
    \caption{Observed relative error of the volume estimate returned by \ttc{} on each benchmark, with the Hit-and-Run walk length fixed to $1000\dm$, where $\dm$ is the dimension of the instance. The shaded region marks the target relative error of at most $0.2$. The boxes benchmarks are arranged in increasing order of $\max_i R_i/r_i$ with median reported for benchmarks with same $\max_i R_i/r_i$, and the sunflower benchmarks in increasing order of number of bodies $M$.}
    \label{fig:walk_length_vs_relative_error}
\end{figure}

\paragraph{Observed estimation errors.}
For each benchmark, we run $\ttc$ with Hit-and-Run walk length fixed at $1000\dm$, following the implementation of \cite{shaw-sarkar-meel25}. We measure relative error $|\hat V-V|/V$, where $V$ is the true union volume. As shown in \Cref{fig:walk_length_vs_relative_error}, error tends to increase with the aspect ratio in the boxes benchmark and with the number of petals in the sunflower benchmark. On some instances, the observed error exceeds the target $0.2$. Thus, a walk length fixed independently of the input geometry cannot be assumed sufficient for every instance.

Crucially, however, these failures do not mean that the extreme sunflower samplers accurately model the usual behavior of Hit-and-Run. The lower bound allows a sampler to ignore every petal completely, whereas practical random walks may explore such regions imperfectly but still with appreciable probability. The experiments therefore support a more nuanced conclusion: the theoretical vulnerability is real, yet the worst-case $\ell_1$ model can be far more pessimistic than the behavior of the sampler actually used by $\ttc$.

\subsection{Towards Geometry-aware, Multiplicative Sampling Guarantees}

The lesson is not that approximate sampling is inherently unsuitable for union-volume estimation, nor that the empirical success of $\ttc$ supplies a correctness guarantee. Rather, $\ell_1$ closeness alone fails to distinguish between benign sampling errors and errors concentrated in precisely those regions that matter most to the union. A useful theory should capture additional properties of practical samplers and translate them into guarantees for the final volume estimate.

This leads to the question driving our work: 
\begin{center}
    \emph{Can we exploit multiplicative near-uniformity on a large good region, together with quantitative control of the exceptional region, to obtain provable union-volume estimates?}
\end{center} 
We seek guarantees that explicitly account for the number of bodies and their geometric conditioning, rather than treating all samplers within a given $\ell_1$ ball as equally representative of practice. 
The experiments above test precisely these two sources of difficulty: accumulation as $m$ grows and deterioration of sampling quality as $R/r$ grows.

\section{Tail-Controlled Approximate Sampling}
\label{sec:tail-controlled}

The preceding section shows that small sampling error for each convex body
need not translate into a reliable estimate of their union volume: errors
concentrated in geometrically significant regions may accumulate across
bodies. Requiring multiplicative closeness to uniformity everywhere would
rule out this behavior, but is stronger than what the analysis needs.
We instead allow the sampling density to deviate arbitrarily on a small
exceptional region (which are essentially tails of this distribution, hence the term \textit{Tail-Controlled}), while controlling its multiplicative distortion
elsewhere. We first formalize this guarantee, compare it with standard
$\ell_1$ and $\ell_\infty$ notions, and then discuss its realization
through Hit-and-Run.

\subsection{Almost-\texorpdfstring{$\ell_\infty$}{} closeness}

Let $(S,\mathcal F,\unif)$ be a probability space and let
$\distr$ be another probability distribution on $S$, absolutely
continuous with respect to $\unif$. Write
\[
    h(x):=\frac{\dd\distr}{\dd\unif}(x)
\]
for the relative density of $\distr$ with respect to $\unif$.

For $\sampgood\in(0,1)$, define the exceptional region
\begin{equation}\label{eq:bad}\bad_\sampgood(\distr,\unif)
    :=
    \left\{
       x\in S:
       |h(x)-1|>\sampgood
    \right\}.
\end{equation}

Outside this region, the sampling distribution satisfies
\[
    1-\sampgood
    \le h(x)\le
    1+\sampgood.
\]
Thus, $h$ is multiplicatively close to $1$ outside
$\bad_\sampgood(\distr,\unif)$.

\begin{definition}[$(\eta, \beta)$-almost-$\ell_\infty$ closeness]
\label{def:almost-linf}
For $\sampgood,\sampbad\in(0,1)$, we say that $\distr$ is
\emph{$(\sampgood,\sampbad)$-almost-$\ell_{\infty}$-close} to $\unif$ if
\[
    \unif(\bad_\sampgood(\distr,\unif))\le\sampbad
    \qquad\text{and}\qquad
    \distr(\bad_\sampgood(\distr,\unif))\le 2\sampbad.
\]
When it is clear from context, we shall defer to $\distr$ simply as a distribution which is $(\sampgood,\sampbad)$-close to $\unif$.
\end{definition}

The two parameters capture distinct aspects of sampling
accuracy. The parameter $\eta$ bounds the multiplicative
distortion of the density outside the exceptional region,
while $\beta$ controls the size of this exceptional region.

In particular, if $G=S\setminus\bad_\sampgood(\distr,\unif)$,
then for every measurable $A\subseteq G$,
\[
    (1-\sampgood)\unif(A)
    \le \distr(A)\le
    (1+\sampgood)\unif(A).
\]
No pointwise density guarantee is imposed on the
exceptional region, but its probability mass is controlled, under both the target distribution and the sampling distribution.

\paragraph{Relevance to union-volume estimation.}
Consider convex bodies $\body_1,\ldots,\body_\length\subseteq\mathbb R^\dm$.
Let $\unif_i$ denote the uniform distribution on $\distr_i$,
and suppose the sampling distribution $\distr_i$ is
$(\sampgood,\sampbad)$-almost-$\ell_{\infty}$-close to $\pi_i$.

Write $B_i=\bad_\sampgood(\distr_i,\unif_i)$.
Since $\unif_i$ is uniform on $\body_i$,
\[
    \vol(B_i)\le\sampbad\vol(K_i).
\]
Consequently,
\[
    \vol\left(\bigcup_{i=1}^{\length} B_i\right)
    \le
    \sampbad\sum_{i=1}^{\length}\vol(\body_i)
    \le \length\sampbad V,
\]
where
\[
    V=\vol\left(\bigcup_{i=1}^{\length} \body_i\right).
\]

Thus, the natural parameter regime for an
$\err$-relative approximation is
$\sampgood=\Theta(\err)$ and
$\sampbad=\Theta(\nicefrac{\err}{\length})$, with sufficiently
small absolute constants. The latter makes the
exceptional regions occupy at most an $\bigO(\err)$
fraction of the union, while the former bounds the
multiplicative distortion on their complements.
For constant $\err$, these become
$\eta=\Theta(1)$ and $\beta=\Theta(1/M)$.

These observations motivate the sampling guarantees
used by our algorithm. The formal correctness analysis,
including the effect of approximate individual-volume
estimates, is given in Section~\ref{sec:correctness}.

\subsection{Obtaining Tail-Controlled Guarantees for Hit-and-Run}
\label{subsec:tail-hit-and-run}

We now turn to the question of whether the almost-$\ell_\infty$
guarantees required by our framework can be realized efficiently
using geometric random walks. We focus on the classical
Hit-and-Run algorithm \citep{lovasz2006fast}, a widely used
Markov chain Monte Carlo method for approximately uniform
sampling from convex bodies.

Let $\body\subseteq\R^\dm$ be a convex body, and let
$\unif$ denote the uniform distribution on $\body$.
Starting from a point $x_0\in\body$, Hit-and-Run generates
a Markov chain $(X_t)_{t\geq 0}$ as follows. At each step,
it chooses a direction $\mathbf{u}$ uniformly at random
from the unit sphere $\mathbb{S}^{\dm-1}$ and selects
$X_{t+1}$ uniformly from the chord
\[
    C_\body(X_t,\mathbf{u})
    :=\body\cap\{X_t+s\mathbf{u}:s\in\R\}.
\]
Let $\distr_t$ denote the distribution of $X_t$.
The uniform distribution $\unif$ is stationary for this
chain, and classical mixing-time results establish
rapid convergence to $\unif$ under suitable geometric
and initialization assumptions.

Our interest, however, is in obtaining the particular
sampling guarantee required for union-volume estimation.
In view of the discussion in Section~\ref{subsec:comparison-closeness},
the central question is the following:

%\medskip
%\noindent
\begin{question}
For $\eta=\Theta(\varepsilon)$ and
$\beta=\Theta(\varepsilon/M)$, how many Hit-and-Run
steps suffice to ensure that $\distr_t$ is
$(\eta,\beta)$-almost-$\ell_{\infty}$-close to $\unif$?
\end{question}
%\medskip

Suppose that $\body$ contains a Euclidean ball of radius
$r$ and is contained in a Euclidean ball of radius $R$.
Assume further that \textit{the initial point $x_0$ is at
distance at least $\empinrad>0$ from the boundary
of $\body$}.

% A classical mixing-time bound of Lov\'asz and
% Vempala~\cite{lovasz2006fast} implies, under the
% corresponding assumptions, that
% \[
%     d_{\mathrm{TV}}(\distr_t,\unif)
%     \leq \frac{\eta\beta}{2}
% \]
% after
% \[
%     t=
%     \bigO\left(
%         \dm^3\frac{R^2}{r^2}
%         \log\frac{R}{\empinrad\eta\beta}
%     \right)
% \]
% steps. Combining this with
% Observation~\ref{obs:l1-to-almost-linf}
% immediately yields the following consequence.

\begin{corollary}[Tail-controlled Hit-and-Run]
\label{cor:tail-hit-and-run}
Under the preceding assumptions, Hit-and-Run
produces an $(\sampgood,\sampbad)$-almost-$\ell_{\infty}$-close distribution
to $\unif$ after
\[
    \bigO\left(
        \dm^3\frac{\circrad^2}{\inrad^2}
        \log\frac{\circrad}{\empinrad\sampgood\sampbad}
    \right)
\]
steps.
\end{corollary}
This is a consequence of the mixing property of \har{}~\citep{lovasz2006fast}.

Although this establishes that the desired sampling
guarantee can be achieved in polynomial time, the
argument proceeds through the substantially stronger
requirement of $\sampgood\sampbad$-$\ell_1$ closeness.
As discussed in the preceding subsection, such
accuracy is sufficient but not necessary for
almost-$\ell_\infty$ closeness. This raises the
natural question of whether the desired guarantee
can be obtained with a smaller number of
random-walk steps.

We somewhat address this question through a refined analysis
of Hit-and-Run, obtaining the following improved
bound.

\begin{theorem}[Improved tail-controlled Hit-and-Run,
informal]
\label{thm:improved-tail-hit-and-run-informal}
Under the geometric and initialization assumptions
of our analysis, Hit-and-Run produces an
$(\sampgood,\sampbad)$-almost-$\ell_{\infty}$-close distribution to $\unif$
after
\[
    t=
    \bigO\left(
        \dm^2\frac{\circrad^2}{\inrad^2}
        \left[
            \dm\log\frac{\circrad}{\empinrad}
            +\log\frac{1}{\sampgood\sampbad}
        \right]
    \right)
\]
steps.
\end{theorem}

In particular, for the parameter regime relevant
to union-volume estimation, namely
$\sampgood=\Theta(\varepsilon)$ and
$\sampbad=\Theta(\varepsilon/\length)$, this yields
\[
    t=
    \bigO\left(
        \dm^2\frac{\circrad^2}{\inrad^2}
        \left[
            \dm\log\frac{\circrad}{\empinrad}
            +\log\frac{\length}{\err}
        \right]
    \right).
\]

Thus, our analysis improves the dependence on the
sampling-accuracy parameters relative to the
preceding TV-based bound, while retaining the
almost-$\ell_\infty$ guarantee required by $\algo{}$.
The formal statement, proof, and resulting
complexity bounds are presented in
Section~\ref{sec:union-complexity}.

In fact, we believe even this dependency is not optimal. Heuristically, the factor of $\log(\length/\err)$ should not have a dependency of $\dm^2$ in front of it. We conjecture that it is possible to get a bound of order

\[t=\bigO\!\left(\dm^2\frac{\circrad^2}{\inrad^2}\left[\dm\log\frac{\dm\circrad^2}{\empinrad\inrad}+\log\frac1\sampgood\right]\right)+\bigO\!\left(\frac{\circrad}{\inrad}\log\frac{1}{\sampbad}\right)\]

which, with $\sampgood=\Theta(\err)$ and
$\sampbad=\Theta(\err/\length)$ yields

\[t=\bigO\!\left(\dm^2\frac{\circrad^2}{\inrad^2}\left[\dm\log\frac{\dm\circrad^2}{\empinrad\inrad}+\log\frac1{\err}\right]\right)+\bigO\!\left(\frac{\circrad}{\inrad}\log\frac{\length}{\err}\right)\]

We elaborate on this conjecture later in \cref{sec:conclusion}.

\section{Algorithmic Framework: \texorpdfstring{$\algo$}{}}\label{sec:union-algorithm}

Using the notion of tail-controlled sampling, we develop the algorithm $\algo{}$ that estimates the volume of union of convex bodies. The analysis of the algorithm is done in the next section. 

\subsection{Overview and intuition}
We are given a collection $\{\body_1,\ldots,\body_\length\}$ of convex bodies in $\R^\dm$ and wish to estimate the volume of their union, i.e. solve Problem~\ref{pr:union-volume-estimation}. We assume that all the bodies, $\body_i$'s, are subsets of $\Omega$, where $\Omega$ is a compact measurable subset of $\mathbb{R}^n$ with volume $|\Omega|$. 

The guiding idea is to maintain a multiset $\bucket$ of representative points at a common sampling intensity $p \in (0,1)$. Here, \textit{intensity} quantifies the contribution made to the random quantity $\bucketsize$ by each representative point. Ideally, a measurable region of volume $v$ would contribute a Poisson number  of representatives with mean $pv$. If this property holds for the union, then $|\bucket|/p$ acts a natural estimator of the union-volume.

The first challenge is \textit{accounting for the overlaps}. While processing a body $\body_i$ from a given arbitrary ordering of the convex bodies, we remove from $\bucket$ every point lying in $\body_i$ before adding new samples from that body. In this way, the representatives retained at the end correspond to disjoint portions of the union, determined by the last body in the chosen ordering that contains each point. To obtain intensity $p$ per unit volume in $\body_i$, one would ideally draw $N_i\sim\pois(p\vol(\body_i))$, and then choose $N_i$ independent points uniformly from $\body_i$. 

The second challenge is \textit{to keep the total number of retained samples $|\bucket|$ bounded}. Whenever the proposed insertion of new samples exceeds a threshold, we independently retain each existing point with probability $1/2$ and replace $p$ by $p/2$. Accordingly, we reduce the proposed sample size from $N_i$ to $\texttt{Binomial}(N_i, 1/2)$. This ensures the thinning property of Poisson random variables: at a fixed intensity, independently thinning a Poisson sample $X$ via $\texttt{Binomial}(X, 1/2)$ produces a Poisson sample at half the intensity.

There is, however, a more fundamental challenge: \emph{the ideal sampling process cannot be implemented exactly with the available efficient algorithms}. Neither can we sample points uniformly, nor do we have the ability to do an exact computation of the volume. For each convex body $\body_i,$ we have only a PAC estimate $\empvol_i$ of $\vol(\body_i)$, and each generated point follows a distribution $\distr_{i}$ that is $(\eta,\beta)$-close to the uniform distribution $\unif_{\body_i}$.

These two approximations affect different aspects of the final sample set over the union. To see this, let us denote $h_i\define \frac{\dd\distr_i}{\dd\unif_{\body_i}}$ to be the Radon-Nikodym derivative. For a measurable set $A\subseteq\body_i$, the expected number of newly generated points in $A$, at a fixed intensity $p$ and conditional on $\empvol_i$, is
\[
 p\empvol_i\,\distr_{i}(A)
 =p\,\frac{\empvol_i}{\vol(\body_i)}\int_A h_i(x)\,\dd x.
\]
Thus, the volume-estimation error changes the \emph{overall scale} of the sampling intensity, whereas the sampling error changes its \emph{spatial distribution}. Even if the total number of generated points is close to the ideal count, they may be distributed unevenly among overlapping and non-overlapping portions of the bodies. This matters because the algorithm removes points according to their locations, not merely their number.  

Let us look back to the particular case of Sunflower families. Note that the key reason for ineffectiveness of $\TV$-close samplers was this sampling bias. It might happen, that the provided samplers are heavily biased towards the overlapping portions, rather than the portions which are unique to the convex bodies. Thus, points from overlapping portions are sampled; only to be thrown out later, leaving no trace of the convex body from which it was sampled. And hence it causes a major collapse in the point sampling based framework for volume estimation.

The almost-$\ell_\infty$ guarantee gives a useful way to separate these effects. Outside the exceptional set $\bad_\sampgood$ defined in \cref{def:almost-linf}, the density ratio $h_i$ lies between $1-\sampgood$ and $1+\sampgood$. Hence, on this good portion of a body, the local sampling intensity differs from the ideal intensity by controlled multiplicative factors, provided the volume estimate is accurate. Since the exceptional portion cannot be treated pointwise in the same way; its contribution must be bounded via the bounds on the exceptional region, which we obtain from the definition of almost-$\ell_\infty$ close sampling. We explain these guarantees with the oracles below and quantify their effect in the correctness proof.

Finally, the sampling intensity $p$ is reduced adaptively, depending on the points already retained and the proposed Poisson counts. Although Poisson thinning explains the operation at a fixed rate, the analysis must also justify concentration for the resulting adaptive procedure. 

The algorithmic framework $\algo$ is described next in details, while its correctness and oracle efficiency are established later in~\Cref{sec:correctness}.

\subsection{Volume and Sampling Oracles}\label{sec:subroutine-descriptions}

The algorithm $\algo$ uses a couple of oracles deployed as subroutines, and the guarantees and implementation of them are crucial both for the analysis of correctness and the overall complexity analysis of \algo. Let us first describe the guarantees given by the subroutines. We will analysis the implementation and analysis of these subroutines later.  

\paragraph{I. PAC Volume Estimation.}
The subroutine $\compvol$ takes a convex body $\body\subseteq\R^\dm$, its geometric parameters $(\inrad,\circrad)$, and accuracy parameters $(\volerr,\volconf)$. It returns an estimate $\empvol$ satisfying
\[
\Pr\!\left[(1-\volerr)\vol(\body)\leq \empvol\leq (1+\volerr)\vol(\body)\right]\geq 1-\volconf.
\]
We use the volume-estimation algorithms discussed in \cite{cousins-vempala18}.
Their implementation and resource requirements are treated separately in the complexity analysis.

\paragraph{II. $(\eta,\beta)$-almost Uniform Point Sampling.}
The subroutine $\gensamp$ takes a convex body $\body\subseteq\R^\dm$, its geometric parameters $(\inrad,\circrad)$, and parameters $(\sampgood,\sampbad)$. Each invocation produces an independent point in $\body$ according to a distribution $\distr$ satisfying the $(\sampgood,\sampbad)$-almost-$\ell_\infty$-uniform guarantee of \cref{def:almost-linf}. Such sampling guarantees can be realised using suitable implementations of \har{} (ref. \cref{sec:prelims-point-sampling}).

% \paragraph{Interpretation of the sampling guarantee.}
% We use the notation of \cref{def:almost-linf, sec:prelims-point-sampling} throughout. In particular, for a sampling distribution $\distr$ absolutely continuous with respect to the uniform distribution $\unif$ on $\body$, write $h=d\distr/d\unif$. The exceptional set $\bad_\sampgood$ is already defined in \cref{eq:bad} by
% \[
% \bad_\sampgood=\{x\in\body:|h(x)-1|>\sampgood\}.
% \]
% The $(\sampgood,\sampbad)$-closeness guarantee asserts that $\unif(\bad_\sampgood)\le\sampbad$, and $\distr(\bad_\sampgood)\le 2\sampbad$. Outside this set, the relative sampling density differs from $1$ by at most $\sampgood$. Thus the exceptional region is small both in uniform measure and in sampling probability, although these are distinct statements.

% For each body $\body_i$, we apply the preceding notation with $\unif=\unif_{i}$ and the corresponding sampling distribution $\distr_{i}$. When the body needs to be made explicit, we write $\bad_{i}$ for its exceptional set (the $\sampgood$ mention is dropped; as the value of $\sampgood$ is fixed throughout the algorithm). No new definition of good or bad regions is needed here; the correctness analysis will refer directly to these sets and to their complements.

\subsection{The Main Algorithm: \texorpdfstring{$\algo$}{}}\label{sec:main-algorithm}
We assume that each body $\body_i$ is $(\inrad,\circrad)$-fat convex body, that is,  it contains an sphere of radius at most $\inrad > 0$ and is circumscribed by a sphere of radius at least $\circrad \ge \inrad > 0$. %We also assume that volume of each $\body_i$. 
The algorithm $\algo{}$ takes the number of bodies $\length$, approximation parameters $(\err,\conf)$, and the geometric specifications $(\inrad,\circrad)$ as input. It maintains a multiset $\bucket$ of sampled points and a common sampling intensity $p$. The threshold $\thresh$ and the accuracy parameters of the two subroutines are fixed as functions of the input parameters. %For precise choices, we refer to the analysis.

\begin{algorithm}
\caption{\algo ($\err, \conf, \length, \inrad, \circrad$)}\label{algo:alpona}
\begin{algorithmic}[1]
\State Initialise $\thresh\leftarrow \threshval$ \label{line:threshold-init}
%\State Initialise $\probthresh \leftarrow \probthreshval$
\State Initialise $p\leftarrow 1$ \label{line:prob-init}
\State Initialise multiset $\bucket \leftarrow \emptyset$ \label{line:bucket-init}
\State Set parameters: $\volerr = \volerrval, \volconf = \volconfval, \sampgood = \sampgoodval, \sampbad = \sampbadval$ \label{line:parameters-init}
\For{$i = 1$ \textbf{to} $\length$}\label{line:forloopstarts}
    \For {all $x \in \bucket$}\label{line:remove-elts}
        \State If $x\in \body_i,$ remove all copies of $x$ from $\bucket$ \label{line:remove-elts2}
        \EndFor
    \State {Query $\empvol_i \leftarrow \textsc{ComputeVolume}(\body_i, \volerr, \volconf, \inrad, \circrad)$}\label{line:compute-volume}
    \State Draw $N_i \sim \pois(\empvol_i\cdot p)$ \label{line:poisson-draw}
    \While{$(|\bucket|+N_i>\thresh)$ }\label{line:halving-routine}
        \State {For each $y\in \bucket,$ throw $y$ from $\bucket$ with probability 1/2}
        \State {$N_i \leftarrow\texttt{Binomial}(N_i, 1/2)$ and $p\leftarrow p/2$}
       % \State {$\textbf{if }p<2^{-L}, \textbf{output } \perp$} \label{line:halving-routine-aborts}
        \EndWhile \label{line:halving-routine-ends}
    \For{$j = 1$ \textbf{to} $N_i$}\label{line:generate-samples}
        \State Query $z\leftarrow\gensamp(\body_i, \sampgood, \sampbad, \inrad, \circrad)$
        \State $\bucket\leftarrow \bucket\cup \{z\}$
        \EndFor\label{line:generate-samples-ends}
    \EndFor \label{line:forloopends}
\State \textbf{return} $\bucketsize / p$ \label{line:output}
\end{algorithmic}
\end{algorithm}

In lines~\ref{line:threshold-init}--\ref{line:parameters-init}, the algorithm sets the storage threshold $\thresh$, initializes the common sampling intensity $p$ to $1$, and starts with an empty multi-set $\bucket$, that would contain the samples from the union of bodies seen so far. It also fixes the accuracy and confidence parameters for the volume and sampling subroutines. The choices of these parameters are justified in the analysis.

Inside the \textbf{for loop} (lines~\ref{line:forloopstarts}--\ref{line:forloopends}), the multi-set $\bucket$ is updated according to the bodies $\body_1$ to $\body_{\length}$. For each body $\body_i$, the algorithm firstly (in lines~\ref{line:remove-elts}--\ref{line:remove-elts2}) removes every occurrence in $\bucket$ whose location belongs to $\body_i$. Thus, if a point belongs to several bodies, all the previous occurrence of the point does not have any influence on its inclusion in $\bucket$, suggesting that every measurable subset of the union is processed by the algorithm exactly once (during it's last occurrence in the ordering). The resulting last-occurrence decomposition of the union is formalized in the correctness proof.

In line~\ref{line:compute-volume}, \algo{} computes the PAC estimate $\empvol_i$. 
In line~\ref{line:poisson-draw}, it draws $N_i\sim\pois(p\empvol_i)$, i.e., the number of representatives from $\body_i$ that would be added to $\bucket$ so that each point in the union $(\body_1 \cup \dots \cup \body_i)$ is equally represented. 
Since $\empvol_i$ is an estimate, the intended number of representatives is proportional to the \emph{estimated} volume, rather than the unknown true volume. This distinction, together with the non-uniformity of the generated points, is explicitly accounted for in the correctness proof.

If inserting $N_i$ new points would exceed $\thresh$, the algorithm independently retains each occurrence already in $\bucket$ with probability $1/2$ and halves $p$ (lines~\ref{line:halving-routine}--\ref{line:halving-routine-ends}), and appropriately reduce the estimated number of points to be sampled from $\body_i$ and resamples the points in $\bucket$. 
%It returns $\perp$ if the prescribed lower cutoff on $p$ is crossed (line~\ref{line:halving-routine-aborts}). (Note that this line is for easier understanding of the algorithm and dropping this line does not change the correctness of the algorithm.) 
%Otherwise i
%It reduce the $p$ and appropriately reduce the estimated number of points to be sampled from $\body_i$ and resamples the points in $\bucket$. 
The loop continues until the proposed count fits within the storage threshold. 
%The cutoff $2^{-L}$ is a safeguard; its choice and the probability of reaching it are analyzed later.
Once the count fits, \algo{} (in lines~\ref{line:generate-samples}--\ref{line:generate-samples-ends}) makes $N_i$ independent calls to $\gensamp$ and inserts the resulting points into $\bucket$. The points follow the almost-$\ell_\infty$-uniform sampling distribution described above, rather than an exactly uniform distribution.

After all bodies have been processed, the algorithm returns $|\bucket|/p$. Informally, $|\bucket|$ is the number of retained representatives at the final sampling intensity $p$, and division by $p$ rescales the count into a volume estimate. Proving that this estimate is accurate despite approximate subroutines and adaptive halving is the subject of the next section.

\subsection{Relation to the Existing \uve{} Algorithms}\label{sec:challenge-related}

While randomized polynomial-time approximation algorithms
for estimating the volume of convex bodies are well
established~\citep{dyer1991random,lovasz1990mixing,cousins-vempala18},
extending these techniques to non-convex regions presents
additional challenges. In particular, a union of convex
bodies need not be convex, and the geometric properties
underlying efficient random-walk-based volume estimation
may no longer hold. An important special case is Klee's
Measure Problem, which asks for the volume of a union
of axis-parallel boxes and has been extensively studied
in computational geometry.

The classical framework of \citet{karp-luby-madras89} provides randomized
approximation algorithms for unions using size,
uniform-sampling, and membership oracles. \citet*{BringmannF10} extended
this approach to unions of high-dimensional geometric
objects, including general convex bodies and settings
with approximate oracles. In a different computational
model, \citet{meel-vinodchandran-chakraborty21, 
meel-chakraborty-vinodchandran22} developed
space-efficient algorithms for estimating unions
of implicitly represented sets, introducing the
framework of Delphic sets and its approximate-oracle
generalization. Their results yield efficient
algorithms for discrete versions of Klee's Measure
Problem and were subsequently improved by~\citet{Nandi24}. More recently,
\citet{BringmannLNRW25}
established an $\Omega(M/\varepsilon^2)$ lower bound
for general oracle-based union-volume estimation,
matching the classical query complexity, and
obtained faster algorithms for Klee's Measure
Problem by exploiting the geometric structure
of axis-parallel boxes.

Our algorithm, $\algo$, addresses a complementary
aspect of this problem. Unlike the algorithms for
Delphic sets, which operate over a discrete universe,
$\algo$ works directly with continuous convex
bodies, without discretizing the underlying space.
Moreover, while the oracle-based framework of~\citet{BringmannF10} already permits 
oracles which may be unrealistic,  our emphasis is on the guarantees required
when these oracles are implemented using randomized
volume-estimation procedures and geometric random
walks. In particular, $\algo$ uses PAC volume
estimates and almost-$\ell_\infty$-uniform samples,
together with Poissonization and adaptive thinning,
to control the combined effects of approximation
errors and overlapping regions. Thus, rather than
improving the general oracle-query lower bound,
our contribution is an end-to-end analysis of a
sampling-based algorithm using concrete approximate
geometric primitives.

\section{Correctness Proof of \texorpdfstring{$\algo$}{}}\label{sec:correctness}

In this section, we formally state and prove the guarantees offered by the algorithm $\algo{}.$ The proof proceeds by first
showing that almost all of this volume is contained in
regions where the sampler behaves approximately uniformly,
and then showing that the algorithm accurately estimates
the volume of these regions.

\begin{theorem}[Correctness and oracle efficiency of \algo]\label{thm:main-theorem}
    Given input parameters $(\err, \conf, \length, \inrad, \circrad)$ and a collection $\{\body_1, \dots, \body_\length\}$ of $\length$ many $(\inrad, \circrad)$-fat convex bodies, the algorithm $\algo{}$ yields an $(\err, \conf)$-correct \uve, i.e., it returns a value $\Tilde{V}$ such that 
    \begin{align*}
        (1 - \err) \vol\left( \bigcup_{i=1}^{\length} \body_i\right) \le \Tilde{V} \le (1 + \err) \vol\left( \bigcup_{i=1}^{\length} \body_i\right)
    \end{align*}
    with probability at least $1 - \conf$, with $\bigO(\length)$ many calls to the PAC volume computation oracle and $\bigO \left( \frac{\length}{\err^2} \log \nicefrac{\length}{\conf} \right)$ many calls to the $(\sampgood, \sampbad)$-tailed sampler oracle.
\end{theorem}

\subsection{Overview of the Proof}\label{sec:proof-overview}

We write $\unionvol:=\vol(\bigcup_{j=1}^{\length}\body_j)$ for the volume of the union. Since the algorithm $\algo{}$ is a random process, it makes sense to talk about it's output from the perspective of distributions. For simplicity of the proof, we consider a variant $\algomod{}$ of the main algorithm. Note that in $\algomod{}$, we are assuming that we know the quantity $\nicefrac{1}{\unionvol}$ beforehand; and that the only difference between $\algo$ and $\algomod$ is in line~\ref{line:halving-routine2}. If $p_*$ denotes the value of $p$ at the end of $\algo$, the event where $p_* > \nicefrac{1}{\goodunionvol}$ should correspond to runs of $\algomod$ with the same output. Thus, we consider an event $\failprob$ where $p$ falls too low. 

Our first claim (\cref{claim:alpona-alpona-mod}) informally tells us that as long as we are capable of proving that $\failprob$ occurs with very low probability, the output distributions of $\algo$ and $\algomod$ are very close to each other w.r.t Total Variation distance. As a consequence, if we are able to prove a correctness guarantee upon $\algomod$, that should give us a correctness guarantee on $\algo$ as well (after accounting for this TV-distance between the two distributions). Our objective moving forward, will therefore be upon trying to prove the correctness of $\algomod$.

Recall the good and bad regions associated with the sampler
$\gensamp$, as defined in
\cref{sec:subroutine-descriptions}.
In particular, $\bad_j$ denotes the bad region of
$\body_j$, and $\good_j:=\body_j\setminus\bad_j$ its good
region. To avoid counting overlapping regions multiple
times, we assign each point in the union to the last body
(in the fixed processing order) that contains it. Formally,
we define
\[
    \last_j:=\body_j\setminus\bigcup_{i>j}\body_i.
\]
These \textit{unique sections} $\last_j$ are pairwise disjoint and their union
is precisely the union of the convex bodies. We write
$\gd{j}$ and $\bd{j}$ for the good and bad parts of
$\last_j$, respectively.

The first step of the proof is to show that the total volume
of the good parts,
\begin{equation}\label{def:F0'-definition}
    \goodunionvol :=
    \vol\left(\bigcup_j(\last_j\setminus\bad_j)\right)
    = \sum_j\vol(\gd{j}),
\end{equation}
is close to $\unionvol$. This is established in
\cref{claim:F0'-F0-equivalence}. Consequently, it suffices
to show that the output of the algorithm is a sufficiently
accurate estimate of $\goodunionvol$.

To analyse the output, we partition the surviving samples
according to whether they originate from the good or bad
parts of the corresponding bodies. Let $\prob$ denote the sampling intensity at the end of a run of $\algomod$. Let $\estgood$ and
$\estbad$ denote the respective numbers of such samples.
Thus, whenever the algorithm returns a numerical estimate,
\[
    \bucketsize=\estgood+\estbad,
    \qquad
    \out=\frac{\estgood+\estbad}{\prob}.
\]
We show that $\estgood/\prob$ is concentrated around
$\goodunionvol$, while $\estbad/\prob$ contributes only
a small additive error. Together, these establish that
$\out$ approximates $\unionvol$.

The analysis must account for four possible sources of
failure: an inaccurate estimate returned by the volume
oracle ($\failvol$), a significant deviation in the
number of good samples ($\failgood$), and an excessive
number of bad samples ($\failbad$). In
\cref{claim:fail-decomposition}, we show that the algorithm
succeeds whenever none of these events occurs. The
remainder of the proof bounds the probabilities of these
four events individually and combines the bounds using
a union bound to obtain the desired correctness guarantee.

\paragraph{Proof of \cref{thm:main-theorem}.} First, we prove \cref{claim:alpona-alpona-mod} to show that the TV-distance between the output distributions of \algo{} and \algomod{} is at most $\frac{\conf}{4}.$ Then, we are left to prove that \algomod{} is an $(\err, \conf)$-\uve{} algorithm with probability at most $\frac{3\conf}{4}$.

In this context, we prove \cref{claim:fail-decomposition}, i.e., the event where the algorithm \algomod{} results in a failure can be decomposed into \textbf{three} events: $\failvol$ due to the approximation errors of PAC volume estimator, $\failgood$ due to deviations of the `good' sample counts, and $\failbad$ due to the `bad' samples. 

In the next subsections, we bound each of their probabilities of occurrence:
\begin{enumerate}
    \item We prove in \cref{claim:fail-volume-computation} that $\pr[\failvol] < \frac{\conf}{4}.$
    \item We prove in \cref{claim:fail-good-estimate} that $\pr[\failgood] < \frac{\conf}{4}.$
    \item We prove in \cref{claim:fail-bad-estimate} that $\pr[\failbad] < \frac{\conf}{4}.$
\end{enumerate}
Combining the three above inequalities tells us that the random process \algomod{} outputs a value outside the range $[(1 - \err)\unionvol, (1 + \err)\unionvol]$ is at most $\frac{3\conf}{4}.$ 

By~\Cref{lem:indistinguish}, then we can conclude that the probability of \algo{} not yielding a value in the range $[(1 - \err)\unionvol, (1 + \err)\unionvol]$ is at most:
\[ \frac{3\conf}{4} + \frac{\conf}{4} = \conf, \]
as desired. \qed

Now, we are going to formally introduce \algomod{}, and prove the claims corresponding to \algomod{} in the next sections. 

\subsection{\algomod{}: A Technical Modification for Analysis}

For simplicity in proving correctness, we define a modified version of \algo{} below strictly for the purpose of analysis. As one can note the key distinction between the two algorithms is the inclusion of the clause $p \ge \nicefrac{1}{\unionvol}$. Notably in this version of the algorithm there is no strict upper bound on $\bucketsize$, because of which we can make observations as demonstrated in \cref{lem:mean-properties}, something we could not do for $\algo$ itself because in the original algorithm there is an implicit upper bound $\bucket < \thr$.

Consider the event $\failprob$, where the value of the sampling intensity in \algo{} decreases below $\nicefrac{1}{\unionvol}$. If the $\failprob$ event does not occur, then the two algorithms are essentially the same; and thus must output the same quantity. In other words, the only situation where the output distributions of the two algorithms would differ is when $\failprob$ would occur. This intuition is presented formally below.

\begin{algorithm}[H]
\caption{\algomod ($\err, \conf, \length, \inrad, \circrad$)}\label{algo:alpona2}
\begin{algorithmic}[1]
\State Initialise $\thresh\leftarrow \threshval$ \label{line:threshold-init2}
%\State Initialise $\probthresh \leftarrow \probthreshval$
\State Initialise $p\leftarrow 1$ \label{line:prob-init2}
\State Initialise multiset $\bucket \leftarrow \emptyset$ \label{line:bucket-init2}
\State Set parameters: $\volerr = \volerrval, \volconf = \volconfval, \sampgood = \sampgoodval, \sampbad = \sampbadval$ \label{line:parameters-init2}
\For{$i = 1$ \textbf{to} $\length$}\label{line:forloopstarts2}
    \For {all $x \in \bucket$}\label{line:remove-elts22}
        \State If $x\in \body_i,$ remove all copies of $x$ from $\bucket$ \label{line:remove-elts222}
        \EndFor
    \State {Query $\empvol_i \leftarrow \textsc{ComputeVolume}(\body_i, \volerr, \volconf, \inrad, \circrad)$}\label{line:compute-volume2}
    \State Draw $N_i \sim \pois(\empvol_i\cdot p)$ \label{line:poisson-draw2}
    \While{$(|\bucket|+N_i>\thresh)$  {\color{blue} and  $(p\geq 1/\unionvol)$}}\label{line:halving-routine2}
        \State {For each $y\in \bucket,$ throw $y$ from $\bucket$ with probability 1/2}
        \State {$N_i \leftarrow\texttt{Binomial}(N_i, 1/2)$ and $p\leftarrow p/2$}
       % \State {$\textbf{if }p<2^{-L}, \textbf{output } \perp$} \label{line:halving-routine-aborts}
        \EndWhile \label{line:halving-routine-ends2}
    \For{$j = 1$ \textbf{to} $N_i$}\label{line:generate-samples2}
        \State Query $z\leftarrow\gensamp(\body_i, \sampgood, \sampbad, \inrad, \circrad)$
        \State $\bucket\leftarrow \bucket\cup \{z\}$
        \EndFor\label{line:generate-samples-ends2}
    \EndFor \label{line:forloopends2}
\State \textbf{return} $\bucketsize / p$ \label{line:output2}
\end{algorithmic}
\end{algorithm}

\begin{cl}[Closeness of \algo{} and \algomod{}]\label{claim:alpona-alpona-mod}
    The total variation distance between the output distribution of the two algorithms \algo{} and \algomod{} is upper bounded by $\Pr(\failprob)$.
\end{cl}
\begin{proof}
Couple the two executions of \algo{} and \algomod{} from shared randomness. Let $O$ and $O_{\mathsf{mod}}$ denotes their respective outputs. During this coupled execution \algo{} differs from \algomod{} when the condition of $p \geq 1/ \unionvol$ in line~\ref{line:halving-routine2} prevents \algomod{} from performing one more halving that \algo{} performs. By the coupling inequality (\cref{lem:coupling-inequality}), therefore the total variation distance is upper bounded by:
    \[\Pr(O \neq O_{\mathsf{mod}}) \leq \Pr(p \leq 1/ \unionvol) \leq \Pr(\prob \leq 1/ \unionvol) = \Pr(\failprob).\]
This completes the proof of the claim. 
\end{proof}

Since each of run \algo{} which results in the final sampling intensity decreasing below $\nicefrac{1}{\unionvol}$ corresponds to a run of \algomod{} resulting in the sampling intensity decreasing below $\nicefrac{1}{\unionvol}$, moving forward we can equivalently consider $\failprob$ to be the event where the sampling intensity of \algomod{} at the end of a run falls below $\nicefrac{1}{\unionvol}$.

    % \begin{enumerate}
    %     \item $\failvol \define$ At least one \compvol query returns an incorrect estimate of the volume.

    %     % \item $\failprob \define$ All the \compvol queries return accurate estimates, but the value $\prob$ at the end of the algorithm falls below $\nicefrac{1}{\unionvol}$.

    %     \item $\failgood \define$ The \compvol queries are all correct, and we have $\prob \ge \nicefrac{1}{\unionvol},$ but $\estgood$ does not lie in $\left(1\pm \frac{15}{32}\err\right)\prob\goodunionvol.$

    %     \item $\failbad \define$ The \compvol queries are all correct, and we have $\prob \ge \nicefrac{1}{\unionvol},$ but $\estbad$ takes a value larger than $\frac{15}{32}\err\prob\goodunionvol.$

    % \end{enumerate}

\subsubsection{Step I. The Decomposition of Failure of \algomod}

% {\color{red}
% Recall the parameters that were set in the algorithm. For notational simplicity, we shall denote $\thresh$ by $\thr$. We have $0 < \err,\conf < 1,$ and:
% \begin{align*}
%     \volerr = \volerrval,\ \volconf = \volconfval,\ \sampgood = \sampgoodval,\ \sampbad = \sampbadval,\ \thr = \threshval,\  \probthresh = \probthreshval.
% \end{align*}
% \Aranya{Hard to read the parameter initializations, make a table or something?}}

As explained in the previous section, our current strategy is to prove the correctness of $\algomod$. We proceed along the lines described in \cref{sec:proof-overview}; firstly we shall formally reduce the problem of estimating $\unionvol$ into the problem of estimating $\goodunionvol$. We do this by strictly upper-bounding the mass stored in exceptional regions of each convex body.

\begin{cl}[Equivalence of $\unionvol$ and $\goodunionvol$]\label{claim:F0'-F0-equivalence}
    Referring to notations defined previously (\ref{def:F0'-definition}), we have:
    \begin{equation*}
        \left(1-\sampbad\right)\unionvol = \left(1-\frac{\err}{32}\right)\unionvol \le \goodunionvol \le \unionvol.
    \end{equation*}
    If the output of the algorithm is in the range $\out\in \left(1\pm\gooderr\right)\goodunionvol,$ we will have $\out\in (1\pm\err)\unionvol.$
\end{cl}

\begin{proof}
    Observe that we can write $\unionvol$ as a sum of $\vol(\last_j)$, since the union of all the sets can be viewed as a disjoint union of the unique sections. We can further break down each $\vol(\last_j)$ into a sum of volumes of good and bad sections, and bound the volume of bad sections by the guarantees of almost-$\ell_\infty$ sampling. 
        \begin{align*}
            \unionvol = \sum_{j=1}^{\length} \vol(\last_j) &= \sum_{j=1}^{\length} \vol(\gd{j}) + \vol(\bd{j})
            \\ &\le \goodunionvol + \sum_{j=1}^{\length} \vol(\bad_j)
            \\ &\le \goodunionvol + \sampbad\unionvol \length = \goodunionvol + \frac{\err}{32}\unionvol.
        \end{align*}
    The second part is simple algebraic manipulation. Note that by default we have $(1 + \gooderr\err) \goodunionvol \le (1 + \err) \unionvol$. For the other direction, we apply the above obtained lower bound on $\goodunionvol$.
    \[ \left(1 - \gooderr\err \right) \goodunionvol \ge \left( 1 - \gooderr\err\right)\left(1 - \sampgoodval\right)\unionvol \ge (1 - \err) \unionvol. \]
    
\end{proof}

\begin{cl}[Decomposition of Failure]\label{claim:fail-decomposition}
    The event where a run of $\algomod$ results in a failure, i.e. the output distribution of $\algomod$ returns a value outside $(1\pm \err)F_0$ is contained in the union of the following events:

    \begin{enumerate}
        \item $\failvol \define$ At least one \compvol query returns an incorrect estimate of the volume.

        % \item $\failprob \define$ All the \compvol queries return accurate estimates, but the value $\prob$ at the end of the algorithm falls below $\nicefrac{1}{\unionvol}$.

        \item $\failgood \define$ The \compvol queries are all correct, and we have $\prob \ge \nicefrac{1}{\unionvol},$ but $\estgood$ does not lie in $\left(1\pm \frac{15}{32}\err\right)\prob\goodunionvol.$

        \item $\failbad \define$ The \compvol queries are all correct, and we have $\prob \ge \nicefrac{1}{\unionvol},$ but $\estbad$ takes a value larger than $\frac{15}{32}\err\prob\goodunionvol.$

    \end{enumerate}

    % Additionally, the event where a run of $\algo$ results in a failure in the union of $\failprob$ in addition to the above events.
\end{cl}

\begin{proof}
    Observe that if a run of $\algomod$ belongs to the intersection of complements of the above mentioned events, then $\out = \frac{\estgood + \estbad}{\prob} \in (1\pm \gooderr) \goodunionvol$ by adding the bounds for $\estgood$ and $\estbad.$ By \cref{claim:F0'-F0-equivalence}, it would therefore suggest the algorithm has resulted in a success. Thus the failure event must be contained within the union of the above four events.

    % Additionally by \cref{claim:alpona-alpona-mod} we know that the output distributions of $\algo$ and $\algomod$ differ by a total variation distance of at most the probability of $\failprob$. Thus, the event where $\algo$ would output result in failure; is contained in the union of the two events: either $\failprob$ occurs, or $\algomod$ fails.
\end{proof}

Rest of the subsections are structured as follows. A section each is dedicated to bounding the probability of each of the events mentioned in the above claim. Through these series of steps, we shall prove that all the four above mentioned events can occur with probability at most $\nicefrac{\conf}{4}$, and thus by a Union bound argument we give a concrete proof to \cref{thm:main-theorem} at the end of this section.

\subsubsection{Step II. Bounding \texorpdfstring{$\failvol:$}{} Error due to Volume Computation}

We first consider the event of failure due to volume computation. This event is relatively easy to bound; because the volume oracle can output incorrect estimates in any of the $\length$ many queries, so a simple application of union bound should restrict the probability of $\failvol$. This intuition is formalised below. 

\begin{cl}[Bounding Error due to Volume Computation]\label{claim:fail-volume-computation}
    Consider the event $\failvol,$ where at least one of the \compvol queries does not return a $\volerr$-factor approximation to the corresponding volume. Then $\pr[\failvol] < \frac{\conf}{4}.$
\end{cl}

\begin{proof}
    The probability that a single \compvol query does not return a $\volerr$-factor approximation to the corresponding volume is bounded at most by $\volconf.$ By applying union bound over the $\length$ many queries and substituting the values for $\volerr$ and $\volconf,$ we get $\pr[\failvol] < \volconfval\cdot \length = \frac{\conf}{4}.$
\end{proof}

\paragraph{Conditioning on the volume queries.}\label{cond:accurate-volume-queries} Since the volume queries use randomness independent of the sampling randomness, from now on we fix an outcome of all the volume queries in which every query returns an accurate estimate. Thus for each $j,$ we have $\empvol_j \in (1\pm \volerrval)\vol(\body_j),$ \label{eq:empvol-bound} where $\empvol_j$ denotes output of the \compvol query done on $\body_j.$

\subsubsection{Step III. Bounding \texorpdfstring{$\failprob:$}{} Error due to Low Intensity of Points}

Restricting the volume computation error beforehand allows us to be more lax moving forward, as we don't have to needlessly worry abound the cases where volume computation might go wrong. To prove $\failprob$ however, we need to do a bit more work. We start by better explaining the behaviour of the random quantity $\bucketsize$.

Let $X_j$ denote the number of surviving samples belonging to $\last_j$. Using this notation, we can say $\sum_j X_j$ is precisely equal to $\bucketsize$, as it represents all the surviving samples. Recall the definitions defined earlier; $\distr_j$ represents the sampling distribution, corresponding to the \gensamp query on $\body_j$. The quantity $\distr_j(\last_j)\cdot \vol(\body_j)$ reflects the \textit{relative mass} stored in the region $\last_j$ w.r.t the distribution $\distr_j$. The following lemma informally tells us that the random variables $X_j$ are mutually independent, and each $X_j$ follows a Poisson distribution of intensity proportional to the relative mass of $\last_j$.

\begin{lemma}[Properties of $X_j$]\label{lem:Xj-properties}
    The random variables $\langle X_j:1\le j\le \length\rangle$ are mutually independent, and for each $j \in [\length],$ we have:
    \begin{equation*}
        X_j\sim\pois(\prob\empvol_j\distr_j(\last_j)).
    \end{equation*}
\end{lemma}

\begin{proof}
    When the body $\body_j$ gets processed, we draw $N_j\sim \pois(p \empvol_j)$ many samples from it, where the $p$ gets adjusted and decreased to $\prob$ over the course of the algorithm. Now consider the measurable partition of $\body_j$ into two parts: $\last_j$ and $\body_j \backslash \last_j.$ Recall \cref{lem:poissonisation}, since we are drawing from a distribution $\distr_j$ on $\body_j,$ the random variable $X_j$ signifying number of surviving samples from $\last_j$ would follow: $X_j \sim \pois(\prob \empvol_j \distr_j(\last_j))$. Additionally, since the value of the random quantity $X_j$ depend only upon the fresh randomness of the \gensamp queries done while processing $\body_j$, the obtained random quantities $\langle X_j : j \in [\length]\rangle$ would therefore be independent.
\end{proof}

\begin{lemma}[Properties of $\mean$]\label{lem:mean-properties}
    Define $\mean \define \sum_j \empvol_j \distr_j(\last_j)$. Then $\bucketsize \sim \pois(\prob\mean)$, and
    \[ \mean < \left( 1 + \frac{\err}{7}\right)\goodunionvol.\]
\end{lemma}

\begin{proof}
    As the random variable $\bucketsize$ can be written as a sum $\sum_j X_j$ of independent Poisson random variables, by the additivity property of Poisson distributions (\cref{lem:poisson-additivity}) we get the first statement. To prove the bound on $\mean$, we would need to better analyse the relative mass $\distr_j(\last_j)$. Recall that $\unif_j$ represents the uniform distribution over $\body_j$.
    \begin{align}
        \distr_j(\last_j) &= \distr_j(\gd{j}) + \distr_j(\bd{j}) \nonumber
        \\&\le (1+\sampgood)\unif_j(\gd{j}) + \distr_j(\bad_j). \label{eq:break-the-distribution}
    \end{align}
    Multiplying the above equation by $\empvol_j$, the left side of the above equation becomes the expression for $\mean.$ For handling the right side, we bound $\empvol_j \le (1+\volerr)\vol(\body_j)$, as mentioned earlier (\ref{eq:empvol-bound}). First term of the right side is easier to handle because we are bounding the good region of $\last_j$. 
    \begin{align}
        \sum_j \empvol_j (1 + \sampgood)\unif_j(\gd{j}) &\le \sum_j (1 + \sampgood)(1 + \volerr) \vol(\body_j) \unif_j (\gd{j}) \nonumber
        \\&= \left(1 + \sampgoodval\right) \left(1 + \volerrval\right) \sum_j \vol(\gd{j}) \nonumber
        \\&\le \left( 1 + \frac{\err}{15}\right) \sum_j \vol(\gd{j}) = \left( 1 + \frac{\err}{15}\right) \goodunionvol \label{eq:mean-good-upper-bound}
    \end{align}
    The last equality follows from \cref{def:F0'-definition}. For bounding the second term, we shall use the tail control guarantee of $\distr_j$.
    \begin{align}
        \sum_j \empvol_j \distr_j(\bad_j) &\le \sum_j (1 + \volerr) \vol(\body_j) \cdot 2\sampbad \nonumber
        \\&\le 2 \left(1 + \volerrval \right)\sampbadval \length \unionvol \nonumber
        \\&\le 2 \left(1 + \volerrval \right)\frac{\err}{32} \frac{\goodunionvol}{1 - \err/32} < \frac{\err}{15}\goodunionvol \label{eq:mean-bad-upper-bound}
    \end{align}
    The bound on $\unionvol$ is obtained from \cref{claim:F0'-F0-equivalence}. Adding up the two different bounds, we get:
    \begin{equation*}
        \mean \le \left(1 + \frac{\err}{15}\right)\goodunionvol + \frac{\err}{15}\goodunionvol < \left(1 + \frac{\err}{7}\right)\goodunionvol.
    \end{equation*}
\end{proof}

In the same spirit as above, consider the quantity $\tempmean{s} \define \sum_{j = 1}^s \empvol_j \distr_j(\last_j)$ for any index $s\in [\length]$. Observe that if $p \leftarrow p_s$ is the sampling intensity and $\tempbucket{s}$ represents state of the bucket just after processing $\body_s$, then by the same idea as \cref{lem:mean-properties}, we can say that 
    \[ |\tempbucket{s}| \sim \pois( p_s \tempmean{s})\qquad\forall s\in[\length].\]

\begin{cl}[Bounding Probability of $\failprob$]\label{claim:fail-low-prob}
    Consider the event $\failprob,$ where the value $\prob$ falls lower than $\nicefrac{1}{\unionvol}$. Then $\pr[\failprob]<\frac{\conf}{4}.$
\end{cl}

\begin{proof}
    Observe that by the bound obtained in \cref{lem:mean-properties}, we can say that: \[ \frac{2\mean}{\thr} < \frac{2(1 + \err/7)}{\thr} \goodunionvol \le \unionvol. \]
    Define the quantity $\probthresh \define \lceil \log_2 \nicefrac{2\mean}{\thr} \rceil.$ By this observation, we can say that $\probthresh \le \log_2 \unionvol$, which implies $2^{-\probthresh} \ge \nicefrac{1}{\unionvol}$. This would mean the event where $p$ falls below $\nicefrac{1}{\unionvol}$ is contained as a subset inside the event where $p$ falls below $2^{-\probthresh}.$ Our strategy will be therefore to bound the probability of this latter event.
    
    Since the value of $p$ can decrease below the above specified threshold while processing any of the $\length$ many convex bodies; we will need to make sure that this event can collectively happen during any of the $\length$ many situations with low probability. To do so, we isolate the event where $p$ falls below the threshold after processing $\body_s$ for each $s$, individually bound these events and apply an union bound argument to finish the proof.

    Define $\tempprobthresh{s} \define \lceil \log_2 \frac{2\tempmean{s}}{\thr} \rceil$. By definition, we know that $\tempmean{s} \le \mean$ for each $s \in [\length].$ This would mean $2^{-\tempprobthresh{s}} \ge 2^{-\probthresh}.$ Thus the event where $p$ falls below $2^{-\probthresh}$ after processing $\body_s$; is a subset of the event where $p$ falls below $2^{-\tempprobthresh{s}}$ after processing $\body_s$. Therefore the event $\failprob$ is contained in the union of the following events.
    \begin{equation*}
        \tempfailprob{s} \define \{p \text{ falls below }2^{-\tempprobthresh{s}}\text{ after processing }\body_s\}\qquad \forall s\in[\length].
    \end{equation*}
    By definition of $\tempprobthresh{s}$, we obtain the following inequalities, which will be helpful in proving the claim. Our strategy moving forward, will be to individually bound the probability of these events, and then continue with our union bound strategy.
    \begin{align}
        \E[|\tempbucket{s}|] &= 2^{-\tempprobthresh{s}} \tempmean{s} \le \frac{\thr}{2\tempmean{s}} \tempmean{s} \le \frac{\thr}{2} \label{eq:temp-bucket-upper-bound}
        \\ \E[|\tempbucket{s}|] &\ge \frac{1}{2} \frac{\thr}{2\tempmean{s}}\tempmean{s} = \frac{\thr}{4} \label{eq:temp-bucket-lower-bound}
    \end{align}

    The value of $p$ decreased below $2^{-\tempprobthresh{s}}$ when processing $\body_s$ suggests that setting $p \leftarrow 2^{-\tempprobthresh{s}}$ when processing $\body_s$ would cause the bucket to spill over. Since $|\tempbucket{s}|$ follows a Poisson distribution, an application of Chernoff bounds (\cref{lem:poisson-chernoff}) gives us the following.
    \begin{align*}
        \pr\left[\tempfailprob{s}\right] &= \pr[|\tempbucket{s}|>\thr:p \leftarrow 2^{-\tempprobthresh{s}}]
        \\&\le \pr\left[|\tempbucket{s}|>2\cdot \E[|\tempbucket{s}|]:p \leftarrow 2^{-\tempprobthresh{s}}\right] \qquad (\text{by }\ref{eq:temp-bucket-upper-bound})
        \\&\le \exp\left(-\frac{1}{2\cdot 2} \E[|\tempbucket{s}|]\right) \le \exp\left(-\frac{\thr}{16}\right) < \frac{\conf}{4\length} \qquad (\text{by }\ref{eq:temp-bucket-lower-bound})
    \end{align*}
    Since the probability of $p$ falling below $2^{-\probthresh}$ when processing $\body_s$ is bounded by $\frac{\conf}{4\length}$, this means the probability of $p$ falling below $\frac{1}{\unionvol}$ is also bounded by $\frac{\conf}{4\length}$, and this holds for each $s\in [\length].$ By application of Union bound, that means $p$ can fall below $\nicefrac{1}{\unionvol}$ at any point during the algorithm, with probability at most $\frac{\conf}{4}$, as claimed.
\end{proof}
    
\paragraph{Conditioning on the high probability events.}\label{cond:high-enough-probability} In the above proof, not only have we proved that $p$ cannot fall below $\nicefrac{1}{\unionvol}$ with high probability; we have also established the value of $p$ falling below $2^{-\probthresh}$ itself is a low probability event. Thus, moving forward we shall condition all the other events on the event where $\prob$ stays above $2^{-\probthresh}.$ Because in situations where we would need a control over $\prob,$ we could simply break those events into two sub-events: one where $\prob \ge 2^{-\probthresh},$ and one where that does not happen. As the latter is already addressed above, we do not need to do so again.

\subsubsection{Step IV. Bounding \texorpdfstring{$\failgood$: Deviations of the `Good' Samples' Count}{}}

In this section, we focus on the $\failgood$ event. If we manage to bound the probability of this event; that would tell us that the quantity $\estgood$ would be a good estimation of $\goodunionvol$, which is one step closer to proving that the entire estimate itself will be a good estimate of $\goodunionvol$. The structure of proof will follow a structure similar to the proof of \cref{claim:fail-low-prob}. We open with formally defining the random quantity $\estgood$, and break it into a sum of independent random variables. Using the conditioning of low-probability as mentioned in the previous section, combined with probabilistic tools such as concentration techniques and union bounds, we offer a bound for $\goodunionvol$.

Recall that $\estgood$ denotes the number of surviving samples from all the good sections. We can decompose this random quantity into a sum of $\length$ many random variables as follows: let $Y_j$ denote the number of samples from $\gd{j}$ for each $j\in [\length]$. Using this notation, we can say that $\sum_j Y_j$ is precisely equal to $\estgood$.

\begin{lemma}[Properties of $Y_j$]\label{lem:Yj-properties}
    The random variables $\langle Y_j:1\le j\le \length\rangle$ are mutually independent, and for each $j \in [\length],$ we have:
    \begin{equation*}
        Y_j\sim\pois(\prob\empvol_j\distr_j(\gd{j})).
    \end{equation*}
\end{lemma}

\begin{proof}
    Recall the proof of \cref{lem:Xj-properties}. Since $\gd{j}$ is a measurable subset of $\body_j$, again by \cref{lem:poissonisation} the random variable $Y_j$ signifying number of surviving sampels from $\gd{j}$ would follow: $Y_j \sim \pois(\prob \empvol_j \distr(\gd{j}))$ as claimed. Independence of $Y_j$'s follow once again since we use mutually independent sources of randomness for obtaining each $Y_j$.
\end{proof}

Using the Poisson properties of $Y_j$ we can better characterize the quantity $\estgood$, and additionally we bound it's expectation in the following lemma.

\begin{lemma}[Properties of $\estgood$]\label{lem:mean-good-properties}
    Define $\meangood \define \sum_j \empvol_j \distr_j(\gd{j})$. Then $\estgood \sim \pois(\prob \meangood).$ Further, the quantity $\meangood$ follows the following inequalities:
    \begin{enumerate}\label{eq:mean-good-bounds}
        \item $\left(1 - \frac{\err}{15}\right) \goodunionvol \le \meangood \le \left(1 + \frac{\err}{15}\right)\goodunionvol$.
        \item $\meangood \ge \frac{\mean}{2}$. \label{eq:meangood-ge-mean}
        \item $\left[ \left(1 - \frac{\err}{8}\right)\meangood , \left(1 + \frac{\err}{8}\right)\meangood \right] \subset \left[ \left(1 - \frac{15}{32}\err\right)\goodunionvol, \left(1 + \frac{15}{32}\err\right)\goodunionvol \right]$.
    \end{enumerate}
\end{lemma}

\begin{proof}
    The Poisson property of $\estgood$ follows from the additivity property of Poisson random variables (\cref{lem:poisson-additivity}). To prove the first inequality, observe that the upper bound can be obtained from the equation obtained earlier (\ref{eq:mean-good-upper-bound}). For the lower bound, we operate similarly, using the $\sampgood$-closeness combined with the lower bound on $\empvol_j$ (\cref{eq:empvol-bound}).
    \begin{align*}
        \meangood &\le \sum_j \empvol_j (1 + \sampgood) \unif_j (\gd{j}) \le \left(1 + \frac{\err}{15}\right) \goodunionvol
        \\ \meangood &\ge \sum_j \empvol_j (1 - \sampgood) \unif_j (\gd{j}) \le \left(1 - \frac{\err}{15}\right) \goodunionvol
    \end{align*}
    For the bound between $\meangood$ and $\mean$, we combine the inequalities obtained in \cref{lem:mean-properties}, in conjunction with the bounds we just proved.
    \begin{align*}
        \mean \le \meangood + \sum_j \empvol_j \distr_j(\bd{j}) < \meangood + \frac{\err}{15}\goodunionvol < 2\meangood,
    \end{align*}
    where the last inequality follows due to the obtained lower bound on $\meangood$, as we know $\frac{\err}{15} < 1 - \frac{\err}{15}$. The third result follows directly from \cref{eq:mean-good-bounds}.
\end{proof}

\begin{cl}[Bounding Probability of $\failgood$]\label{claim:fail-good-estimate}
    Consider the event $\failgood$, where:
    \[ \estgood \not\in \left[\left(1 - \frac{15}{32}\err\right)\prob\goodunionvol, \left(1 + \frac{15}{32}\err\right)\prob\goodunionvol \right]. \] 
    Then $\pr[\failgood] < \frac{\conf}{4}$.
\end{cl}

\begin{proof}
    Since the definition of $\failgood$ is dependent upon the random quantity $\prob$, we need to apply an union bound over all the possible values of $\prob$, which we know to be all the values $2^{-t}$, where $t$ is any non-negative integer at most $\probthresh$. (The reasoning is the conditioned, mentioned previously in \cref{cond:high-enough-probability}). Our strategy will therefore be to prove the claim for each of these possible values of $\prob$, and then tie all of these bounds together via an union bound argument.

    Say $\prob \leftarrow 2^{-t}$ for some non-negative integer, at most $\probthresh$. Since we know $\estgood$ to follow a Poisson distribution, an application of Chernoff bounds give us the following.
    \begin{align*}
        \pr[\failgood : \prob \leftarrow 2^{-t}] &\le \pr\left[\estgood \not\in \left(1 \pm \frac{\err}{8}\right)\prob\meangood : \prob \leftarrow 2^{-t}\right]\qquad (\text{by }\cref{eq:mean-good-bounds})
        \\&\le 2\exp\left(-\frac{\err^2}{16(8+\err)}2^{-t}\meangood\right) = 2\exp\left(-\frac{\err^2}{16(8+\err)}2^{-\probthresh}\meangood\cdot 2^{\probthresh - t}\right)
        \\&\le 2\exp\left(-\frac{\err^2}{16(8+\err)}\frac{\thr}{2\mean}\frac{\mean}{2}\cdot 2^{L-t}\right) \qquad (\text{by }\cref{eq:meangood-ge-mean})
        \\&= 2\exp\left(-\frac{\err^2\thr}{64(8+\err)}\cdot 2^{\probthresh-t}\right). 
    \end{align*}
    Denote the quantity $c \define \frac{\err^2\thr}{64(8+\err)}.$ Note that by substituting the value of $\thr$, we shall obtain $c \ge \ln 2$, i.e. $e^{-c} < 1$. Additionally we will also have $c > \ln \frac{16\length}{\conf}$. All that remains for us to apply union bound over the above bounds for each possible value of $t$.
    \begin{align*}
        \pr[\failgood] &\le \sum_{t = 0}^{\probthresh} \pr[\failgood : \prob \leftarrow 2^{-t}] = \sum_{t=0}^{\probthresh} 2\exp(-c\cdot 2^{\probthresh - t})
        \\&\le \sum_{s = 0}^{\probthresh} 2\exp\left(-c\cdot 2^{s}\right)\qquad (\text{substitute }s\leftarrow \probthresh - t)
        \\&\le 2\sum_{s=0}^{\probthresh} (e^{-c})^{2^s} < 2\sum_{u\ge 1} (e^{-c})^u \qquad (\text{well defined since }e^{-c} < 1)
        \\&\le 2\exp(-c) \sum_{u \ge 0} \exp(-c)^{s} = \frac{2e^{-c}}{1 - e^{-c}}
        \\&< 4\exp(-c) \le \frac{\conf}{4\length} < \frac{\conf}{4}.
    \end{align*}
\end{proof}

\subsubsection{Step V. Bounding \texorpdfstring{$\failbad$: Contribution of the `Bad' Samples}{}}

We continue our routine of better understanding the contribution to the net estimate coming from the exceptional regions. Recall that $\estbad$ denotes the number of surviving samples from all the bad sections. We again decompose this random quantity into a sum of $\length$ many random variables as follows: let $Z_j$ denote the number of samples from $\bd{j}$ for each $j \in [\length]$. Using this notation, we can say that $\sum_j Z_j$ is precisely equal to $\estbad$.

\begin{lemma}[Properties of $Z_j$]\label{lem:Zj-properties}
    The random variables $\langle Z_j : 1 \le j \le \length \rangle$ are mutually independent, and for each $j \in [\length]$, we have:
    \begin{align*}
        Z_j\sim\pois(\prob\empvol_j\distr_j(\bd{j})).
    \end{align*}
\end{lemma}

\begin{proof}
    Follows the same structure as \cref{lem:Yj-properties}, since $\bd{j}$ is a measurable subset of $\body_j$.
\end{proof}

\begin{lemma}[Properties of $\estbad$]\label{lem:mean-bad-properties}
    Define $\meanbad \define \sum_j \empvol_j \distr_j(\bd{j})$. Then $\estbad \sim \pois(\prob \meanbad)$. Further, the quantity $\meanbad$ obeys the following inequalities:
    \begin{enumerate}
        \item $\meanbad \le \frac{\err}{15}\goodunionvol$.
        \item $\left(1 + \frac{\err}{8}\right) \meanbad \le \frac{15}{32}\err\goodunionvol$.
    \end{enumerate}
\end{lemma}

\begin{proof}
    Follow along the argument of \cref{lem:mean-good-properties} to obtain the Poisson properties of $\estbad$. The inequalities have also been proved previously: the first one has been proved in \cref{eq:mean-bad-upper-bound}, and the second one can be obtained by substituting the upper bound for $\meanbad$ in the left side expression.
\end{proof}

\begin{cl}[Bounding Probability of $\failbad$]\label{claim:fail-bad-estimate}
    Consider the event $\failbad$, where the quantity $\estbad$ takes a value larger than $\frac{15}{32}\err \prob \goodunionvol$. Then $\pr[\failbad] < \frac{\conf}{4}$.
\end{cl}

\begin{proof}
    Following the proof structure of \cref{claim:fail-good-estimate}, we would like to apply an union bound argument by individually bounding the probability of $\failbad$ conditioned under each of the possible values for $\prob$. 

    Fix a value of $p \leftarrow 2^{-t},$ where $t$ is a non-negative integer at most $\probthresh$. We shall apply Chernoff bounds, with an additional layer of Poisson additivity (\cref{lem:poisson-additivity}) applied beforehand. We replace $\estbad$, which follows $\pois(\prob \meanbad)$, by a random variable $W \sim \pois\left(\prob \frac{\err}{15}\goodunionvol \right).$ The monotonicity bound is applicable here, since we have proved that $\meanbad \le \frac{\err}{15}\goodunionvol$ above (\cref{lem:mean-bad-properties}).

    \begin{align*}
        \pr[\failbad : \prob \leftarrow 2^{-t}] &= \pr \left[ \estbad > \frac{15}{32}\err \prob \goodunionvol : \prob \leftarrow 2^{-t}] \right]
        \\&\le \pr\left[W > \frac{15^2}{32} \E[W] : W \sim \pois\left(2^{-t}\frac{\err}{15} \goodunionvol\right)\right]
        \\&\le \pr\left[W > 7 \E[W] : W \sim \pois\left(2^{-t}\frac{\err}{15} \goodunionvol\right)\right]
        \\&\le \exp\left(-\frac{6^2}{2\cdot 7} 2^{-t}\frac{\err}{15} \goodunionvol \right) = \exp \left( -\frac{6\err}{35}2^{-\probthresh}\goodunionvol \cdot 2^{\probthresh - t} \right)
        \\&\le \exp\left(-\frac{6\err}{35}\frac{\thr}{2\mean}\goodunionvol\cdot 2^{\probthresh - t}\right)
        \\&\le \exp\left(-\frac{3\err\thr}{35} \frac{1}{\mean} \frac{\mean}{1 + \nicefrac{\err}{7}} \cdot 2^{\probthresh - t}\right) \qquad (\text{by }\mbox{\Cref{lem:mean-properties}})
        \\&\le \exp\left(-d\cdot 2^{\probthresh - t} \right),
    \end{align*}
    where $d \define \frac{3\err\thr}{5(\err + 7)}$. Again by the choice of value for $\thr,$ we have $d \ge \ln 2$, i.e. $e^{-d} < 1.$ Additionally we will also have $d \ge \ln \frac{8}{\conf}$. We use an union bound argument, similar to the one used in \cref{claim:fail-good-estimate}.
    \begin{align*}
        \pr[\failbad] &\le \sum_{t = 0}^{\probthresh} \pr[\failbad : \prob \leftarrow 2^{-t}] \le \exp\left(-d\cdot 2^{\probthresh - t} \right)
        \\&\le \sum_{s=0}^{\probthresh} \exp(-d\cdot 2^s) = \sum_{s=0}^{\probthresh} (e^{-d})^{2^s}\qquad (\text{substituting }s\leftarrow \probthresh -t)
        \\&< \sum_{u \ge 1} (e^{-d})^u
        \\&\le e^{-d}\sum_{u\ge 0} (e^{-d})^s = \frac{e^{-d}}{1 - e^{-d}}
        \\&< 2\exp(-d) < \frac{\conf}{4}.
    \end{align*}

\end{proof}

\section{Time and Space Complexities of \texorpdfstring{\algo{}}{}}\label{sec:union-complexity}

In this section, we analyze the costs of the volume-estimation and point-sampling oracles and then combine them with the operations of \cref{algo:alpona}. 
Throughout, time complexity denotes the number of calls to the membership oracles of the convex bodies, each counted at unit cost. The layout of the section is as follows. We first derive the complexity of \algo{} as a function of the complexities of volume-estimation and point-sampling oracles (\cref{sec:alpona-complexity}). In subsequent subsections, we instantiate the oracles with their actual complexities (\cref{sec:prelims-compute-volume,sec:prelims-point-sampling}), combine them to obtain the complexity of \algo{} under this instantiation (\cref{sec:alpona-instantiated}). And finally discuss the main ideas behind our improved walk length for \har{} (\cref{sec:walk-length-discussion}).

% \Aranya{Do we really take arithmetic as unit cost? Then everything needs to get an extra factor of $n$, because say for example, one step of hit and run requires us to update a $n$ length vector, which takes $n$ time. And similarly for volume estimation. This is just too tedious in my opinion, can we do a geometric oracle model?}

With the above in mind, we let $T_{\mathrm{vol}}$, $S_{\mathrm{vol}}$
denote the time and auxiliary space of one invocation of $\compvol$, respectively,
and $T_{\mathrm{samp}}$, $S_{\mathrm{samp}}$ denote the time and auxiliary space of one invocation of \gensamp, respectively.

% let $T_{\mathrm{vol}}(\volerr,\volconf,\dm,\inrad,\circrad)$ and $S_{\mathrm{vol}}(\volconf,\dm)$ denote the time and auxiliary space of one invocation of $\compvol$, respectively.  The algorithm invokes $\compvol$ once for each of the $\length$ bodies, with parameters $\volerrval$ and $\volconfval$. Consequently, the total time invested by \algo{} towards for volume estimation is
% \[
% O\left(\length\,T_{\mathrm{vol}}\left(\volerrval,\volconfval,\dm,\inrad,\circrad\right)\right).
% \]
% Since the calls are sequential, their workspace can be reused, and the corresponding auxiliary space is at most
% \[
% S_{\mathrm{vol}}\left(\volconfval,\dm\right).
% \]

% Similarly, we let $T_{\mathrm{samp}}(\eta,\beta,\dm,\inrad,\circrad)$ and $S_{\mathrm{samp}}$ be the time and auxiliary space of one invocation of \gensamp respectively.

\subsection{Generic Complexity of \texorpdfstring{\algo}{}}\label{sec:alpona-complexity}

We deduce the complexity of \algo{} by segregating the algorithm's workflow in the following four essential components. 

\paragraph{Removal.}
Recall that the algorithm maintains at most $\thresh=\threshval$ points in $\bucket$. In each of its $\length$ iterations, the removal step (Line~\ref{line:remove-elts2} of \cref{algo:alpona}) tests at most $\thresh$ stored points for membership in the current body. Hence all removal steps together require $O(\length\cdot\thresh)$ membership queries. 

\paragraph{Volume Estimation.}
For the volume-estimation step (Line~\ref{line:compute-volume}), the algorithm invokes $\compvol$ once for each of the $\length$ bodies, with parameters $\volerrval$ and $\volconfval$. Consequently, the total time invested by \algo{} towards for volume estimation is
\[
O\left(\length\,T_{\mathrm{vol}}\left(\volerrval,\volconfval,\dm,\inrad,\circrad\right)\right).
\]
% And since the calls are sequential, their workspace can be reused, and the corresponding auxiliary space is at most
% \[
% S_{\mathrm{vol}}\left(\volconfval,\dm\right).
% \]

\paragraph{Halving.}
Each execution of the halving step (Lines~\ref{line:halving-routine}--\ref{line:halving-routine-ends}) examines at most $\thresh$ stored points and requires $O(\thresh)$ operations. By the bound on the event $\failprob$ in the correctness analysis, with probability at least $1-\conf/4$ (\Cref{claim:fail-low-prob}) the total number of halvings is at most $\log \unionvol$. Thus, on this event, their total cost is $O(\thresh \cdot \log \unionvol)$. 

\paragraph{Sampling.}
Finally, at most $\thresh$ points are inserted during any one iteration (Lines~\ref{line:generate-samples}--\ref{line:generate-samples-ends}), so the algorithm makes at most $\length\cdot\thresh$ calls to $\gensamp$ in total, with parameters $\sampgoodval,\sampbadval$. Therefore the total time invested by \algo{} towards for volume estimation is
\[
O\left(\length\cdot \thresh\cdot T_{\mathrm{samp}}\left(\sampgoodval,\sampbadval,\dm,\inrad,\circrad\right)\right).
\]

\paragraph{Total time complexity.}
Combining all these contributions, with probability at least $1-\conf/4$, the total running time is
\begin{align}
\label{eq:alponacomplexity}
O\Bigl(&\length \cdot \thresh + \thresh \cdot \log \unionvol
+\length \cdot T_{\mathrm{vol}}\left(\volerrval,\volconfval,\dm,\inrad,\circrad\right)\nonumber\\
&+\length\cdot \thresh\cdot T_{\mathrm{samp}}\left(\sampgoodval,\sampbadval,\dm,\inrad,\circrad\right)\Bigr).
\end{align}
Recall from Lines~\ref{line:threshold-init}--\ref{line:prob-init} of \cref{algo:alpona} that
\[
\thresh=\threshval
\qquad\text{and}\qquad
\unionvol \leq |\Omega|,
\]
where each convex body $\body_i$ is assumed to be a subset of $\Omega \subseteq \mathbb{R}^n$.
%\Eeshan{using $\Omega$ as a harsh bound on $\unionvol$, need to define rigorously - will need it later as well in section 6.2.3}
Since $T_{\mathrm{samp}}\ge1$, the term $\length\cdot\thresh$ is dominated by $\length\cdot\thresh\cdot T_{\mathrm{samp}}$. Substituting these values in \cref{eq:alponacomplexity} and collecting the terms that contain $\thresh$, the total running time of \algo{} is
\begin{align}
\label{eq:alponacomplexity-explicit}
\bigO\Bigl(&\frac{1}{\err^2}\log\frac{\length}{\conf}\cdot\Bigl[\log |\Omega|
+\length\cdot T_{\mathrm{samp}}\left(\sampgoodval,\sampbadval,\dm,\inrad,\circrad\right)\Bigr]\nonumber+\length \cdot T_{\mathrm{vol}}\left(\volerrval,\volconfval,\dm,\inrad,\circrad\right)\Bigr).
\end{align}

% Here the dependence of $T_{\mathrm{samp}}$ on its arguments is as given above. Substituting the choices of $\thresh$, $L$, and the subroutine parameters gives a polynomial bound under the stated membership-oracle and normalization assumptions, provided the volume-estimation cost and the per-step Hit-and-Run cost are polynomial in the relevant parameters.

\paragraph{Space complexity.}
For space, the bucket contains at most $\thresh$ points of $\R^\dm$, and hence occupies $O(\dm\cdot\thresh)$ real-coordinate storage locations. The volume and sampling oracles are called sequentially, so their workspaces may be reused. The overall auxiliary space is therefore
\[
O\Bigl(S_{\mathrm{vol}}\left(\volconfval,\dm\right) + S_{\mathrm{samp}} + \frac{n}{\err^2}\log\frac{\length}{\conf}\Bigr).
\]
real words.

% \deb{Eeshan join and rewrite next three sections}

\subsection{Complexity of a Practical Instantiation of \texorpdfstring{\algo{}}{}}
\label{sec:alpona-instantiated}

First, we derive the time and space complexities for each of the PAC volume estimation and approximate sampling oracles, namely Gaussian cooling and \har{} used to instantiate \algo. Then, we derive the final time and space complexity for this instantiation.

\subsubsection{Complexity of the PAC Volume Estimation Oracle: Gaussian Cooling}\label{sec:prelims-compute-volume}

We instantiate \(\compvol\) with the Gaussian-cooling approach of \cite{cousins-vempala18}, under the geometric normalization assumed for our $(\inrad,\circrad)$-fat bodies. The implementation requires the appropriate interior-point and radius information.

\paragraph{Time complexity.}
By \cite[Theorem~1.1]{cousins-vempala18}, $\compvol$ returns a
$(1\pm\volerr)$-approximation with probability at least
$1-\volconf$, using
\begin{equation}\label{eq:CousinsVempalaVolumeTime}
 T_{\mathrm{vol}}(\volerr,\volconf,\dm,\inrad,\circrad)=\bigO\!\left(
 \frac{\max\{\dm^3,\dm^2(\circrad/\inrad)^2\}}{\volerr^2}
 \log^2\dm\,
 \log^2\frac1{\volerr}\,
 \log^2\frac{\dm}{\volerr}\,
 \log\frac1{\volconf}
 \right)
\end{equation}
membership-oracle queries.

\paragraph{Space complexity.}
%\uddalok{If space bound is not very important I would suggest to keep it off. Otherwise we need to walk through the entire backend algorithm here, to convince a reader about the space bound of Counsin-Vempala as well as Hit and Run---since they do not hand us with the corresponding bounds.}
%\Aranya{I believe brief accounts of how to account for the space as we have done should suffice? It's a call we have to take}
For the space bound, observe that the next temperature in
Step~2(b) of \cite[Figure~1]{cousins-vempala18} can be computed before generating the points in Step~2(a). We generate these points one at a time, retaining the current point of the walk, and update the sum in Step~2(c) as each point is generated. We also update the product in Step~3 after each cooling phase. Thus each run uses $\bigO(\dm)$ real words. Running the
confidence-boosting repetitions sequentially and storing their final estimates therefore gives
\[
 S_{\mathrm{vol}}(\volconf,\dm)
 =\bigO\!\left(\dm+\log\frac1{\volconf}\right)
\]
real words.

\subsubsection{Complexity of the Approximate Point Sampling Oracle: \har}\label{sec:prelims-point-sampling}

We instantiate $\gensamp$ using Hit-and-Run, started from an interior point $x_0$ at distance $\empinrad=\dist(x_0,\partial\body)>0$ from the boundary. 

\paragraph{Space complexity.}
Let us show what $S_{\mathrm{samp}}$ is first. A walk retains only its current point, direction, proposed point, and a few scalars, so it is just $O(n)$ real words.

\paragraph{Time complexity.}
Now we move on to $T_{\mathrm{samp}}$. First, we recall the almost-$\ell_\infty$ guarantee and the mixing result that determine the number of steps required.
Note that the algorithm uses $S=\body$, a convex body, and the conditions\footnote{the subscripts denote the space under consideration, we will drop it when the context is clear}
\begin{equation}\label{eq:exceptional-sampler-requirements}
 \unif_\body(\bad_\sampgood)\le\sampbad,
 \qquad \distr_\body(\bad_\sampgood)\le2\sampbad.
\end{equation}
along with the following parameter assignments in line~\ref{line:parameters-init}:
\begin{equation}\label{eq:exceptional-application-parameters}
 \sampgood=\sampgoodval,\qquad \sampbad=\sampbadval.
\end{equation}
Here, $\bad_\sampgood=\bad_{t,\sampgood}$ denotes the exceptional set of the distribution $\distr_t$ after $t$ Hit-and-Run steps. We drop $t$ when it is clear from the context. We now present three bounds on $t$.

\paragraph{(i) The LV bound.}
Thus, we retrieve the
sufficient bound (we call this the LV bound) on the number of mixing step as follows:
\begin{equation}\label{eq:exceptional-LV-application}
 t>10^{11}\dm^3\left(\frac{\circrad}{\inrad}\right)^2
 \left[\log\frac{2\circrad}{\empinrad}
       +2\log\frac{32}{\err}+\log\length\right].
\end{equation}

\paragraph{(ii) The KV bound.}

Now, for our required values of $\sampgood$ and $\sampbad$, the LV bound carries a factor $\dm^3$ in the accuracy and stream-length terms. The most natural question then is, 
\begin{center}
    \textit{can we improve their dependence on the dimension?}
\end{center}

The spectral-gap bound of Kook and Vempala gives us a way to investigate this. Let us first see what follows directly from their result. We show in \cref{sec:exceptional-comparisons} that using the spectral-gap bound of Kook and Vempala~\cite[Theorem~1.1]{KookVempala2026}, combined with the convex-domain diameter bound~\cite[Theorem~5.10]{milman09} and truncation of the first-step density\footnote{This is the usual point-start reduction in Hit and Run analysis. Take a small ball around $x_0$ and split the density after the first step into a piece inside the ball and a piece outside the ball. The complement has bounded density, so one can do spectral contraction or other techniques on this, the other piece is controlled using its small mass which the walk preserves.}, we can get the sufficient point-start bound (we call this the KV bound)
\begin{equation}\label{eq:exceptional-KV-application}
 t\ge1+\bigO\!\left(\dm^3\left(\frac{\circrad}{\inrad}\right)^2\left[
 \log\frac{32\circrad}{\empinrad}
 +\log\frac{32}{\err}+\log\length\right]\right).
\end{equation}
Note that this also gives both bounds in \cref{eq:exceptional-sampler-requirements}. 
% Plugging this in, we get the time complexity to be

% \begin{multline*}
%     \bigO\Big( \frac{1}{\err^2} \log \frac{\length \log \volmax}{\conf} \cdot \left[ \log_2 \length\volmax + \length \timesampkv \right] \\ + \length \cdot \timevolcv \Big)  
% \end{multline*}

\paragraph{(iii) Our bound.}

It turns out that the KV bound still gives the same complexity bounds of our algorithm for our parameters as the LV bound (see comparison below). In \Cref{sec:generate-samples}, we show that for our choice of parameters for $\sampgood$ and $\sampbad$, we can obtain the following better bound by doing a more careful analysis of the whole walk.

\begin{theorem}[Bound on the Walk length for the Sampling Subroutine of \algo]\label{thm:exceptional-application}
With $\sampgood=\sampgoodval$ and $\sampbad=\sampbadval$, both bounds in \cref{eq:exceptional-sampler-requirements} can be obtained in
\begin{equation}\label{eq:exceptional-geometric-application}
 t=1+\lfloor\dm/2\rfloor+\bigO\!\left(\dm^2\left(\frac{\circrad}{\inrad}\right)^2
 \left[\dm\log\frac{32\circrad}{\empinrad}
       +\log\frac{32}{\err}+\log\length\right]\right)
\end{equation}
Hit and Run steps.
\end{theorem}
We describe the technical challenges and new techniques developed to prove this result in Section~\ref{sec:walk-length-discussion}.

\subsubsection{Final Time and Space Complexity for \uve}

We now substitute the subroutine costs of \cref{sec:prelims-compute-volume,sec:prelims-point-sampling} into \cref{eq:alponacomplexity}. By substituting the LV bound for $T_{\mathrm{samp}}$ and using \Cref{eq:CousinsVempalaVolumeTime}) for $T_{\mathrm{vol}}$ in \Cref{eq:alponacomplexity} we get the following running time for \algo{}: 

\begin{multline*}
  \bigO\Big( \frac{1}{\err^2} \log \frac{\length}{\conf} \cdot \left[ \log |\Omega| + \length \cdot \timesamplv \right] \\ + \length \cdot \timevolcv \Big)  
\end{multline*}

Plugging in the KV bound instead, we get the time complexity to be

\begin{multline*}
    \bigO\Big( \frac{1}{\err^2} \log \frac{\length}{\conf} \cdot \left[ \log |\Omega| + \length \timesampkv \right] \\ + \length \cdot \timevolcv \Big)  
\end{multline*}

Finally, substituting the walk length of \cref{thm:exceptional-application} back into the complexity calculations, this gives us a runtime requirement of:

\begin{multline*}
    \bigO\Big( \frac{1}{\err^2} \log \frac{\length}{\conf} \cdot \left[ \log |\Omega| + \length\dm + \length \timesampus \right] \\ + \length \cdot \timevolcv \Big)  
\end{multline*}

\paragraph{Overall complexity.}
Instantiating \compvol and \gensamp with the algorithm from \citep{cousins-vempala18} and \har{} respectively, and using the better walk length bound from \Cref{thm:exceptional-application}, we can now state a concrete space and time complexity for \algo. We also replace some multiplicative constants inside logs with additive ones.

\paragraph{Time Complexity: }

% \[\boxed{\begin{aligned}
% \bigO\Big( \frac{1}{\err^2} \left(\log \frac{\length}{\conf} + \log \dm + \log \log \circrad \right) &\cdot \left[ \log_2 \length + \log_2 \dm + \log_2 \log \circrad + \length\dm + \length \dm^2 (\circrad / \inrad)^2 \left[ \dm\log\frac{1 + \circrad}{\empinrad} + \log\frac{1}{\err} + \log\length \right] \right] \\ 
% &+ \length \cdot \frac{\max\left\{\dm^3, \dm^2 (\circrad / \inrad)^2\right\}}{\err^2} \log^2\dm \log^2 \frac{1}{\err} \log^2\frac{1 + \dm}{\err} \log\frac{\length}{\conf} \Big) 
% \end{aligned}}\]

\[\boxed{\begin{aligned}
\bigO\Big( \frac{1}{\err^2} \log \frac{\length}{\conf} &\cdot \left( \log |\Omega| + \length \dm^2 (\circrad / \inrad)^2 \left[ \dm\log\frac{1 + \circrad}{\empinrad} + \log\frac{1}{\err} + \log\length \right]\right) \\ 
&+ \length \cdot \frac{\max\left\{\dm^3, \dm^2 (\circrad / \inrad)^2\right\}}{\err^2} \log^2\dm \log^2 \frac{1}{\err} \log^2\frac{1 + \dm}{\err} \log\frac{\length}{\conf} \Big) 
\end{aligned}}\]

\paragraph{Space Complexity: }

\[\boxed{O\Bigl(\frac{\dm}{\err^2}\log\frac{\length}{\conf} + \log\frac{4\length}{\conf}\Bigr)\text{ real words}}\]

\paragraph{Improved runtime of \algo{} using \Cref{thm:exceptional-application}: }Let us see what improvements this gives for the running time of \algo.
% Here we use $\volerr=\err/32$ and $\volconf=\conf/(4\length)$ for the volume queries, so that together they have failure probability at most $\conf/4$, and use \cref{eq:exceptional-application-parameters} for the sampling queries.
Suppose $\circrad/\inrad$ and $\circrad/d_0$ are both $\mathcal O(1)$, and fix $\err$, $\conf$. Since we are using the volume computation algorithm of~\citet[Theorem~1.1]{cousins-vempala18}, note that after rescaling the inball to have radius one, the constant aspect ratio gives the second-moment bound required by their theorem. Then with the parameters above, its cost is
\[
 T_{\mathrm{vol}}=\bigO\!\left(\dm^3\log^4\dm\log\length\right)\,.
\]
%while our parameter choices give $\thresh=\bigO(\log\length)$ and $\probthresh=\bigO(\log\length)$.

Then both the LV bound in \cref{eq:exceptional-LV-application} and the KV bound in \cref{eq:exceptional-KV-application} give a total running-time bound of
\[
 \bigO\!\left(\length\left[
 \dm^3\log^4\dm\log\length+\dm^3(\log\length)^2\right]\right),
\]
while our result via \cref{eq:exceptional-geometric-application} gives
\[
 \bigO\!\left(\length\left[
 \dm^3\log^4\dm\log\length+\dm^3\log\length
       +\dm^2(\log\length)^2\right]\right).
\]
The first term in each bound is the common volume computation cost. Let us now make the improvement a little more concrete. Whenever
\[
 \log\length\ge\dm\log^4\dm
\]
these bounds simplify to
\[
 \bigO\!\left(\length\dm^3(\log\length)^2\right)
 \qquad\text{and}\qquad
 \bigO\!\left(\length\dm^2(\log\length)^2\right)
\]
respectively. More concretely, for $\length=e^{\dm^2}$, we get the following comparison.

\begin{center}
\begin{tabular}{lcc}
\hline
Sampling bound & Steps per sample & Total running time of $\algo$ \\
\hline
LV/KV & $\bigO(\dm^5)$ & $\bigO(\length\dm^7)$ \\
Ours & $\bigO(\dm^4)$ & $\bigO(\length\dm^6)$ \\
\hline
\end{tabular}
\captionof{table}{Comparison table when the stream length is \texorpdfstring{$\length=e^{\dm^2}$}{}.}
\end{center}

Thus, in this regime, the improvement in the sampling bound gives a factor of order $\dm$ in the total running-time bound for $\algo$.

\subsection{Discussion on The Improved Walk Length of \har{} (Theorem~\ref{thm:exceptional-application})}
\label{sec:walk-length-discussion}

In this subsection, we talk a little about the result of \Cref{thm:exceptional-application}, its main proof ideas, and some interesting results that go into it (all a bit heuristically). This result and the proof techniques (specifically the construction and use of the explicit radial vector field) may be independently useful to the Hit and Run community.

For the sake of exposition, we ignore the functional analytical details about domains and constants here.

As we show in the appendix, the result of \Cref{thm:exceptional-application} follows from two results:
\begin{enumerate}
    \item For any $0<\sampgood,\sampbad<1$ taking at least
    \[1+\dm/2+O\left(\frac{\dm\log(32\circrad/d)+\log(1/(\sampgood^2\sampbad))}{\gamma_{\mathrm{HR}}}\right)\]
    suffices to guarantee that $\unif(\bad_{t,\sampgood})\le\sampbad$ and $\distr_t(\bad_{t,\sampgood})\le2\sampbad$.
    \item A lower bound on $\gamma_{\mathrm{HR}}$ and contraction of the Hit and Run operator $P$, namely for functions $f$ with mean $0$,
    \[\norm{P^sf}_{L^2(\unif)}\le e^{-\gamma_{\mathrm{HR}}s}\|f\|_2\]    
\end{enumerate}

Let us talk first about item~(2). This can be proven in a myriad of ways, one of them provided by (a corollary of) \cite[Theorem 1.1]{KookVempala2026}. We give a different geometric proof for it using an explicit radial vector field construction. The idea is (after some technicalities) to consider a scaled copy $C$ of the body such that $C$ has half the volume of $K$. After some technicalities, we define a radial vector field which satisfies
\[\int_\body f\dd\unif-2\int_{2^{-1/\dm}\body}f\dd\unif=-2\int_\body F\cdot\nabla f\dd\unif\]
This comes out from the fact that scaling moves the points in the original body and the field $F$ says what the combined flow through each point is. In particular, our particular $F$ can be viewed as the accumulated mass flow (up to a factor of $2$) across a ``shell", generated by the scaling itself. In effect, this gives
\[\underbrace{\mathbb E_{\pi_C}f-\mathbb E_{\pi_K}f}_{\text{change in the average}}=
2\int_K F\cdot\nabla f\,d\pi_K\]
Geometrically, the change is the amount of mass passing through each location, multiplied by how much $f$ changes there. We then use the estimate $\norm{F(x)}\lesssim(\circrad/\inrad)\dist(x,\partial\body)$ (which we build inside the vector field definition by decaying it suitably), which gives the ``flux" term the desired weight (heuristically, across a surface just beneath the boundary, we only need to carry the mass lying in the thin shell outside that surface. As that shell becomes thinner, there is less mass to carry.). The smaller copy stays at depth $\gtrsim\inrad/\dm$, so its Cheeger's inequality controls the remaining error by a gradient bound similar to that in \cite[Theorem~3.1]{KookVempala2026}.

In contrast, their proof compares a function with its values after moving points slightly towards the centre
\[
 f(x)-f\!\left(\left(1-\frac1{2\dm}\right)x\right)
 =\int_{1-1/(2\dm)}^1 x\cdot\nabla f(tx)\dd t.
\]
Then they average over $x$, turning the radial integral into a weighted gradient bound (weights being the distance from the boundary, which appears because a location very close to the boundary can belong to only a small range of the radial segments. Its gradient therefore receives a small weight.), while the contracted copy has depth at least $\inrad/(2\dm)$. They then use the ordinary Poincare inequality on this copy to control the remaining error~\cite[proof of Theorem~3.1]{KookVempala2026}.

Now item~(1) and this is much more interesting. To prove this, we do two things (again):

\begin{enumerate}[label=1.\arabic*.]
    \item First use a few steps of Hit and Run to make the density square-integrable.
    \item Then use the remaining steps to reduce its (squared) error by spectral contraction (that is, item~(2)).
\end{enumerate}

We then show that item~(1.2) suffices to get the $\unif(\bad_{t,\sampgood})\le\sampbad$ and $\distr_t(\bad_{t,\sampgood})\le2\sampbad$ guarantees. Once done, all that remains is to prove item~(1.1). To the best of our knowledge, we have not found this in the literature and looks like a very interesting result.

The idea for item~(1.1) is as follows. For a point start, the density is singular at the starting point $x_0$, and the singularity is captured by the term $|y-x_0|^{1-n}$, where $y$ is some other point (from standard Hit and Run theory, see \cite[equation~(5)]{LV06}). We follow this singularity near $x_0$ through the initial steps and keep track of it inside the fixed ball $U=B(x_0,d/2)$. Suppressing constants and scale factors, we follow the progression (ignoring the part outside $U$ which we can bound)
\[|y-x_0|^{1-n} \to  |y-x_0|^{2-n} \to |y-x_0|^{3-n} \to\cdots\]
Now let $q$ be a small distance scale. To reach $y$, we average over possible intermediate points $x$. Points at that scale occupy volume of order $q^n$. Combining them gives
\[\underbrace{q^{s-n}}_{\text{old spike}}\times\underbrace{q^{1-n}}_{\text{transition kernel}}\times\underbrace{q^n}_{\text{volume of intermediate points}}=q^{s+1-n}\] This is essentially what we formally do in the proof. This helps us because a spike of order $q^{j-n}$ has squared integral near $x_0$ behaving like
\[\int_0^{d/2} \underbrace{q^{2(j-n)}}_{\text{squared density}}\times\underbrace{q^{n-1}\,dq}_{\text{volume of a thin shell}}=\int_0^{d/2} q^{2j-n-1}\,dq\] This becomes finite precisely when $2j>n$, predicting that $\lfloor n/2\rfloor+1$ should suffice for getting a square-integrable density. In fact, we show that we get
\[\chi^2(\mathcal D_t\Vert\pi)\le\left(\frac{32R}{d}\right)^n e^{-2\gamma_{\mathrm{HR}}(t-\lfloor n/2\rfloor-1)}\]
Choosing the remaining walk length so that this is at most $\sampgood^2\sampbad$ gives both sampling guarantees and recovers our required bound.

So the first few steps spread out the mass near $x_0$, weakening the singularity (the spike) until the squared density has a finite integral. This is what is known as \emph{regularization}. What happens if we do not do this regularization? Well, if we kept bounding every subsequent density by the original singularity of the first-step bound, namely $q^{1-n}$, the same calculation would give
\[\int_0^{d/2}\underbrace{q^{2(1-n)}}_{\text{squared first-step bound}}\underbrace{q^{n-1}\,dq}_{\text{volume of a thin shell}}=\int_0^{d/2}q^{1-n}\,dq\] 
which diverges for $n\ge2$. The walk would still regularise though, but that estimate would never give us a finite $L^2$ norm. Another approach is to separate off the first-step mass inside a small ball around $x_0$, and bound the density outside this ball. We would then control the contribution from inside using its small probability. The ball must be chosen small enough for our required $\sampgood,\sampbad$, so these parameters enter the resulting density bound. By following the weakening spike as above, we obtain a finite bound for the whole density after about $n/2$ steps, with this initial number of steps depending only on the dimension.
The detailed proof is in Appendix~\ref{sec:generate-samples}.
% \deb{very cryptic!}
% \Aranya{Can you elaborate on what you mean? Do you want me to expand on it a bit more?}

% \input{Analysis-and-complexity/Challenges}

% \section{Special Cases}
\section{A Special Case: Bounded Frequency Streams}\label{sec:bounded-frequency}
% Should consider and discuss the special case of sunflowers itself when discussing the algorithm. And then draw reference to here, where we consider the special case of frequency bounded sets.

In this section, we consider a special case of the union volume estimation. Observe that in full generality, each $\body_j$ can intersect with every other convex body in the collection. This type of input is slightly inefficient for the algorithm to process, as it would mean elements from the intersection region, if drawn, would get thrown out repeatedly, and our strategy of storing samples from the unique sections would not be achieved. This is why we considered the special case, where each point of the union has a bounded frequency. This special case of bounded frequency sets has been studied extensively, particularly in the streaming literature (ref. \citet{mcgregor21,eden22,Woodruff26}). We formally define the setup below.

\begin{definition}[$\tau$-bounded Frequence Streams]\label{def:bounded-freq-streams}
    A stream of convex bodies $\stream := \langle \body_j \subset \R^\dm : j \in [\length] \rangle,$ has \textit{frequency at most} $\freqbound$ if every point $x \in \R^\dm$ lies in at most $\freqbound$ of the bodies $\body_j,$  i.e.
    \begin{equation*}
        \sum_{j = 1}^{\length} \textbf{1}_{\body_J} (x) \le \freqbound  \textbf{1}_{\cup_j \body_j}(x)~\forall~x\in\R^\dm.
    \end{equation*}
    By definition, we shall always have $1 \le \freqbound \le \length.$
\end{definition}

\begin{lemma}[Connecting Union Volume and Frequency]
    For a stream $\stream$ of frequency at most $\freqbound,$ we have:
    \begin{equation}\label{eq:F0-harsh-bound}
        \sum_j \vol(\body_j) \le \freqbound \unionvol.
    \end{equation}
\end{lemma}

\begin{proof}
    In the finite setting, observe that if each element can appear in at most $\freqbound$ many different sets, then the sum of sizes of all the sets can contain $\freqbound$ many copies of each distinct element. Hence the sum of their sizes will be bounded by $\freqbound$ times the union size. This intuition is mathematically described below.
    \begin{align*}
        \sum_j \vol(\body_j) = \sum_j \int_{\R^\dm} 1_{\body_j}(x) dx \le \int_{\R^\dm} \freqbound\cdot 1_{\cup_j \body_j}(x) dx = \freqbound \unionvol.
    \end{align*}
\end{proof}

\begin{theorem}[Correctness of \algo{} for Bounded Frequency Streams]\label{thm:bounded-frequency}
    Suppose that \algo{} takes as input, a collection of convex bodies with frequency at most $\freqbound,$ and suppose $\freqbound$ is known to the algorithm beforehand. If we run \algo{} with $\sampbad \leftarrow \frac{\err}{32\freqbound}$ and keep all the other parameters unchanged, then with probability at least $1-\conf$ the output lies in the range $(1\pm \err)\unionvol.$
\end{theorem}

\begin{proof}
    The quantity $\sum_j \vol(\body_j)$ is used in exactly two cases, and in the following argument we shall show that the above chosen value for $\sampbad$ suffices.
    \begin{enumerate}
        \item In the proof of \cref{claim:F0'-F0-equivalence}, where if we apply the bound \cref{eq:F0-harsh-bound} and substitute the chosen value of $\sampbad$, we get the same output saying $\goodunionvol \ge \left(1 - \frac{\err}{32}\right)\unionvol.$ Thus we continue our agenda of proving a $\gooderr$-factor estimation of $\goodunionvol$.
        \item The second place where this modification affects us is in the proof of \cref{lem:mean-properties}, where we try to bound $\distr_j(\last_j)$, and the term $\distr_j(\bad_j)$ comes up. To be precise, the quantity we bound is:
        \begin{align*}
            \sum_j \distr_j(\bad_j)\empvol_j \le 2(1 + \volerr)\sampbad \sum_j \vol(\body_j) \le 2\left(1+\volerrval \right)\frac{\err}{32\freqbound}\cdot \freqbound \unionvol.
        \end{align*}
        Plugging in the values for $\volerr$ and bounding $\unionvol$ by $\goodunionvol/\left(1 - \frac{\err}{32}\right),$ we shall get the same bound as obtained in \cref{eq:mean-bad-upper-bound}.
    \end{enumerate}
    Since the two intermediate assertions which require the value of $\sampbad$ and a bound on $\sum_j \vol(\body_j)$ remain unchanged, the proof follows through.
\end{proof}

Observe that our proof of the general theorem simply requires $\freqbound$ to be set to $\length$. But this does not, however, rid us of the factor $\length$ as we require it regardless when applying the union bound argument to bound the probability of $\failprob$, as demonstrated in \cref{claim:fail-low-prob}.

\section{Discussions and Future Work}\label{sec:conclusion}

We studied union-volume estimation with approximate volume and sampling oracles. Our sunflower construction shows that sampling-based union-volume estimators can fail when their samplers provide only a fixed per-body total-variation guarantee. Such samplers can miss a small portion of each body even when these portions taken together may account for a substantial fraction of the union. This leads to a large error in the volume estimate.

We addressed this issue through a new concept of \emph{tail-controlled samplers}, whose density is multiplicatively close to uniform outside an exceptional region of small uniform measure. Our algorithm, \algo, uses such tail-controlled samplers with PAC volume estimates to return an $(\err,\conf)$-approximation of the union volume. Its analysis bounds the contribution of the bad samples from the exceptional region, while accounting for errors in the individual volume estimates. Along the way, \algo{} also bypasses the discretization step of the earlier work of \cite{shaw-sarkar-meel25}, thereby improving the complexity. 

Finally, we proved that Hit-and-Run provides the required tail control from an interior point start. This leads to a concrete implementation of \algo{} with Hit-and-Run as the sampling subroutine. We also give a slightly better analysis of Hit and Run that improves the dimension dependence of the accuracy and stream-length terms in the walk-length bound from $\dm^3$ to $\dm^2$. 

The significance of this investigation extends beyond union-volume estimation. Our work illustrates how theoretical analysis can identify the limitations of an existing implementation, determine the approximation guarantees that are actually required, and develop algorithms whose correctness can be established using realizable computational primitives, such as, \algo{} with Hit-and-Run. More broadly, we view this as a step toward bridging the gap between idealized randomized algorithms and their practical implementations: using theoretical analysis not merely to justify an abstract algorithm, but to understand and improve algorithms that have already demonstrated empirical promise.

%%%%%%%%%%% shall we put thi folloing here:
% Recent advances in convex-body sampling have made a shift towards other distance guarantees stronger than TV distance: for example, the In-and-Out sampler of \citet{kook2024and} has a guarantee on finite-order R\'enyi divergence, \citet{kook2025renyi} established a R\'enyi-infinity guarantee later on, and \citet{KookVempala2026} gives a $\chi^2$ guarantee.
% We can easily show that, similar to TV distance, these distance guarantees also imply an $(\sampgood, \sampbad)$-guarantee when the divergence is sufficiently small. This is due to the fact that R\'enyi divergence of order $q\in[1,\infty]$ satisfies $\mathcal R_q(\distr\|\unif)\geq\mathrm{KL}(\distr\|\unif)$, and Pinsker's inequality gives $\mathrm{KL}(\distr\|\unif)\geq 2d_{\TV}(\distr,\unif)^2$. Similarly $\chi^2(\distr\Vert\unif)\ge 4d_{\TV}(\distr,\unif)^2$.
% Therefore, if two distributions are sufficiently close in $\mathcal R_q$, they must be close in TV, and by \Cref{lem:tv-to-almost-linfty} they satisfy an $(\sampgood, \sampbad)$-guarantee.
%%%%%%%%%%%%%%%%%%%%%%%%%%%%%%%%%%%%%%%%%%%%%%%

% \paragraph{Future work}
% Our parameters assume that $\length$ is known in advance. For streams of unknown length, an \emph{anytime-valid} variant of \algo{} could replace the current budget splitting of error and confidence on the go, instead of uniformly across the $\length$ bodies.

\subsection{Future Work}
Our $(\sampgood,\sampbad)$-notion is a first step towards capturing the geometric behavior of approximate samplers, with substantial scope for refinement. It separates local density accuracy from exceptional mass, but does not directly describe how sampling errors are distributed across the body. A promising direction is to develop a geometric sampling guarantee that relates the exceptional region to a sub-Gaussian tail within the convex body. Establishing these guarantees for practical samplers such as Ball Walk or Hit-and-Run could help explain why relatively short walks often suffice in implementations, providing a stronger theoretical basis for their empirical success and potentially yielding less conservative sampling requirements for union-volume estimation.

Heuristics suggest that the dependence on the stream length can be improved further at a slight cost of the geometric term. We conjecture the following strengthening of our Hit and Run analysis:
\begin{conjecture}
    With $\sampgood=\sampgoodval$ and $\sampbad=\sampbadval$, both bounds in \cref{eq:exceptional-sampler-requirements} can be obtained after $t$ steps, where
    \[t=1+\dm+\bigO\!\left(\dm^2\left(\frac{\circrad}{\inrad}\right)^2\left[\dm\log\frac{\dm\circrad^2}{\empinrad\inrad}+\log\frac1\err\right]\right)+\bigO\!\left(\left(\frac{\circrad}{\inrad}\right)\log\frac{\length}{\err}\right)\]
\end{conjecture}
This would remove the dependence on the dimension from the term attached to the stream length $\length$. In words, this conjecture says that \emph{after paying the geometric and accuracy costs, making the exceptional region progressively smaller should be much cheaper than what we have now}.

We also believe the dependence on $\log\length$ is not optimal and can possibly be pushed down to $\log\log \length$ with possibly worse factors of $\circrad/\inrad$~\footnote{Note that the dependence on $M$ cannot be removed because of \Cref{thm:worst-case-benchmarks}.}. This would require a different geometric analysis of the Hit and Run Markov chain's tail bounds. 

\paragraph{AI Disclosure.} No generative AI or AI-assisted tools were used in the conception of research ideas or the development of proofs. Their use was limited exclusively to assistance with basic writing, grammatical corrections, and language polishing of the manuscript.

\bibliographystyle{plainnat}
\bibliography{files/bibliography}

@inproceedings{shaw-sarkar-meel25,
 title     = {{Efficient Volume Computation for SMT Formulas}},
    author    = {Shaw, Arijit and Sarkar, Uddalok and Meel, Kuldeep S.},
    booktitle = {{Proceedings of the 22nd International Conference on Principles of Knowledge Representation and Reasoning}},
    pages     = {544--554},
    year      = {2025},
    month     = {10},
    doi       = {10.24963/kr.2025/53},
    url       = {https://doi.org/10.24963/kr.2025/53},}

@inproceedings{meel-vinodchandran-chakraborty21,
 author = {Meel, Kuldeep S. and Vinodchandran, N.V. and Chakraborty, Sourav},
title = {Estimating the Size of Union of Sets in Streaming Models},
year = {2021},
publisher = {Association for Computing Machinery},
doi = {10.1145/3452021.3458333},
booktitle = {Proceedings of the 40th ACM SIGMOD-SIGACT-SIGAI Symposium on Principles of Database Systems {PODS}},
pages = {126–137},
numpages = {12},
series = {PODS'21}}

@inproceedings{meel-chakraborty-vinodchandran22,
 author = {Meel, Kuldeep S. and Chakraborty, Sourav and Vinodchandran, N. V.},
title = {Estimation of the Size of Union of Delphic Sets: Achieving Independence from Stream Size},
year = {2022},
publisher = {Association for Computing Machinery},
doi = {10.1145/3517804.3526222},
booktitle = {Proceedings of the 41st ACM SIGMOD-SIGACT-SIGAI Symposium on Principles of Database Systems {PODS}},
pages = {41–52},
numpages = {12},
series = {PODS '22}}

@INPROCEEDINGS{karp-luby-1983,
  author={Karp, Richard M. and Luby, Michael},
  booktitle={24th Annual Symposium on Foundations of Computer Science}, 
  title={Monte-Carlo algorithms for enumeration and reliability problems}, 
  year={1983},
  pages={56-64},
  doi={10.1109/SFCS.1983.35}
  }

@article{cousins-vempala18, title={Gaussian Cooling and $O^*(n^3)$ Algorithms for Volume and Gaussian Volume}, volume={47}, ISSN={0097-5397, 1095-7111}, DOI={10.1137/15M1054250}, number={3}, journal={SIAM Journal on Computing}, author={Cousins, Ben and Vempala, Santosh}, year={2018}, month=jan, pages={1237–1273}, language={en} }

@article{soos-sarkar-24, 
 title={Engineering an Efficient Approximate {DNF}-Counter}, url={http://arxiv.org/abs/2407.19946}, 
 journal={arXiv preprint}, 
 author={Soos, Mate and Sarkar, Uddalok and Aggarwal, Divesh and Chakraborty, Sourav and Meel, Kuldeep S. and Obremski, Maciej}, year={2024}
 }

@inproceedings{lovasz1990mixing,
  title={The mixing rate of Markov chains, an isoperimetric inequality, and computing the volume},
  author={Lov{\'a}sz, L{\'a}szl{\'o} and Simonovits, Mikl{\'o}s},
  booktitle={Proceedings [1990] 31st annual symposium on foundations of computer science},
  pages={346--354},
  year={1990},
  organization={IEEE}
}

@inproceedings{applegate1991sampling,
  title={Sampling and integration of near log-concave functions},
  author={Applegate, David and Kannan, Ravi},
  booktitle={Proceedings of the twenty-third annual ACM symposium on Theory of computing},
  pages={156--163},
  year={1991}
}

@article{dyer1991random,
  title={A random polynomial-time algorithm for approximating the volume of convex bodies},
  author={Dyer, Martin and Frieze, Alan and Kannan, Ravi},
  journal={Journal of the ACM (JACM)},
  volume={38},
  number={1},
  pages={1--17},
  year={1991},
}

@article{kannan1997random,
  title={Random walks and an o*(n5) volume algorithm for convex bodies},
  author={Kannan, Ravi and Lov{\'a}sz, L{\'a}szl{\'o} and Simonovits, Mikl{\'o}s},
  journal={Random Structures \& Algorithms},
  volume={11},
  number={1},
  pages={1--50},
  year={1997},
  publisher={Wiley Online Library}
}

@inproceedings{lovasz2006fast,
  title={Fast algorithms for logconcave functions: Sampling, rounding, integration and optimization},
  author={Lov{\'a}sz, L{\'a}szl{\'o} and Vempala, Santosh},
  booktitle={2006 47th Annual IEEE Symposium on Foundations of Computer Science (FOCS'06)},
  pages={57--68},
  year={2006},
  organization={IEEE}
}

@article{lovasz2007geometry,
  title={The geometry of logconcave functions and sampling algorithms},
  author={Lov{\'a}sz, L{\'a}szl{\'o} and Vempala, Santosh},
  journal={Random Structures \& Algorithms},
  volume={30},
  number={3},
  pages={307--358},
  year={2007},
  publisher={Wiley Online Library}
}

@inproceedings{kannan2009random,
  title={Random walks on polytopes and an affine interior point method for linear programming},
  author={Kannan, Ravi and Narayanan, Hariharan},
  booktitle={Proceedings of the forty-first annual ACM symposium on Theory of computing},
  pages={561--570},
  year={2009}
}

@article{mangoubi2022sampling,
  title={Sampling from log-concave distributions over polytopes via a soft-threshold Dikin walk},
  author={Mangoubi, Oren and Vishnoi, Nisheeth K},
  journal={arXiv preprint arXiv:2206.09384},
  year={2022}
}

@inproceedings{kook2025sampling,
  title={Sampling and integration of logconcave functions by algorithmic diffusion},
  author={Kook, Yunbum and Vempala, Santosh S},
  booktitle={Proceedings of the 57th Annual ACM Symposium on Theory of Computing},
  pages={924--932},
  year={2025}
}

@article{laddha2023convergence,
  title={Convergence of Gibbs sampling: Coordinate Hit-and-Run mixes fast},
  author={Laddha, Aditi and Vempala, Santosh S},
  journal={Discrete \& Computational Geometry},
  volume={70},
  number={2},
  pages={406--425},
  year={2023},
  publisher={Springer}
}

@article{polyak2014billiard,
  title={Billiard walk-a new sampling algorithm for control and optimization},
  author={Polyak, Boris T and Gryazina, EN},
  journal={IFAC Proceedings Volumes},
  volume={47},
  number={3},
  pages={6123--6128},
  year={2014},
  publisher={Elsevier}
}

@article{karp-luby-madras89,
  title={Monte-Carlo approximation algorithms for enumeration problems},
  author={Karp, Richard M and Luby, Michael and Madras, Neal},
  journal={Journal of algorithms},
  volume={10},
  number={3},
  pages={429--448},
  year={1989},
  publisher={Elsevier}
}

@article{khachiyan1989,
    author = {Khachiyan, Leonid},
    title = {The problem of computing the volume of polytopes is \#P-hard},
    journal = {Uspekhi Mat. Nauk},
    year = {1989},
    volume = {44},
    number = {3},
    pages= {199--200}
}

@article{klee1977can,
  title={Can the measure of be computed in less than $O(n\log n)$ steps?},
  author={Klee, Victor},
  journal={The American Mathematical Monthly},
  volume={84},
  number={4},
  pages={284--285},
  year={1977},
  publisher={Taylor \& Francis}
}

@article{bhore2026dynamic,
  title={Dynamic and Streaming Algorithms for Union Volume Estimation},
  author={Bhore, Sujoy and Bringmann, Karl and Chan, Timothy M and Wang, Yanheng},
  journal={arXiv preprint arXiv:2602.16306},
  year={2026}
}

@article{LV06,
    author={L{\'{a}}szl{\'{o}} Lov{\'{a}}sz and
                  Santosh S. Vempala},
  title        = {Hit-and-Run from a Corner},
  journal      = {SIAM Journal on Computing},
  volume       = {35},
  number       = {4},
  pages        = {985--1005},
  year         = {2006},
  doi          = {10.1137/S009753970544727X},
  url          = {https://doi.org/10.1137/S009753970544727X}
}

@article{milman09,
  title        = {On the role of convexity in isoperimetry, spectral gap and concentration},
  author={Milman, Emanuel},
  journal={Inventiones mathematicae},
  volume={177},
  number={1},
  pages={1--43},
  year={2009},
  publisher={Springer}
}

@article{Zsuppan2020,
  author       = {S{\'a}ndor Zsupp{\'a}n},
  title        = {Connections Between Optimal Constants in Some Norm Inequalities for Differential Forms},
  journal      = {Zeitschrift f{\"u}r Analysis und ihre Anwendungen},
  volume       = {39},
  number       = {2},
  pages        = {171--184},
  year         = {2020},
  doi          = {10.4171/ZAA/1656},
  url          = {https://doi.org/10.4171/ZAA/1656}
}

@inbook{Costabel_2017, title={Inequalities of Babuška–Aziz and Friedrichs–Velte for Differential Forms}, ISBN={9783319470795}, ISSN={2296-4878}, url={http://dx.doi.org/10.1007/978-3-319-47079-5_4}, DOI={10.1007/978-3-319-47079-5_4}, booktitle={Recent Trends in Operator Theory and Partial Differential Equations}, publisher={Springer International Publishing}, author={Costabel, Martin}, year={2017}, pages={79–88} }

@book{AmbrosioFuscoPallara2000,
  author       = {Luigi Ambrosio and Nicola Fusco and Diego Pallara},
  title        = {Functions of Bounded Variation and Free Discontinuity Problems},
  publisher    = {Oxford University Press},
  series       = {Oxford Mathematical Monographs},
  year         = {2000},
  isbn         = {9780198502456},
  doi          = {10.1093/oso/9780198502456.001.0001},
  url          = {https://doi.org/10.1093/oso/9780198502456.001.0001}
}

@book{Schneider2013,
  title={Convex bodies: the Brunn-Minkowski theory},
  author={Schneider, Rolf},
  volume={151},
  edition={2},
  year={1993},
  publisher={Cambridge university press, Cambridge}
}

@misc{KookVempala2026,
      title={Spectral Gaps of Hit-and-Run and Coordinate Hit-and-Run}, 
      author={Yunbum Kook and Santosh S. Vempala},
      year={2026},
      eprint={2608.16878},
      archivePrefix={arXiv},
      primaryClass={cs.DS},
      url={https://arxiv.org/abs/2608.16878}, 
}

@article{RudolfUllrich2013,
   title={Positivity of hit-and-run and related algorithms},
   volume={18},
   ISSN={1083-589X},
   url={http://dx.doi.org/10.1214/ECP.v18-2507},
   DOI={10.1214/ecp.v18-2507},
   number={none},
   journal={Electronic Communications in Probability},
   publisher={Institute of Mathematical Statistics},
   author={Rudolf, Daniel and Ullrich, Mario},
   year={2013},
   month=Jan, pages={1–8} }

@book{Stein1970SingularIntegrals,
  title={Singular integrals and differentiability properties of functions},
  author={Stein, Elias M},
  volume={2},
  year={1970},
  publisher={Princeton university press}
}

@article{dyer1988complexity,
  title={On the complexity of computing the volume of a polyhedron},
  author={Dyer, Martin E. and Frieze, Alan M.},
  journal={SIAM Journal on Computing},
  volume={17},
  number={5},
  pages={967--974},
  year={1988},
  publisher={SIAM}
}

@article{sinclair1989approximate,
  title={Approximate counting, uniform generation and rapidly mixing Markov chains},
  author={Sinclair, Alistair and Jerrum, Mark},
  journal={Information and computation},
  volume={82},
  number={1},
  pages={93--133},
  year={1989},
  publisher={Elsevier}
}

@article{meel2017hashing,
  title={On hashing-based approaches to approximate DNF-counting},
  author={Meel, Kuldeep S and Shrotri, Aditya A and Vardi, Moshe Y},
  journal={arXiv preprint arXiv:1710.05247},
  year={2017}
}

@article{bringmann2024approximating,
  title={Approximating Klee's Measure Problem and a Lower Bound for Union Volume Estimation},
  author={Bringmann, Karl and Larsen, Kasper Green and Nusser, Andr{\'e} and Rotenberg, Eva and Wang, Yanheng},
  journal={arXiv preprint arXiv:2410.00996},
  year={2024}
}

@book{ross2023,
  title     = {Introduction to Probability Models},
  author    = {Ross, Sheldon M.},
  edition   = {13th},
  year      = {2023},
  publisher = {Academic Press},
  address   = {London},
  isbn      = {978-0443187612}
}

@inproceedings{Nandi24,
  author       = {Mridul Nandi and
                  N. V. Vinodchandran and
                  Arijit Ghosh and
                  Kuldeep S. Meel and
                  Soumit Pal and
                  Sourav Chakraborty},
  title        = {Improved Streaming Algorithm for the Klee's Measure Problem and Generalizations},
  booktitle    = {Approximation, Randomization, and Combinatorial Optimization. Algorithms
                  and Techniques, {APPROX}},
  series       = {LIPIcs},
  volume       = {317},
  pages        = {26:1--26:21},
  year         = {2024},
  doi          = {10.4230/LIPICS.APPROX/RANDOM.2024.26}
}

@article{BringmannF10,
  author       = {Karl Bringmann and
                  Tobias Friedrich},
  title        = {Approximating the volume of unions and intersections of high-dimensional
                  geometric objects},
  journal      = {Comput. Geom.},
  volume       = {43},
  number       = {6-7},
  pages        = {601--610},
  year         = {2010},
  doi          = {10.1016/J.COMGEO.2010.03.004}
}

@inproceedings{BringmannLNRW25,
  author       = {Karl Bringmann and
                  Kasper Green Larsen and
                  Andr{\'{e}} Nusser and
                  Eva Rotenberg and
                  Yanheng Wang},
  title        = {Approximating Klee's Measure Problem and a Lower Bound for Union Volume
                  Estimation},
  booktitle    = {41st International Symposium on Computational Geometry, (SoCG)},
  volume       = {332},
  pages        = {25:1--25:16},
  year         = {2025},
  doi          = {10.4230/LIPICS.SOCG.2025.25},
}

@article{payne1960optimal,
  title={An optimal Poincar{\'e} inequality for convex domains},
  author={Payne, Lawrence E and Weinberger, Hans F},
  journal={Archive for Rational Mechanics and Analysis},
  volume={5},
  number={1},
  pages={286--292},
  year={1960},
  publisher={Springer}
}

@article{bebendorf2003note,
  title={A note on the Poincar{\'e} inequality for convex domains},
  author={Bebendorf, Mario},
  journal={Zeitschrift f{\"u}r Analysis und ihre Anwendungen},
  volume={22},
  number={4},
  pages={751--756},
  year={2003}
}

@inproceedings{Woodruff26,
  author       = {William Swartworth and
                  David P. Woodruff and
                  Samson Zhou},
  title        = {High Probability Streaming Lower Bounds for {$F_2$} Estimation},
  booktitle    = {Approximation, Randomization, and Combinatorial Optimization. Algorithms
                  and Techniques, {APPROX}},
  volume       = {392},
  pages        = {76:1--76:23},
  year         = {2026},
  doi          = {10.4230/LIPICS.APPROX/RANDOM.2026.76},
}

@article{den2012probability,
  title={Probability theory: The coupling method},
  author={Den Hollander, Frank},
  journal={Lecture notes available online (http://websites. math. leidenuniv. nl/probability/lecturenotes/CouplingLectures. pdf)},
  volume={3},
  pages={60},
  year={2012}
}

@inproceedings{mcgregor21,
  author       = {Andrew McGregor and
                  David Tench and
                  Hoa T. Vu},
  title        = {Maximum Coverage in the Data Stream Model: Parameterized and Generalized},
  booktitle    = {24th International Conference on Database Theory, {ICDT}},
  volume       = {186},
  pages        = {12:1--12:20},
  year         = {2021},
  doi          = {10.4230/LIPICS.ICDT.2021.12},
}

@inproceedings{eden22,
  author       = {Omri Ben{-}Eliezer and
                  Talya Eden and
                  Krzysztof Onak},
  title        = {Adversarially Robust Streaming via Dense-Sparse Trade-offs},
  booktitle    = {5th Symposium on Simplicity in Algorithms, {SOSA}},
  pages        = {214--227},
  publisher    = {{SIAM}},
  year         = {2022},
  doi          = {10.1137/1.9781611977066.15},
}

\newpage
\appendix

\section{Additional Related Work on Point Sampling}

Although sampling from log-concave distributions is \#P-Hard, a long line of research has been devoted to designing randomized approximate samplers for log-concave distributions. These algorithms generate samples whose distribution is $\varepsilon$-close (in total variation distance) to a target log-concave distribution $\pi$ supported on a convex body $K \subseteq \mathbb{R}^n$.

The initial approaches towards this direction used \emph{grid walk} on a discretization of the log-concave function on the assumption that the function is sufficiently smooth (e.g. Lipschitz continuous)~\citep{applegate1991sampling}. 
This large polytime dependence on the dimension $n$ was soon brought down to $O^*(n^2R^2/r^2)$ queries to the membership oracle by \emph{ball walk}~\citep{kannan1997random}. The idea of ball walk is to propose a point sampled uniformly from a ball of small radius around the current point, and accept the proposal if does not leave the body. Starting from $W$-warm distribution, the Ball walk was later analyzed~\citep{lovasz2007geometry} to have a query complexity of $O^*\left(\frac{W^4}{\err^4} \cdot \frac{n^2R^2}{r^2} \log \frac{W}{\err}\right)$where $W$ is the warmness of the starting distribution with respect to the target. 

Another style of algorithm, namely 
\emph{Hit-and-Run} was showed to mix as fast as the \emph{Ball-walk}. \citet[Theorem1.1]{lovasz2006fast} showed the dependence on $W$ to be polylogarithmic for Hit-and-Run: $O\left(\frac{n^2 R^2}{r^2} \log^{2} \frac{nWR}{\err r}\log^{3} \frac{W}{\err}\right)$. This algorithm draws a uniformly random line $\ell$ through the current point $x$ and moves to a new point sampled proportional to $\pi$ conditioned on $\ell \cap K$. 
This approach became influential, inspiring many new variants: a coordinate variant that restricts the 
random line to coordinate directions was proposed by~\citet{laddha2023convergence}, and a version with reflections off the boundary of $K$, called the 
\emph{Billiard Walk}~\citep{polyak2014billiard}, was later introduced.

Sampling becomes easier when then convex bodies are polytopes. \citet{kannan2009random} proposed \emph{Dikin walk} for polytopes that were later generalized for log-concave distributions by~\citet{mangoubi2022sampling}. More recently, attention has shifted toward guarantees in distances 
stronger than TV, such as R\'{e}nyi and $\mathcal{R}_\infty$ 
divergences~\citep{kook2025sampling}.

\iffalse
\highlight{\paragraph{On Streaming and Dynamic Union Volume Estimation.} There are recent works which address scenarios where the sets arrive in a data stream, or are dynamically inserted and deleted. Works such as \citet{meel-vinodchandran-chakraborty21,meel-chakraborty-vinodchandran22} address the union of sets problem in the streaming model, which also gives algorithms for Klee's Measure Problem and the DNF counting problem.} Additionally, the problem of volume estimation for convex bodies has also been studied in dynamic setups, such as \citet{bhore2026dynamic}, where they developed efficient dynamic approximation algorithms for convex bodies in constant dimension $\dm$ under insertions and deletions.

Works of \citet{sinclair1989approximate} first introduced the notion of Rapidly mixing markov chains, and since then random walk based approaches have been heavily used to give algorithms for extracting geometric properties of convex bodies, such as volume estimation and random sample generation.
\fi

\section{Useful Technical Results}

We recall a few standard probabilistic tools that will be used
throughout our analysis. They are available in any standard probability text book like \cite{ross2023}. We begin with Markov's inequality,
followed by some basic properties and concentration bounds
for Poisson random variables.

\begin{lemma}[Markov's Inequality]\label{thm:markov}
    Let $X$ be a non-negative random variable. Then, for
    every $a>0$,
    \[
        \pr(X\ge a)\le \frac{\E[X]}{a}.
    \]
    In particular, for any real-valued random variable $Y$
    with mean $\mu$ and any $\err>0$,
    \[
        \pr(|Y-\mu|>\err)
        \le \frac{\E[|Y-\mu|]}{\err}.
    \]
\end{lemma}

\paragraph{Poissonization.}
Poisson random variables play a central role in our algorithm.
Unlike sampling schemes that require distinct representatives,
Poisson sampling naturally accommodates repeated samples.
More importantly, the additivity and splitting properties
of Poisson distributions allow us to analyze sample counts
over disjoint regions of a continuous space. This enables
us to work directly with volumes, avoiding the discretization
step employed in~\cite{shaw-sarkar-meel25}.

We recall the relevant properties below.

\begin{lemma}[Additivity of Poisson Random Variables]
\label{lem:poisson-additivity}
    Let $X_1,\ldots,X_L$ be independent random variables
    with $X_i\sim\pois(\lambda_i)$, where $\lambda_i>0$.
    Then
    \[
        X:=\sum_{i=1}^{L}X_i
        \sim\pois(\lambda),
        \qquad
        \text{where }\lambda:=\sum_{i=1}^{L}\lambda_i.
    \]
\end{lemma}

The next lemma states the Poisson splitting property:
when a Poisson number of independent samples is drawn
from a distribution, the numbers falling in disjoint
measurable regions are independent Poisson random variables.

\begin{lemma}[Poissonization]\label{lem:poissonisation}
    Let $(S,\nu)$ be a probability space, and let $\distr$
    be a probability distribution on $S$. Consider a
    partition
    \[
        S=\bigsqcup_{j=1}^{L}S_j
    \]
    into disjoint measurable subsets. Suppose that
    $N\sim\pois(\lambda)$ and, conditional on $N$,
    we draw $N$ independent samples from $\distr$.

    For each $j\in[L]$, let $N_j$ denote the number
    of samples belonging to $S_j$. Then
    $N_1,\ldots,N_L$ are mutually independent, and
    \[
        N_j\sim\pois\bigl(\lambda\distr(S_j)\bigr)
        \qquad\text{for every }j\in[L].
    \]
\end{lemma}

Additionally we state the monotonicity property of Poisson random variable; which essentially states that the tail bound of a Poisson distribution increases when the Poisson parameter is increased.

\begin{lemma}[Monotonicity of Poisson Random Variables]\label{lem:poisson-monotonicity}
    For positive reals $0 < \mu_1 \le \mu_2$ and a quantity $a > 0,$ Then the following inequality holds:
    \begin{align*}
        \pr(X > a : X\sim \pois(\mu_1)) \le \pr(Y > a: Y \sim \pois(\mu_2)).
    \end{align*}
\end{lemma}

\begin{proof}
    We proceed by a coupling argument. Consider independent random variables $X \sim \pois(\mu_1)$, and $Z\sim \pois(\mu_2 - \mu_1)$. Since $\mu_2 > \mu_1$, this construction is valid. By \cref{lem:poisson-additivity}, the sum $Y' \define X + Z \sim \pois(\mu_2)$. Thus it remains to prove the claimed inequation, with $Y'$ in place of $Y$ (since both follow the same distribution, and all we care about are their tail bounds).

    Observe that whenever the event $\{X > a\}$ occurs, by our construction this naturally forces $Y$ to be larger than $a.$ Hence the event $\{ X > a\}$ is contained within the event $\{ Y > a \}$, and thus the inequality holds.
\end{proof}

Finally, we recall a standard concentration inequality
for Poisson random variables.

\begin{lemma}[Chernoff Bound for Poisson Random Variables]
\label{lem:poisson-chernoff}
    Let $X\sim\pois(\lambda)$. Then, for every
    $\err>0$,
    \[
        \pr(|X-\lambda|>\err\lambda)
        \le
        2\exp\left(
            -\frac{\err^2\lambda}{2(1+\err)}
        \right).
    \]
\end{lemma}

\begin{definition}[Coupling]\label{def:coupling}
A \emph{coupling} of two probability distributions $\mu$ and $\nu$ is a pair of random variables $(X,Y)$ defined on a common probability space such that
$X\sim\mu$ and $Y\sim\nu$. That is, the marginals of the joint distribution of $(X,Y)$ is exactly $\mu$ and $\nu$ respectively. 
\end{definition}

\begin{lemma}[Coupling inequality]
\label{lem:coupling-inequality}
\emph{(See~\cite[Theorem 2.4]{den2012probability}.)}
Let $(X,Y)$ be any coupling of $\mu$ and $\nu$. Then
$\TV(\mu,\nu)\le\pr(X\ne Y)$.
\end{lemma}

\begin{lemma}[Indistinguishability lemma]
\label{lem:indistinguish}
    Let $P$ and $Q$ be two algorithms with total variation distance $\delta$. Then if $P$ fails with probability at most $\rho$, then $Q$ fails with probability at most $\delta + \rho$. 
\end{lemma}

%\Eeshan{Cannone - Topics and Techniques in Distribution Testing, Pg. 149, Theorem A.8}

\subsection{Property of Almost-\texorpdfstring{$\ell_\infty$}{} Closeness}
\label{subsec:comparison-closeness}

We compare the proposed guarantee with the two familiar
notions of approximate sampling. Throughout, $h=dD/d\pi$
is as defined above.

\paragraph{$\ell_\infty$ closeness.}
We say that $D$ is $\eta$-$\ell_\infty$ close to $\pi$
if
\[
    \|h-1\|_{L^\infty(\pi)}
    :=\operatorname*{ess\,sup}_{x\in S}|h(x)-1|
    \leq\eta.
\]
Equivalently,
\[
    1-\eta\leq h(x)\leq 1+\eta
    \qquad\text{for $\pi$-almost every }x\in S.
\]

This is a stronger requirement than almost-$\ell_\infty$
closeness, since it demands multiplicative control of
the sampling density over the entire domain, without
allowing any exceptional region. Indeed, if $D$ is
$\eta$-$\ell_\infty$ close to $\pi$, then
\[
    \pi(\bad_\eta(D,\pi))
    =D(\bad_\eta(D,\pi))=0.
\]
Consequently, $D$ is $(\eta,\beta)$-close to $\pi$
for every $\beta>0$.

\paragraph{$\ell_1$ closeness.}
Recall that the $\ell_1$ distance between $D$ and $\pi$
is defined as
\[
    \|D-\pi\|_1
    :=\int_S |h(x)-1|\,d\pi(x)
    =2d_{\mathrm{TV}}(D,\pi).
\]
Unlike $\ell_\infty$ closeness, an $\ell_1$ guarantee
controls only the average deviation of the sampling
density. Consequently, $\eta$-$\ell_1$ closeness does
not necessarily imply $(\eta,\beta)$-almost-$\ell_\infty$
closeness when $\beta$ is small.

For example, let $\pi$ be uniform on $[0,1]$ and
consider a distribution $D$ with relative density
\[
    h(x)=
    \begin{cases}
        1+2\eta, & x\in[0,1/4],\\[2mm]
        1-\dfrac{2\eta}{3}, & x\in(1/4,1],
    \end{cases}
\]
where $0<\eta<1/2$. Clearly, $h$ is a valid probability
density, and
\[
\begin{aligned}
    \|D-\pi\|_1
    &=\frac14(2\eta)
      +\frac34\left(\frac{2\eta}{3}\right)\\
    &=\eta.
\end{aligned}
\]
However, $[0,1/4]\subseteq\bad_\eta(D,\pi)$, implying
\[
    \pi(\bad_\eta(D,\pi))\geq\frac14.
\]
Therefore, $D$ is not $(\eta,\beta)$-close to $\pi$
for any $\beta<1/4$. In particular, an $\ell_1$
guarantee of order $\eta$ is insufficient for the
regime $\beta=\Theta(\varepsilon/M)$ required by
our application.

Nevertheless, a substantially stronger $\ell_1$
guarantee does imply almost-$\ell_\infty$ closeness,
as established by the following lemma.

\begin{properties} Let $D\ll\pi$ and $\eta,\beta\in(0,1)$.
\begin{enumerate}
    \item If $D$ is $(\eta,\beta)$-close to $\pi$, $\|D-\pi\|_1\leq\eta+3\beta$.
    \item If $\|D-\pi\|_1\leq\eta\beta,$ then $D$ is $(\eta,\beta)$-close to $\pi$.
\end{enumerate}
\end{properties}

% \begin{observation}[From $\ell_1$ to almost-$\ell_\infty$ closeness]
% \label{obs:l1-to-almost-linf}
% Let $D\ll\pi$ and $\eta,\beta\in(0,1)$.
% If
% \[
%     \|D-\pi\|_1\leq\eta\beta,
% \]
% then $D$ is $(\eta,\beta)$-close to $\pi$.
% \end{observation}

% \begin{proof}
% Let $B=\bad_\eta(D,\pi)$. By the definition of $B$,
% \[
%     \eta\pi(B)
%     \leq\int_B|h(x)-1|\,d\pi(x)
%     \leq\|D-\pi\|_1
%     \leq\eta\beta.
% \]
% Hence $\pi(B)\leq\beta$.

% Furthermore, using the relation between $\ell_1$
% and total variation distance,
% \[
% \begin{aligned}
%     D(B)
%     &\leq\pi(B)+d_{\mathrm{TV}}(D,\pi)\leq\beta+\frac{\eta\beta}{2}\leq 2\beta.
% \end{aligned}
% \]
% Thus, $D$ is $(\eta,\beta)$-close to $\pi$.
% \end{proof}

The comparison clarifies the strength of our requirement.
Global multiplicative $\ell_\infty$ closeness implies
$(\eta,\beta)$-closeness for every $\beta>0$, whereas
an $\ell_1$ error of order $\eta$ need not suffice when
$\beta$ is small. The conversion above instead requires
$\ell_1$ error at most $\eta\beta$; for our parameter regime,
this is $\bigO(\varepsilon^2/M)$. This is a sufficient condition,
not a necessary one: almost-$\ell_\infty$ closeness allows
larger average error provided that the density distortion is
localized and its exceptional mass is controlled. %In fact,
% $(\eta,\beta)$-closeness itself implies
% $\|D-\pi\|_1\leq\eta+3\beta$.

\section{Exceptional Sampling Guarantees for Hit and Run}\label{sec:generate-samples}

% \subsection{Sampling requirements and walk-length bounds}\label{sec:exceptional-requirements}\label{sec:exceptional-application}

In this appendix, we provide a proof of \Cref{thm:exceptional-application}. We restate it below too after recalling the relevant notions.

Let us first recall the point sampling subroutine described in \cref{sec:subroutine-descriptions}. For a convex body $\body\subset\R^\dm$, we require independent samples from a distribution $\distr$ that is $(\sampgood,\sampbad)$-close to the uniform distribution $\unif_\body$. If $\density=d\distr/d\unif_\body$ and
\[
 \bad_\sampgood=\{x\in\body:|\density(x)-1|>\sampgood\},
\]
then the implementation used in the analysis uses the two bounds
\begin{equation}\label{eq:exceptional-sampler-requirements-appendix}
 \unif_\body(\bad_\sampgood)\le\sampbad,
 \qquad \distr(\bad_\sampgood)\le2\sampbad.
\end{equation}
In particular, in line~\ref{line:parameters-init} of \cref{algo:alpona}, we assign the parameters
\begin{equation}\label{eq:exceptional-application-parameters-appendix}
 \sampgood=\sampgoodval,\qquad \sampbad=\sampbadval.
\end{equation}

We recall the geometric setup and Hit and Run terminology here again for convenience. Some other standard notions will be recalled as we move through the section. Let $\body$ contain a Euclidean ball of radius $\inrad$ (called the inball) and be contained in a Euclidean ball of radius $\circrad$ (called the outball). We start from an interior point $x_0$ and let $d=\dist(x_0,\partial\body)>0$. Let $\distr_t$ denote the distribution after $t$ Hit and Run steps. We use $\unif=\unif_\body$ for the uniform distribution and $\density_t=d\distr_t/d\unif$ for the relative density. These densities exist for $t\ge1$ by the transition-density formula of Lov\'asz and Vempala~\cite[equation~(5)]{LV06}. The exceptional set at time $t$ (so after step $t$) is precisely the set $\bad_{t,\sampgood}$ introduced in \cref{sec:prelims-point-sampling}.

The theorem we want to prove is this.

% \begin{theorem}[Walk length for the sampling subroutine]\label{thm:exceptional-application-appendix}
% With $\sampgood=\sampgoodval$ and $\sampbad=\sampbadval$, both bounds in \cref{eq:exceptional-sampler-requirements} can be obtained in
% \begin{equation}\label{eq:exceptional-geometric-application-appendix}
%  t=1+\lfloor\dm/2\rfloor+\bigO\!\left(\dm^2\left(\frac{\circrad}{\inrad}\right)^2
%  \left[\dm\log\frac{32\circrad}{d}
%        +\log\frac{32}{\err}+\log\length\right]\right)
% \end{equation}
% Hit and Run steps.
% \end{theorem}

\begin{theorem}[Walk length for the sampling subroutine]\label{thm:exceptional-application-appendix}
With $\sampgood=\sampgoodval$ and $\sampbad=\sampbadval$, both bounds in \cref{eq:exceptional-sampler-requirements} can be obtained in
\begin{equation}\label{eq:exceptional-geometric-application-appendix}
 t=1+\lfloor\dm/2\rfloor+\bigO\!\left(\dm^2\left(\frac{\circrad}{\inrad}\right)^2
 \left[\dm\log\frac{32\circrad}{d}
       +\log\frac{32}{\err}+\log\length\right]\right)
\end{equation}
Hit and Run steps.
\end{theorem}

For the proof, we would require to invoke the convex-body diameter bound of~\cite[Equation~4.3]{payne1960optimal}, whose proof was corrected by~\cite{bebendorf2003note}. We will use the formulation of~\cite[Theorem 5.10]{milman09} consistently.

Let us now proceed towards proving \Cref{thm:exceptional-application-appendix}.

\subsection{General sampling and contraction bounds}\label{sec:exceptional-main-result}

We write $\gamma_{\mathrm{HR}}$ for the spectral gap of Hit and Run. Also since in dimension one, a single Hit and Run step samples exactly uniformly from the interval $\body$, we assume from now on that $\dm\ge 2$.

What we will prove abstractly is the following theorem, which gives a sufficient number of steps required to run Hit and Run for, starting from any interior point, so that we can satisfy the guarantees of \Cref{eq:exceptional-application-parameters-appendix}.

\begin{theorem}\label{thm:exceptional-point-sampler}
Let $0<\sampgood,\sampbad<1$. Then every integer $t$ satisfying
\begin{equation}\label{eq:exceptional-generic-steps}
 t\ge 1+\lfloor\dm/2\rfloor+\left\lceil
 \frac{\dm\log(32\circrad/d)+\log(1/(\sampgood^2\sampbad))}
      {2\gamma_{\mathrm{HR}}}
 \right\rceil
\end{equation}
satisfies the guarantees
\[
 \unif(\bad_{t,\sampgood})\le\sampbad,
 \qquad \distr_t(\bad_{t,\sampgood})\le2\sampbad.
\]
\end{theorem}

Given this theorem, if we have a lower bound on $\gamma_{\mathrm{HR}}$, then we can complete the proof of \Cref{thm:exceptional-application-appendix}. Towards that, let us define the following quantity. We define $C_{\mathrm{PI}}(\unif)$ as the least constant for which
\[
 \Var_\unif(f)\le C_{\mathrm{PI}}(\unif)
              \int_\body\norm{\nabla f}^2\dd\unif
\]
holds for every $f\in H^1(\operatorname{int}\body)$. Here $H^1$ consists of square-integrable functions with square-integrable weak derivatives\footnote{A function $f\in L^2$ has weak derivative $\partial_jf$ if for all test functions $\varphi\in C_c^\infty$ (the space of all compactly supported smooth functions) one has $\int f\partial_j\varphi=-\int (\partial_jf)\varphi$.}.
Then we have the following.
\begin{lemma}[Contraction for Hit and Run {Corollary and Extension of \cite[Theorem~1.1]{KookVempala2026}}]\label{lem:exceptional-contraction}
There is a universal constant $c>0$ such that
\begin{equation}\label{eq:exceptional-KV-gap}
 \gamma_{\mathrm{HR}}\ge
 c\,\frac{\inrad^2}{\dm^2C_{\mathrm{PI}}(\unif)}.
\end{equation}
 In particular
\begin{equation}\label{eq:exceptional-geometric-gap}
 \gamma_{\mathrm{HR}}\ge c\,\frac{\inrad^2}{\dm^2\circrad^2}.
\end{equation}
Moreover, for every mean-zero $f\in L^2(\unif)$ and every integer $s\ge0$,
\begin{equation}\label{eq:operator-contraction-appendix}
    \norm{P^sf}_{L^2(\unif)}\le e^{-\gamma_{\mathrm{HR}}s}\|f\|_2\le \exp\!\left(-c\frac{\inrad^2s}{\dm^2C_{\mathrm{PI}}(\unif)}\right)\norm f_{L^2(\unif)}
\end{equation}
\end{lemma}

Now we prove \Cref{thm:exceptional-application-appendix}.

\begin{proof}[Proof of \Cref{thm:exceptional-application-appendix}]
Substituting the parameter values gives
\[\sampgood^2\sampbad=\frac{\err^3}{32^3\length}\]
and thus
\[ \log\frac1{\sampgood^2\sampbad}
 =3\log\frac{32}{\err}+\log\length \]
By \cref{thm:exceptional-point-sampler}, both sampling guarantees hold whenever
\begin{equation}\label{eq:exceptional-application-steps}
 t\ge1+\lfloor\dm/2\rfloor+\left\lceil
 \frac{\dm\log(32\circrad/d)+3\log(32/\err)+\log\length}
 {2\gamma_{\mathrm{HR}}}
 \right\rceil.
\end{equation}
Plugging \cref{eq:exceptional-geometric-gap} in this bound gives \cref{eq:exceptional-geometric-application-appendix}.
\end{proof}

With this settled, we move on to the proofs of \Cref{thm:exceptional-point-sampler} and \Cref{lem:exceptional-contraction}.

\subsection{Regularising the point-start density}\label{sec:exceptional-regularisation}

We now prove \cref{thm:exceptional-point-sampler}. The idea of the proof is as follows:

\begin{enumerate}
      \item First use a few steps of Hit and Run to make the density square-integrable
      \item Then use the remaining steps to reduce its (squared) error by spectral contraction.
\end{enumerate}
Let us first see how a squared-error bound gives both sampling guarantees in \Cref{eq:exceptional-sampler-requirements}. This will justify why we want item~(2).

\begin{lemma}\label{lem:exceptional-L2-interface}
Let $0<\sampgood,\sampbad<1$, and let $\distr$ have relative density $\density$ with respect to $\unif$. If
\[
 \chi^2(\distr\Vert\unif)
 :=\int_\body|\density-1|^2\dd\unif
 \le\sampgood^2\sampbad,
\]
then $\unif(\bad_\sampgood)\le\sampbad$ and $\distr(\bad_\sampgood)\le(1+\sampgood)\sampbad\le2\sampbad$.
\end{lemma}

\begin{proof}
For the first guarantee, we can use Markov's inequality to get
\[
 \unif(\bad_\sampgood)
 \le\frac1{\sampgood^2}\int_\body|\density-1|^2\dd\unif
 \le\sampbad.
\]
For the second guarantee, we can use Cauchy-Schwarz on the exceptional set to get
\begin{align*}
 \distr(\bad_\sampgood)
 &=\unif(\bad_\sampgood)
    +\int_{\bad_\sampgood}(\density-1)\dd\unif\\
 &\le\unif(\bad_\sampgood)
    +\left(\unif(\bad_\sampgood)
             \int_\body|\density-1|^2\dd\unif\right)^{1/2}\\
 &\le\sampbad+\sampgood\sampbad.
\end{align*}
\end{proof}

Now we proceed to towards item~(1), that is, bound the number of steps required for the first square-integrable step. For that we need a bound on the density after the first few steps. Let us write $\kappa_\dm=\vol(B_2^\dm)$, where $B_2^\dm$ is the Euclidean unit ball.

\begin{lemma}[The first square-integrable step]\label{lem:exceptional-regularisation}
Let $k=\lfloor\dm/2\rfloor+1$. Then
\begin{equation}\label{eq:exceptional-regularisation}
 \int_\body\density_k^2\dd\unif
 \le32^\dm\frac{\vol(\body)}{\kappa_\dm d^\dm}
 \le\left(\frac{32\circrad}{d}\right)^\dm.
\end{equation}
and $\density_k$ is the first square-integrable density.
\end{lemma}

Let us first see how this estimate completes the proof of \cref{thm:exceptional-point-sampler}.

\begin{proof}[Proof of \cref{thm:exceptional-point-sampler}]
Let $k=1+\lfloor\dm/2\rfloor$. Since $\rho_k$ is a probability density, $\density_k-1$ has mean zero and so by \cref{lem:exceptional-regularisation}, we have
\[
 \norm{\density_k-1}_2^2=\int_\body\density_k^2\dd\unif-1
 \le\left(\frac{32\circrad}{d}\right)^\dm.
\]
Applying \Cref{eq:operator-contraction-appendix} from \cref{lem:exceptional-contraction} then gives, for $t\ge k$,
\[\chi^2(\mathcal D_t\Vert\pi)=\|P^{t-k}(\rho_k-1)\|_2^2\le e^{-2\gamma_{\mathrm{HR}}(t-k)}\|\rho_k-1\|_2^2\le e^{-2\gamma_{\mathrm{HR}}(t-k)}\left(\frac{32R}{d}\right)^n
\]\label{eq:exceptional-chi-contraction}
Then plugging in the value in \cref{eq:exceptional-generic-steps} bounds the right-hand side by $\sampgood^2\sampbad$ and hence both sampling guarantees now follow from \cref{lem:exceptional-L2-interface}.
\end{proof}

\begin{remark}\label{rem:chi-square-improvement}
Let us also mention what the proof gives for convergence from a point start in standard Hit and Run terms. With $k=1+\lfloor\dm/2\rfloor$, for every $t\ge k$, note that we have
\[\chi^2(\distr_t\Vert\unif)\le\left(\frac{32\circrad}{d}\right)^\dm e^{-2\gamma_{\mathrm{HR}}(t-k)}\]
Consequently, for any $0<\err<1$, we have $\chi^2(\distr_t\Vert\unif)\le\err$ whenever
\[t\ge k+\left\lceil\frac{\dm\log(32\circrad/d)+\log(1/\err)}{2\gamma_{\mathrm{HR}}}\right\rceil\]
We can also get total variation error at most $\err$ by replacing $\err$ by $\err^2$ since $\TV(\distr_t,\unif)\le\frac12\sqrt{\chi^2(\distr_t\Vert\unif)}$. Let us compare this with the bound Lovasz-Vempala got. Their original point-start result gives total variation error at most $\err$ in
\[\bigO\!\left(\dm^3\left(\frac{\circrad}{\inrad}\right)^2\left[\log\frac{\circrad}{d}+\log\frac1\err\right]\right)\]
steps~\cite[Corollary~1.2]{LV06}. The same also follows from \cite[Theorem~1.1 and Corollary~1.2]{KookVempala2026}, using the usual first-step truncation together with the convex-domain diameter bound~\cite[Theorem~5.10]{milman09} for the Poincare. On the other hand, Lovasz-Vempala's conductance bound~\cite[Theorem~4.2]{LV06} with our regularisation estimate gives the same total variation guarantee in
\[\bigO\!\left(\dm^3\left(\frac{\circrad}{\inrad}\right)^2\log\frac{32\circrad}{d}+\dm^2\left(\frac{\circrad}{\inrad}\right)^2\log\frac1\err\right)\]
steps. Thus our result improves the coefficient of $\log(1/\err)$ by a factor of order $\dm$.
\end{remark}

Let us now prove \cref{lem:exceptional-regularisation}. We recall some standard facts about Hit and Run. For distinct $x,y\in\body$, let $\ell(x,y)$ be the length of the chord on the line through $x$ and $y$. The Lebesgue transition density is~\cite[equation~(5)]{LV06}
\begin{equation}\label{eq:exceptional-kernel}
 p(x,y)=\frac{2}{\dm\kappa_\dm\ell(x,y)|x-y|^{\dm-1}}.
\end{equation}
Note that this formula is symmetric in $x,y$ (so we have a reversible Markov chain). Consequently densities evolve by $\density_{j+1}=P\density_j$, where
\[
 Pf(y)=\int_\body p(x,y)f(x)\dd x.
\]
For nonnegative integrable functions these identities follow by Tonelli's theorem.

The ball $B(x_0,d)$ lies inside $\body$, so every chord through $x_0$ has length at least $2d$. It follows that
\begin{equation}\label{eq:exceptional-first-density}
 \density_1(y)\le
 \frac{\vol(\body)}{\dm\kappa_\dm d}|y-x_0|^{1-\dm}.
\end{equation}
Observe that the factor $|y-x_0|^{1-\dm}$ is due to the singularity near the starting point. The idea for the proof of \Cref{lem:exceptional-regularisation} is to follow this singularity near $x_0$ through the initial steps and keep track of it inside the fixed ball $U=B(x_0,a)$ with $a=d/2$.
Since every $x\in U$ has boundary distance at least $d/2$, the kernel in \cref{eq:exceptional-kernel} satisfies
\begin{equation}\label{eq:exceptional-local-kernel}
 p(x,y)\le\frac{2}{\dm\kappa_\dm d}|x-y|^{1-\dm}
 \qquad \text{for all } x\in U,\ y\in\body.
\end{equation}
The next lemma is a small technicality that explains why applying this kernel shaves off a power from the singularity, and we will then use this to prove \cref{lem:exceptional-regularisation}.

\begin{lemma}\label{lem:exceptional-convolution}
For $\phi_s(y)=(|y-x_0|/a)^{s-\dm}$ and $3/4\le s\le\dm/2-1/4$,
\[
 P(\phi_s\one_U)(y)\le16\phi_{s+1}(y)
 \quad\text{for almost every }y\in\body.
\]
\end{lemma}

Let us now use this estimate to prove \cref{lem:exceptional-regularisation} and then prove \Cref{lem:exceptional-convolution} itself.

\begin{proof}[Proof of \cref{lem:exceptional-regularisation}]
We split the density at each step into a singular part supported on $U$ and a bounded part. Let us define the sequence\footnote{The value $c_1$ bounds the first density outside $U$. Each application of \cref{lem:exceptional-convolution} multiplies the coefficient of the singular part by at most $16$, so $c_j$ captures this growth.}
\[
 c_1=\frac{\vol(\body)}{\dm\kappa_\dm d}a^{1-\dm},
 \qquad c_j=16^{j-1}c_1,
\]
and initially let $\density_1=b_1+u_1$, where $u_1=\density_1\one_U$ and $b_1=\density_1\one_{U^c}$. By \cref{eq:exceptional-first-density}, we have
\[
 b_1\le c_1,\qquad u_1\le c_1\phi_{3/4}\one_U.
\]

For $1\le j<k$, we define\footnote{The idea for this definition is to understand what happens when we stay in $U$ vs when we go outside of $U$. $u_j$ keeps track of when we have stayed inside $U$ throughout the first $j$ steps, and this lets us repeatedly apply \cref{lem:exceptional-convolution}. The $b_j$ on the other hand tracks what happens if we venture out and also if we venture out and come back to $U$.}
\begin{equation}\label{eq:exceptional-density-decomposition}
 u_{j+1}=\one_U Pu_j,
 \qquad b_{j+1}=Pb_j+\one_{U^c}Pu_j.
\end{equation}
Both $u_{j+1}$ and $b_{j+1}$ are nonnegative and sum to $\density_{j+1}$. Thus each component has $\unif$-integral at most one.

We now claim that for every $j\le k$,
\[
 u_j\le c_j\phi_{j-1/4}\one_U,
 \qquad \norm{b_j}_\infty\le\sum_{i=1}^j c_i\le\frac{16}{15}c_j.
\]
The claim holds at $j=1$. If it holds at $j<k$, then
$3/4\le j-1/4\le\dm/2-1/4$, so \cref{lem:exceptional-convolution} gives
\[
 Pu_j\le16c_j\phi_{j+3/4}=c_{j+1}\phi_{j+3/4}.
\]
This profile is at most one on $U^c$ because $j+3/4<\dm$. Consequently the second term in the expression for $b_{j+1}$ is bounded by $c_{j+1}$. Since $P$ is a Markov operator, it contracts $L^\infty$ and hence we have
\[
 \norm{b_{j+1}}_\infty\le\norm{b_j}_\infty+c_{j+1}
 \le\sum_{i=1}^{j+1}c_i.
\]
This proves the claim by induction.

Now at step $k$, the exponent $k-1/4$ satisfies $2(k-1/4)-\dm\ge1/2$. Moving to polar coordinates therefore gives
\[
 \int_U\phi_{k-1/4}^2\dd\unif
 =\frac{\dm\kappa_\dm a^\dm}
 {\vol(\body)(2k-1/2-\dm)}
 \le\frac{2\dm\kappa_\dm a^\dm}{\vol(\body)}.
\]
Using this and the mass bound $\int_\body b_k\dd\unif\le1$, we obtain the estimates (where all norms are with respect to $\pi$)
\[ \norm{b_k}_2^2\le\norm{b_k}_\infty\le\frac{16}{15}c_k\]
and
\[ \norm{u_k}_2^2\le
 \frac{2\dm\kappa_\dm a^\dm}{\vol(\body)}c_k^2
 =\frac{2^{\dm-1}\vol(\body)}{\dm\kappa_\dm d^\dm}16^{2k-2} \]
Now since $1\le k-1=\lfloor\dm/2\rfloor\le\dm/2$, we have
\[\norm{\density_k}_2^2
 \le2\norm{b_k}_2^2+2\norm{u_k}_2^2
 \le\frac{2^\dm\vol(\body)}{\dm\kappa_\dm d^\dm}
 \left[\frac{16}{15}16^{k-1}+16^{2k-2}\right]
 \le32^\dm\frac{\vol(\body)}{\kappa_\dm d^\dm}
 \le\left(\frac{32\circrad}{d}\right)^\dm\]
This proves \cref{eq:exceptional-regularisation}. All that remains to show is that $\rho_k$ is really the density at the first square-integrable step (or equivalently, $k$ is the first square-integrable step). For this, observe that every chord has length at most $2\circrad$. Thus \cref{eq:exceptional-kernel} gives
\begin{equation}\label{eq:exceptional-kernel-lower}
 p(x,y)\ge\frac1{\dm\kappa_\dm\circrad}|x-y|^{1-\dm}.
\end{equation}
We claim that for some positive constants $a_j$, we have
\[
 \density_j(y)\ge a_j|y-x_0|^{j-\dm}
 \quad\text{for almost every }0<|y-x_0|<d,
 \quad 1\le j\le\lfloor\dm/2\rfloor.
\]
To show the claim, we again use induction on $j$. The lower bound in \cref{eq:exceptional-kernel-lower}, together with $\density_1(y)=\vol(\body)p(x_0,y)$, proves the case of $j=1$. For the inductive step, we let $s=|y-x_0|<d$ and integrate over $B(x_0,s)$. On that ball, $|x-y|\le2s$ and $|x-x_0|^{j-\dm}\ge s^{j-\dm}$, giving
$\density_{j+1}(y)\ge a_j2^{1-\dm}s^{j+1-\dm}/(\dm\circrad)$.
For $2j\le\dm$, the squared-density integral consequently contains the divergent radial integral
\[
 \int_0^d s^{2j-\dm-1}\dd s=\infty.
\]
Thus together with \cref{eq:exceptional-regularisation}, this shows $\int_\body \rho_j^2\ \mathrm d\pi=\infty$ for $1\le j<k$ while $\rho_k$ is square-integrable. Hence this proves that $k$ is the first square-integrable step.
\end{proof}

\begin{remark}\label{rem:regularization-step-independent-epsilon}
    It is interesting to note that the number of steps required for regularisation is completely independent of the required accuracy and depends only on the dimension. To the best of our knowledge, this exact threshold for square-integrability from an interior point has not been stated previously for Hit and Run.
\end{remark}

\begin{remark}\label{rem:kook-vempala-intro-remark}
    We also want to remark the following. The introduction of \citet{KookVempala2026} mentions applying convergence after one Hit and Run step and their $\chi^2$ convergence estimate uses a square-integrable initial relative density~\cite[Corollary~1.2 and equation~(1.2)]{KookVempala2026}. For a point start, note that $\chi^2(\distr_1\Vert\unif)=\infty$ for $\dm\ge2$, but still the usual first-step truncation gives a total variation guarantee~\cite[Corollary~1.2 and Section~5]{LV06}. In contrast, our \Cref{lem:exceptional-regularisation} proves that $1+\lfloor\dm/2\rfloor$ steps are necessary and sufficient for the full density to actually acquire finite $\chi^2$.
\end{remark}

We finish this subsection by proving \Cref{lem:exceptional-convolution}. For that we will use the Riesz convolution formula~\cite[Chapter~V, Section~1]{Stein1970SingularIntegrals}, for $0<s<\dm-1$ and $y\ne0$, which states that
\[
 \int_{\R^\dm}|x|^{s-\dm}|x-y|^{1-\dm}\dd x
 =C_{\dm,s}|y|^{s+1-\dm}
\]
where
\[
 C_{\dm,s}=
 \frac{\pi^{(\dm+1)/2}\Gamma(s/2)\Gamma((\dm-s-1)/2)}
 {\Gamma((\dm-s)/2)\Gamma((\dm-1)/2)\Gamma((s+1)/2)}.
\]

\begin{proof}[Proof of \cref{lem:exceptional-convolution}]

Using \cref{eq:exceptional-local-kernel}, enlarging the integration domain to $\R^\dm$, and recalling that $2a=d$, we obtain
\[
 P(\phi_s\one_U)(y)
 \le\frac{C_{\dm,s}}{\dm\kappa_\dm}\phi_{s+1}(y).
\]

All that is left is to bound $C_{\dm,s}/(\dm\kappa_\dm)$. For that, we can use the fact that $\Gamma$ is log-convex on the positive reals and $(\dm-s-1)/2\ge\dm/16\ge1/8$ and $s/2\ge3/8$ to get
\[
 \frac{C_{\dm,s}}{\dm\kappa_\dm}
 =\frac{\sqrt\pi}{2}
 \frac{\Gamma(\dm/2)}{\Gamma((\dm-1)/2)}
 \frac{\Gamma((\dm-s-1)/2)}{\Gamma((\dm-s)/2)}
 \frac{\Gamma(s/2)}{\Gamma((s+1)/2)}
 \le\frac23\sqrt{140\pi}<16.
\]

This proves the lemma.
\end{proof}

\subsection{Contraction for Hit and Run}\label{sec:exceptional-radial-flux}

We now give the geometric proof of \cref{lem:exceptional-contraction}, following the half-volume comparison described before.

Without loss of generality, let us translate the centre of an inball to the origin, so $\inrad B_2^\dm\subseteq\body$. Let $h(\body)$ be the functional Cheeger constant~\cite[Lemma~2.2]{milman09}, which is the largest constant such that every Lipschitz function $f$ satisfies
\begin{equation}\label{eq:exceptional-cheeger-definition}
 h(\body)\inf_{z\in\R}\int_\body|f-z|\dd\unif
 \le\int_\body\norm{\nabla f}\dd\unif.
\end{equation}
For succinctness, let $\delta_\body(x)=\dist(x,\partial\body)$ and let $R_\body=\max_{x\in\body}|x|$ be the radius measured from the chosen centre.

We will use this slightly technical lemma that bounds the deviation of any Lipschitz $f$ over the body $\body$ using the (norm of the) gradient of $f$ and $h(\body)$.

\begin{lemma}\label{lem:exceptional-radial-flux}
For every Lipschitz function $f$ on $\body$,
\begin{equation}\label{eq:exceptional-flux-L1}
 \inf_{z\in\R}\int_\body|f-z|\dd\unif
 \le\frac{8\dm}{\inrad h(\body)}
          \int_\body\delta_\body\norm{\nabla f}\dd\unif.
\end{equation}
\end{lemma}
Towards proving this, it will help to break down slowly what we really need. One can consider this discussion to be part of the proof itself. Since $\inrad B_2^\dm\subseteq\body$, it makes sense to consider the scaled version $C=2^{-1/\dm}\body$ of $K$ so that $\pi(C)=1/2$ (technically speaking, any scaling that makes the measure a constant less than $1$ works, but $2^{-1/n}$ is convenient because taking half the volume makes life a lot easier as we will later see). Let $p_\body(x)=\inf\{s>0:x\in s\body\}$ be the gauge\footnote{The gauge of a point $x$ in a convex body (containing the origin inside it) is the smallest possible (infimum technically) scaling of the body so that $x$ lies on the boundary.} of $\body$. Then using the convexity of $\body$ and the inball $\inrad B_2^\dm$, we have
\[
 x+(1-p_\body(x))\inrad B_2^\dm\subseteq\body,
 \qquad \delta_\body(x)\ge\inrad(1-p_\body(x)).
\]
In particular, every point $x\in C$ is at least $\inrad(1-2^{-1/\dm})$ from the boundary, that is
\begin{equation}\label{eq:exceptional-core-distance}
 \delta_\body(x)\ge\inrad(1-2^{-1/\dm})
\end{equation}
The Cheeger inequality in \Cref{eq:exceptional-cheeger-definition}, applied after scaling $\body$ to $C$, will control the error on $C$ once we choose $z$ as a median of $f$ on $C$ there. To pass from $C$ to $\body$, we want to bound the following difference by something:
\begin{equation}\label{eq:vector-field-intuition-equation}
 \int_\body|f-z|\dd\unif-2\int_C|f-z|\dd\unif
 =2\int_\body\left(\tfrac12-\one_C\right)|f-z|\dd\unif
\end{equation}
Now we observe that for any Lipschitz vector field $F$, we have the identity 
\[2\int_\body F\cdot\nabla\vert{}f-z\vert{}\dd\unif=-2\int_\body \operatorname{div}(F)\vert{}f-z\vert{}\dd\unif+\frac2{\vol(\body)}\int_{\partial\body}\vert{}f-z\vert{}(F\cdot \nu)\dd s\]
where $\nu$ is the outward unit normal vector to $\partial\body$ and $s$ is the boundary measure.

So it is natural to ask for a vector field $F$ such that $\operatorname{div}F=\tfrac12-\one_C$ almost everywhere and $F|_{\partial\body}=0$ so that the right hand side of \Cref{eq:vector-field-intuition-equation} turns into\footnote{Use $\operatorname{div}(F|f-z|)
=|f-z|\operatorname{div}F+F\cdot\nabla|f-z|$ and note that the integral of the left side vanishes since it becomes a boundary integral by the divergence theorem and $F=0$ on $\partial\body$. Then rearrange the left hand side terms and use $\operatorname{div}F=\tfrac12-\one_C$.}
\[-2\int_\body F\cdot\nabla|f-z|\dd\unif\]
which is much easier to handle. For the actual bound in \Cref{lem:exceptional-radial-flux}, we note that
\[\left\vert{}-2\int_\body F\cdot\nabla\vert{}f-z\vert{}\dd\unif\right\vert{} \le2\int_\body \vert{}F(x)\vert{}\Vert{}\nabla f(x)\Vert{}\dd\unif\]
and hence we also want $|F(x)|\le(R_\body/\inrad)\delta_\body(x)$. This turns out to be a boundary value problem, which has as (one of its) solutions
\[
 F(x)=\phi(p_\body(x))x,\qquad
 \phi(q)=\begin{cases}
 -1/(2\dm),&0\le q\le2^{-1/\dm},\\
 (1-q^{-\dm})/(2\dm),&2^{-1/\dm}<q\le1.
 \end{cases}
\]
So we will work with this. Let us quickly verify that this actually works.

The gauge $p_\body$ is easily seen to be Lipschitz and satisfies $x\cdot\nabla p_\body(x)=p_\body(x)$ almost everywhere, and so directly differentiating the vector field gives\footnote{This is the payoff for choosing the particular scaling of $2^{-1/n}$ for $C$, as now $\int_{\body}\left(\tfrac12-\mathbf1_C\right)\dd\unif=0$.}
\[
 F|_{\partial\body}=0,
 \qquad \operatorname{div}F=\tfrac12-\one_C.
\]
Also note that $q|\phi(q)|\le1-q$ for $0\le q\le1$ and $|x|\le R_\body p_\body(x)$, so
\begin{equation}\label{eq:exceptional-radial-field-bound}
 |F(x)|\le R_\body(1-p_\body(x))
          \le\frac{R_\body}{\inrad}\delta_\body(x).
\end{equation}
This is the core of the geometric argument, the rest is going through the motions to get the final bounds. So let us now begin the formal proof of \Cref{lem:exceptional-radial-flux} assuming the above context and notions.

\begin{proof}[Proof of \Cref{lem:exceptional-radial-flux}]
Choose a median $z$ of $f$ on $C$. Integration by parts gives
\begin{equation}\label{eq:exceptional-flux-identity}
 \int_\body|f-z|\dd\unif
 =2\int_C|f-z|\dd\unif
       -2\int_\body F\cdot\nabla|f-z|\dd\unif.
\end{equation}
To control the first term, we have
\[ \int_C|f-z|\dd\unif
 \le\frac{2^{-1/\dm}}{h(\body)}\int_C\norm{\nabla f}\dd\unif
 \le\frac{1}{\inrad h(\body)(2^{1/\dm}-1)}
       \int_\body\delta_\body\norm{\nabla f}\dd\unif \]
where the first inequality comes from applying the Cheeger inequality in \cref{eq:exceptional-cheeger-definition} to $x\mapsto f(2^{-1/\dm}x)$ and changing variables from $\body$ to $C$, and the second inequality follows from the distance bound in \cref{eq:exceptional-core-distance}. Substituting this and \cref{eq:exceptional-radial-field-bound} into \cref{eq:exceptional-flux-identity} gives
\begin{align}
 \int_\body|f-z|\dd\unif
 &\le\left[\frac{2}{\inrad h(\body)(2^{1/\dm}-1)}
              +\frac{2R_\body}{\inrad}\right]
       \int_\body\delta_\body\norm{\nabla f}\dd\unif\notag\\
 &\le\left[\frac{2\dm}{\inrad h(\body)\log2}
              +\frac{2R_\body}{\inrad}\right]
       \int_\body\delta_\body\norm{\nabla f}\dd\unif,
       \label{eq:exceptional-flux-coefficient}
\end{align}
where we used $2^{1/\dm}-1\ge(\log2)/\dm$.

We now need to bound $2R_\body/\inrad$. For that, we use the elementary estimate $R_\body h(\body)\le2\dm$. To see this, choose a unit vector $e$ with $R_\body e\in\body$, and let $q$ be the density of $e\cdot X$ for $X\sim\unif$, with support $[a,b]$ and median $m$. By the Brunn-Minkowski theorem~\cite[Theorem~7.1.1]{Schneider2013}, $q^{1/(\dm-1)}$ is concave and hence $q(t)\ge q(m)((b-t)/(b-m))^{\dm-1}$ for $m\le t\le b$. Integrating this bound gives
\[ \frac12\ge\frac{q(m)(b-m)}\dm \]
Analogously
\[ \frac12\ge\frac{q(m)(m-a)}\dm \]
Thus in particular, we have $q(m)\le\dm/(b-a)$. To turn this into a bound on $h(\body)$, we show $h(\body)\le2q(m)$ by testing the Cheeger inequality on functions approaching the indicator of the median halfspace $\{e\cdot x>m\}$. For $s>0$, consider the function
\[
 \psi_s(x)=\min\left\{1,\frac{(e\cdot x-m)_+}{s}\right\}
\]
This function has median zero, so \cref{eq:exceptional-cheeger-definition} gives
\[
 h(\body)\int_\body\psi_s\dd\unif
 \le\int_\body\norm{\nabla\psi_s}\dd\unif
 =\frac1s\int_m^{m+s}q(t)\dd t.
\]
Letting $s\downarrow0$ gives $h(\body)/2\le q(m)$. Since $b-a\ge R_\body+\inrad$, we obtain
\begin{equation}\label{eq:exceptional-radius-cheeger}
 h(\body)\le\frac{2\dm}{R_\body+\inrad}.
\end{equation}
Using \cref{eq:exceptional-radius-cheeger} in \cref{eq:exceptional-flux-coefficient}, we have the bound
\[
 \frac{2\dm}{\inrad h(\body)\log2}+\frac{2R_\body}{\inrad}
 \le\left(4+\frac2{\log2}\right)\frac{\dm}{\inrad h(\body)}
 <\frac{8\dm}{\inrad h(\body)}.
\]
Taking the infimum over constants proves the lemma.
\end{proof}

We can now finish the proof of \Cref{lem:exceptional-contraction}. Also, we recall that the Babu\v{s}ka-Aziz constant $C_{\mathrm{BA}}$ is the least constant such that every mean-zero $f\in L^2(\unif)$ admits a vector field $G\in H_0^1(\operatorname{int}\body;\R^\dm)$ with $\operatorname{div}G=f$ and \[\int\norm{\nabla G}_{\mathrm F}^2\dd\unif \le C_{\mathrm{BA}}\norm f_2^2 \]

\begin{proof}[Proof of \cref{lem:exceptional-contraction}]
Let us first turn the inequality from \Cref{lem:exceptional-radial-flux} into a variance bound. For a Lipschitz function $f$, choose a median $z$ under $\unif$ and apply \cref{eq:exceptional-flux-L1} separately to $(f-z)_+^2$ and $(z-f)_+^2$, each of which has median zero. Applying Cauchy-Schwarz and adding the bounds gives
\begin{equation}\label{eq:exceptional-flux-variance}
 \Var_\unif(f)\le
 \frac{256\dm^2}{\inrad^2h(\body)^2}
       \int_\body\delta_\body^2\norm{\nabla f}^2\dd\unif.
\end{equation}
The inequality in \cref{eq:exceptional-flux-variance} is the improved Poincare inequality~\cite[Definition~3.1]{Zsuppan2020} and thus the improved Poincare constant is $\frac{256\dm^2}{\inrad^2h(\body)^2}$.

To connect the improved Poincare constant with the Babu\v{s}ka-Aziz constant $C_{\mathrm{BA}}$, we would need \Cref{eq:exceptional-flux-variance} to hold on a larger class than just Lipschitz functions. To that end, let us show that we can extend the inequality to $f\in L^2(\unif)\cap H^1_{\mathrm{loc}}(\operatorname{int}\body)$ as follows\footnote{$H^1_{\mathrm{loc}}$ is the space of all functions that are locally in $H^1(U)$ for every open subset $U$ with a compact closure contained in the domain}. For $0<s<1$, contract $f$ and let $f_s(x)=f(sx)$. The function $f_s$ is $H^1$ on a neighbourhood of $\body$, and convexity gives $\delta_\body(sx)\ge s\delta_\body(x)$. Thus its weighted energy\footnote{For $f_s$, this means $\int_\body\delta_\body^2\norm{\nabla f_s}^2\dd\unif$,
where we weight the squared gradient by the squared distance
from the boundary.} is at most $s^{-\dm}$ times that of $f$. Convolving $f_s$ with mollifiers gives smooth functions converging to it in
$H^1(\operatorname{int}\body)$~\cite[Section~2.2]{AmbrosioFuscoPallara2000}. Since $\delta_\body$ is bounded, we can pass to the limit in \cref{eq:exceptional-flux-variance} to obtain it for $f_s$.
Finally, $f_s\to f$ in $L^2(\unif)$ as $s\uparrow1$, so
$\Var_\unif(f_s)\to\Var_\unif(f)$. Using the weighted-energy bound above and $s^{-\dm}\to1$ now gives \cref{eq:exceptional-flux-variance} for $f$.

We now use \cref{eq:exceptional-flux-variance} to bound the Babu\v{s}ka-Aziz constant $C_{\mathrm{BA}}$, and then apply the comparison with the Hit and Run spectral gap $\gamma_{\mathrm{HR}}$.

Applying the Costabel-Zsupp\'an comparison~\cite[Theorem~2.1]{Costabel_2017}\cite[Lemma~3.4 and Remark~3.6]{Zsuppan2020} to \cref{eq:exceptional-flux-variance} gives\footnote{Costabel bounds $C_{\mathrm{BA}}$ by $1+$the Friedrichs-Velte constant, and Zsuppan bounds the Friedrichs-Velte constant by $4$ times the improved Poincare constant.}
\[
 C_{\mathrm{BA}}\le1+\frac{1024\dm^2}{\inrad^2h(\body)^2}.
\]

Next we can use ~\cite[Proposition~2.1]{KookVempala2026} to get
\[
 \gamma_{\mathrm{HR}}\ge\frac{\dm+2}{\dm(2+C_{\mathrm{BA}})}.
\]
Using \cref{eq:exceptional-flux-variance} and $\inrad h(\body)/\dm\le1$ from \cref{eq:exceptional-radius-cheeger}, we have
\[
 \gamma_{\mathrm{HR}}\ge
 \frac1{1027}\frac{\inrad^2h(\body)^2}{\dm^2}.
\]
The Cheeger-Poincare equivalence for convex bodies then gives
$h(\body)^2\ge c/C_{\mathrm{PI}}(\unif)$ for a universal $c>0$~\cite[Theorem~1.5]{milman09}. This proves \cref{eq:exceptional-KV-gap}.

For the geometric bound of \cref{eq:exceptional-geometric-gap}, we use Milman's convex-domain diameter estimate~\cite[Theorem~5.10]{milman09} to get
\begin{equation}\label{eq:exceptional-poincare-radius}
 C_{\mathrm{PI}}(\unif)
 \le\frac{\diam(\body)^2}{\pi^2}
 \le\frac{4\circrad^2}{\pi^2}.
\end{equation}
Substituting this in \cref{eq:exceptional-KV-gap} proves \cref{eq:exceptional-geometric-gap}.

Finally, let us prove the $L^2$ contraction claim. For each fixed direction, averaging along parallel chords is a conditional-expectation operator and hence an orthogonal projection. Averaging these projections over the sphere gives a self-adjoint positive operator $P$ that preserves $\unif$~\cite[Section~2.1]{KookVempala2026}, see also \cite[Section~3.1]{RudolfUllrich2013}. Its spectral gap is
\[
 \gamma_{\mathrm{HR}}=
 \inf_{\substack{f\in L^2(\unif),\ \int f\dd\unif=0\\f\ne0}}
 \frac{\langle f,(I-P)f\rangle_{L^2(\unif)}}{\norm f_{L^2(\unif)}^2}.
\]

On the $P$-invariant subspace $\left\{f\in L^2(\unif):\int_\body f\dd\unif=0\right\}$, the spectrum of $P$ lies in $[0,1-\gamma_{\mathrm{HR}}]$ and thus the spectral theorem gives
\begin{align*}
 \norm{P^sf}_{L^2(\unif)}
 \le(1-\gamma_{\mathrm{HR}})^s\norm f_{L^2(\unif)}\le e^{-\gamma_{\mathrm{HR}}s}\norm f_{L^2(\unif)}\le\exp\!\left(-c\frac{\inrad^2s}{\dm^2C_{\mathrm{PI}}(\unif)}\right)
       \norm f_{L^2(\unif)},
\end{align*}
where the last inequality follows from \cref{eq:exceptional-KV-gap}.
\end{proof}

\subsection{The one-step truncation bound}\label{sec:exceptional-comparisons}

Finally, in this subsection, we justify the one-step deduction used in \cref{eq:exceptional-KV-application}. The idea is to separate the mass in a neighborhood around $x_0$ after the first step and apply the contraction bound in \cref{lem:exceptional-contraction} to the density outside this neighbourhood. For arbitrary $0<\sampgood,\sampbad<1$, both sampling guarantees hold whenever
\begin{equation}\label{eq:exceptional-one-step-comparison}
 t\ge1+\frac1{2\gamma_{\mathrm{HR}}}
 \left[\dm\log\frac{8\circrad}{d}
       +(\dm+1)\log\frac1{\sampgood}
       +\dm\log\frac1{\sampbad}\right].
\end{equation}
To see this, let $a=\sampgood\sampbad/4$ for simplicity of notations and consider the neighborhood $S=B(x_0,ad)$. Conditional on the line chosen in the first step of Hit and Run, the fraction of the chord that lies inside $S$ is at most $a$. Hence the first-step probability of being inside $S$ is at most $a$. By \cref{eq:exceptional-first-density}, the density on $S^c$ is bounded by
\[
 \density_1\one_{S^c}\le
 \frac1{\dm a^{\dm-1}}\left(\frac{\circrad}{d}\right)^\dm.
\]
We want to understand what happens to the error that we controlled in \cref{lem:exceptional-contraction} for this neighborhood. So define
\[ f=P^{t-1}(\density_1\one_{S^c})-(1-\distr_1(S))\]
where the first term captures the part of the first-step density outside $S$ after another $t-1$ steps, and the second term is the mean. Then $f$ has mean zero by definition and
\[
 \density_t-1=f+P^{t-1}(\density_1\one_S)-\distr_1(S).
\]
The contribution from $S$ is nonnegative, and its total mass is preserved
\[
 \int_\body P^{t-1}(\density_1\one_S)\dd\unif
 =\distr_1(S)\le a.
\]
and thus \cref{lem:exceptional-contraction} gives
\[
 \norm f_2^2\le
 \frac1{\dm a^{\dm-1}}\left(\frac{\circrad}{d}\right)^\dm
       e^{-2\gamma_{\mathrm{HR}}(t-1)}.
\]
Now suppose this squared norm is at most $\sampgood^2\sampbad/8$. Each of the sets
\[
 \{|f|>\sampgood/2\}
 \qquad\text{and}\qquad \{P^{t-1}(\density_1\one_S)>\sampgood/2\}
\]
has $\unif$-measure at most $\sampbad/2$ by Markov's inequality, and hence their union has measure at most $\sampbad$. Outside this union, the upper deviation of $\density_t$ from one is at most $\sampgood$, and the lower deviation is at most $\sampgood/2+a\le\sampgood$.

Finally, to bound the probability of sampling from the exceptional set, observe that
\[
 \distr_t(\bad_{t,\sampgood})
 \le\sampbad+\sqrt{\sampbad}\norm f_2+a
 \le\sampbad+\frac{\sampgood\sampbad}{\sqrt8}
                +\frac{\sampgood\sampbad}{4}
 \le2\sampbad.
\]
Taking logarithms shows that \cref{eq:exceptional-one-step-comparison} gives $\norm f_2^2\le \sampgood^2\sampbad/8$, as required. Substituting \cref{eq:exceptional-application-parameters-appendix} in this threshold and using \cref{eq:exceptional-KV-gap,eq:exceptional-poincare-radius} gives \cref{eq:exceptional-KV-application}.
%%%
% \begin{remark}
%       Note that the cutoff $a=\sampgood\sampbad/4$ enters the density bound to the power $1-\dm$. This is where the dimension enters the accuracy and stream-length terms in \cref{eq:exceptional-KV-application}.
% \end{remark}
\end{document}